\documentclass[preprint,12pt]{elsarticle}
\usepackage[utf8]{inputenc}
\usepackage{amsmath}
\usepackage{amssymb}
\usepackage{graphicx}
\usepackage[american]{babel}
\usepackage[amssymb]{SIunits}
\usepackage{subfigure}
\usepackage{multicol}
\usepackage{notoccite}
\usepackage{afterpage}
\usepackage{tikz}
\usepackage{ulem}
\usetikzlibrary{calc}
\usetikzlibrary{plotmarks}
\usepackage{etoolbox}
\usepackage{bm}
\usepackage{setspace}
\usepackage{float}
\biboptions{numbers,sort&compress}
\patchcmd{\normalsize}{13.6}{13}{}{}
\usepackage{array}

\begin{document}

\begin{frontmatter}
 
\title{Formation of solid-state dendrites under the influence of coherency stresses: A diffuse interface approach}
\author[label1]{Bhalchandra Bhadak\corref{correspondingauthor}}
\ead{bhalchandrab@iisc.ac.in}

\author[label2]{Tushar Jogi\corref{correspondingauthor}}
\ead{ms14resch11003@iith.ac.in}

\author[label2]{Saswata Bhattacharya}
\ead{saswata@iith.ac.in}

\author[label1]{Abhik Choudhury}%
\ead{abhiknc@iisc.ac.in}

\address[label1]{Department of Materials Engineering, Indian Institute of Science Bangalore- 560 012, India}
\address[label2]{Department of Materials Science and Metallurgical Engineering, Indian Institute of Technology Hyderabad, 
Sangareddy- 502285, India}

\cortext[correspondingauthor]{Corresponding authors}

\begin{keyword}
Phase field; misfit strain; anisotropy; solid-state dendrites 
\end{keyword}

\begin{abstract}
In this paper, we have formulated a phase-field model based on the 
grand-potential functional for the simulation of precipitate growth 
in the presence of coherency stresses. In particular, we study the development 
of dendrite-like patterns arising out of diffusive instabilities during the 
growth of a precipitate in a supersaturated matrix. Here, we characterize the 
role of elastic energy anisotropy and its strength on the selection of a 
dendrite tip radius and velocity. We find that there is no selection of a unique
tip shape as observed in the case of solidification, and the selection 
constant $\sigma^{*}=2d_0D/R_{tip}^{2}V_{tip}$ increases linearly with 
simulation time for all the simulation conditions (where $R_{tip}$ and 
$V_{tip}$ are the tip radius and velocity). Therefore, structures derived in 
solid-state in the presence of elastic anisotropy may only be referred to as 
dendrite-like. 

\end{abstract}

\end{frontmatter}
\doublespacing. 
\section{Introduction}
The formation of dendritic structures during solidification processes is a 
well-known phenomenon. Thermal supercooling in the case of pure metals and 
constitutional supercooling in the case of alloys have traditionally been used 
to explain the instability that gives rise to such structures, and a detailed 
review of the theory of dendritic growth during solidification may be found in 
the reviews~\cite{Kurz1984,langer1980,langer1987,kessler1988,pelce2012,kurz1990,
trivedi1994,trivedi1994solidification}.
Hence, while the study of dendritic growth during solidification has been 
studied quite extensively leading to an almost complete understanding of the 
parameters that lead to the selection of a unique 
dendritic tip radius and velocity, the same is not true for the case of 
solid-state dendritic growth, wherein the number of experimental observations of
such growth is itself limited. Yoo et al.~\cite{Henry1995} observed that during 
the precipitation of $\gamma'$-phase in Ni-base superalloys, 
after giving a specified heat treatment to the alloy, i.e., when it is cooled to 
the ageing temperatures very close to the solvus temperature, 
the coherent precipitates grow into dendritic structures, 
where the arms of the dendrite grow along the $\langle111\rangle$ directions.
Doherty~\cite{Doherty1983} discussed various factors that appear to explain the 
shape instabilities in some solid-state reactions, such as the solid-state 
dendrites of $\gamma$-brass precipitates in the $\beta$-brass matrix in the Cu-Zn 
alloy. Husain et al.~\cite{Husain1999} reported that the precipitation of the
$\gamma_2$-phase in Cu-Al $\beta$-phase alloys yield dendritic morphologies. It 
is also observed that the rapid bulk diffusion and fast interfacial 
reaction kinetics would promote the formation of such morphologies. 
They have predicted the occurrence of dendritic morphologies during solid-state 
precipitation in many binary systems such as Cu-Zn, Cu-Al, Ag-Al, Ag-Zn, Cu-Ga, 
Au-Zn, Ni-Zn, Cu-An, Ag-Cd, and Cu-Sn. 
The authors attribute crystallographic similarities between the parent phase and
the precipitates to be the dominating factor that gives rise to the formation of
the dendritic structures. In a later study, Yoo et al.~\cite{Yoo2005} have 
observed spherical precipitates changing into flower-like structures as the 
supersaturation in the matrix is increased. They have shown that the results are
concurrent with the Mullins-Sekerka theory, where they 
comment that the morphological instabilities should be evident if the point 
effect of the diffusion is significant in the vicinity of the precipitate-matrix
interface regardless of the nature of the matrix phase and the ageing stage.
Khan et al.~\cite{AKhan2004} have also noticed the formation of solid-state 
dendritic structures during the elevated temperature treatment of maraging 
steel. There, the oxy-nitride phase acquires dendritic morphology in the 
martensitic matrix.

At a fundamental level, the presence of a supersaturation in the matrix makes 
the interface unstable to diffusive instabilities of the Mullins-Sekerka type, 
and therefore the observation of such dendritic structures during solid-state 
reactions is not surprising. However, there is an important difference 
from the case of solidification, wherein the solid-solid interface 
in precipitation reactions is typically coherent in the initial stages of 
precipitation, and thereby the precipitate and the matrix are also 
elastically stressed. Since the elastic-effects scale with the size of the 
precipitate, the presence of coherency will therefore naturally influence the 
selection of length scales during dendritic evolution.

Leo and Sekerka~\cite{Leo1989} have extended the linear stability analysis 
performed by Mullins and Sekerka~\cite{Mullins1963} by including the influence 
of elastic coherency stresses and have investigated the morphological stability 
of the precipitate grown from the solid solution. 
The analysis shows that the elastic-effects manifest through the boundary 
conditions based on local equilibrium that determines the compositions 
at a coherent solid-solid interface. The elastic-effects increase in importance 
relative to capillarity as the scale of the system increases, and so will become
very important at small initial supersaturation, 
as the supersaturation is inversely proportional to the critical nucleus radius. 
Elastic fields can also influence the selection of the fastest-growing harmonic,
with stabilizing elastic fields favoring long-wavelength harmonics while 
destabilizing elastic fields favoring short-wavelength
harmonics. The stability analysis reveals the dependence of the different 
elastic parameters in either amplification or decay of the perturbations at the 
solid-solid interface.  

While such a linear stability analysis gives an idea about the characteristics of
the diffusive instability, they cannot explain the selection of a unique
dendritic tip radius and velocity, as is known from solidification experiments. 
The selection requires the presence of anisotropy in either the interfacial 
energy or the attachment kinetics, and the selected growth shape of the dendrite 
may be predicted using the well established microsolvability 
theory~\cite{kessler1986,kessler1986steady,barbieri1987,barbieri1989}. 
Formulation of such a theory is presently lacking for solid-state
dendritic growth, and a possible strategy here is the use of dynamical 
phase-field simulations that naturally lead to the selection of a unique tip 
radius and velocity. Previously, Greenwood and coworkers~\cite{greenwood2009} have 
proposed a phase-field model for microstructural evolution which includes the 
effects of the elastic strain energy. The authors show that the solid-state 
dendritic structures undergo a transition from surface energy anisotropy 
dominated $\langle 10 \rangle$ growth directions to elastically driven 
$\langle 11 \rangle$ growth directions induced by 
changes in the elastic anisotropy, the surface anisotropy, 
and the supersaturation in the matrix. Our work, in this paper, 
will be an extension of this investigation where we will study  
dendritic tip selection using a phase-field model as a function of 
the different elastic and interfacial parameters as well as the 
supersaturation. Similar to dendrites formed during the 
solidification experiments, our studies will aim at identifying the 
variation of the classical selection constant ($\sigma^{*}$) as a 
function of the material and process parameters, and in particular, 
the strength of anisotropy in the stiffness matrix and the 
interfacial energies. Thus, our motivation in this paper
is to determine the effect of several physical properties 
on the formation of the solid-state dendritic structures 
and characterize the influence on the selection constant 
($\sigma^{*}$). Here, we will utilize a diffuse interface 
formulation where we solve the Allen-Cahn equation along with the 
evolution of the diffusion potential of the relevant species that 
ensure mass conservation. While solving the evolution equations 
of the order parameter and the diffusion potential, 
we ensure that the mechanical equilibrium is satisfied
in all parts of the system. The simulations show the evolution 
of a precipitate morphology in the supersaturated matrix 
with imposed anisotropies in the elastic energy as well as the 
interfacial energy. We will systematically study the effect of 
change in different properties of the phases on the selection of radius of 
the dendrite tip and velocity. 

\section{Model formulation}
We formulate a phase-field model based on a grand-potential 
functional which includes the interfacial properties as well as the 
thermodynamics of the bulk phases in the system,
\begin{align}
  \Psi(\mu,\mathbf{u},\phi) &= \int_V \Bigg[\gamma W a^{2}\left(\mathbf{n}\right)|\nabla \phi|^{2} 
		   + \dfrac{1}{W} w\left(\phi\right) + \psi(\mu,\phi) \Bigg] dV \nonumber\\
		  &+ \int_V f_{el}\left(\mathbf{u},\phi\right)dV,
\end{align}
where $V$ is the total volume of the system undergoing 
solid-state phase transformation. 
The phase-field order parameter $\phi$ that describes the presence 
and absence of precipitate ($\alpha$ phase) and matrix ($\beta$ 
phase), acquires value $\phi=1$ in the precipitate phase 
whereas $\phi=0$ in the matrix phase. 
The double obstacle potential which writes as, 
\begin{align}
    w\left(\phi\right) &= \dfrac{16}{\pi^2} \gamma \phi\left(1-\phi\right) \qquad  \phi \in [0,1],\nonumber\\
                           &= \infty \qquad \qquad \qquad \textrm{otherwise},
\end{align}
provides a potential barrier between two phases. Here, the term 
$\gamma$ controls the interfacial energy density, while $W$ 
influences the diffuse interface width separating the precipitate and
the matrix phases. Anisotropy in the interfacial energy is 
incorporated using the function $a\left(\mathbf{n}\right)$, which 
further writes as~\cite{karma1998}:
\begin{align}
 \gamma &= \gamma_0 a(\mathbf{n}), \nonumber \\
 \gamma &= \gamma_0 \left(1 - \varepsilon\left(3 -4\dfrac{{\phi}^{'4}_{x} + {\phi}^{'4}_{y}}{({\phi}^{'2}_{x} 
	    + {\phi}^{'2}_{y})^2} \right) \right),
\label{Eqn_anisotropy}
\end{align}
where $\varepsilon$ is the strength of anisotropy in the interfacial 
energy, $a(\mathbf{n})$ is the anisotropy function
of the interface normal, $\mathbf{n}=-\dfrac{\nabla\phi}{|\nabla\phi|}$.
$\psi(\mu,\mathbf{u},\phi)$ is the grand potential density that 
is a function of the diffusion potential $\mu$. Finally, 
$f_{el}\left(\mathbf{u},\phi\right)$ is a function of the 
displacement field $\mathbf{u}$.

By taking the variational derivative of the functional with respect 
to the phase-field order parameter, using Allen-Cahn dynamics, 
we get the evolution of order parameter $\phi$, which 
writes as,
\begin{align}
 \tau W \dfrac{\partial \phi}{\partial t} &= - \dfrac{\delta F}{\delta \phi},
\end{align}

 \begin{align}
  \tau W \frac{\partial\phi}{\partial t} &= 2 \gamma W \nabla \cdot \left[a\left(\mathbf{n}\right)
					      \left[\dfrac{\partial a\left(\mathbf{n}\right)}{\partial \nabla \phi}|\nabla \phi|^{2} 
						  + a\left(\mathbf{n}\right) \nabla \phi\right]\right] \nonumber\\
                                                &- \dfrac{16}{\pi^2} \dfrac{\gamma}{W} \left(1-2\phi\right) 
						  - \Delta\psi \frac{\partial h(\phi)}{\partial\phi} 
						  + \dfrac{\partial f_{el}\left(\mathbf{u},\phi\right)}{\partial \phi},
 \label{phi_evolve_dend}
 \end{align}
where $\tau$ is a relaxation constant for the evolution of 
$\phi$. The last two terms in Eqn.~\ref{phi_evolve_dend} 
contributes to the driving force necessary for the 
evolution of the precipitate phase. Here the term $\Delta \psi$ 
is the difference between the grand-potential densities of the 
$\alpha$ and the $\beta$ phases, and the corresponding term in the
update of the order parameter derives from 
$\psi\left(\mu, \phi\right)
=\psi_\alpha\left(\mu\right)h\left(\phi\right) + 
\psi_\beta\left(\mu\right)(1-h\left(\phi\right))$, where 
$h(\phi) = \phi^2 (3 - 2\phi)$ is an interpolation polynomial.
Further, the difference of the grand-potential densities between 
the phases may be written as a function of the departure of the 
diffusion potential from its equilibrium values as, 
$\psi_\alpha-\psi_\beta = \left(c^{\alpha}_{eq} - 
c^{\beta}_{eq}\right)\left(\mu-\mu_{eq}\right)$, 
where $c^{\alpha}_{eq}$ and $c^{\beta}_{eq}$ are the equilibrium 
compositions of the bulk precipitate and matrix phases 
respectively. We note that this is the driving force at leading 
order for an arbitrary description of the free-energies while it 
is the exact driving force for a situation where the free-energy 
composition relations of the matrix and precipitate are 
parabolic with equal curvatures.

The second term in the driving force for phase transition 
(Eqn.~\ref{phi_evolve_dend})  
comes from the derivative of the elastic free-energy density that
writes as, 
\begin{align}
  f_{el}(\mathbf{u}, \phi) &= \dfrac{1}{2}C_{ijkl}(\phi)(\epsilon_{ij} - \epsilon^*_{ij}(\phi))(\epsilon_{kl} 
			    - \epsilon^*_{kl}(\phi)),
\end{align} 
where the total strain can be computed from the displacement 
field $\mathbf{u}$, which writes as,
\begin{align}
 \epsilon_{ij} &= \frac{1}{2}\left(\frac{\partial u_i}{\partial x_j} 
 + \frac{\partial u_j}{\partial x_i}\right)
 \label{strain},
\end{align}
while the elastic constants $C_{ijkl}$ and misfit-strain 
$\epsilon^*_{ij}$ can be expressed as,
\begin{align}
 C_{ijkl}(\phi) &= C^{\alpha}_{ijkl}\phi + C^{\beta}_{ijkl}(1-\phi), \nonumber\\
 \epsilon^*_{ij}(\phi) &= \epsilon^{* \alpha}_{ij}\phi + \epsilon^{* \beta}_{ij}(1-\phi). 
\end{align}
To simplify the equations, without any loss of generality, we 
additionally impose that the misfit-strain exists only in 
the precipitate phase ($\alpha$ phase), 
which makes $\epsilon^{* \beta}_{ij}=0$.
Thereafter, the elastic energy density can be recast as,  
\begin{align}
 f_{el}(\phi) &= Z_3 (\phi)^3 + Z_2 (\phi)^2 + Z_1 \phi + Z_0,
 \label{fel_phi_dend}
\end{align}
where, in Eqn.~\ref{fel_phi_dend}, we segregate the terms in 
powers of $\phi$, i.e., $Z_3$, $Z_2$, $Z_1$, and $Z_0$. Each pre-factor is a 
polynomial of $\phi$, elastic constants, and the 
misfit strains. The expansion of these pre-factors is illustrated
in the~\ref{appendix_f_elast}. 
Therefore, the term corresponding to the elastic
energy in the evolution equation of the order parameter can be 
computed as, 
\begin{align}
 \dfrac{\partial f_{el}\left(\mathbf{u},\phi\right)}{\partial \phi} 
 &=  3Z_3 (\phi)^2 + 2Z_2 (\phi) + Z_1. 
\end{align}
Eqn.~\ref{phi_evolve_dend} is solved along with the update of 
the diffusion potential that follows the equation,
    \begin{align} 
	 \dfrac{\partial \mu}{\partial t} 
	= \left(\sum_\alpha h(\phi) \dfrac{\partial c_\alpha}{\partial \mu}\right)^{-1}
	\Bigg[\nabla \cdot M \nabla \mu - \sum_\alpha c_\alpha(\mu) \dfrac{\partial h}{\partial \phi} 
	  \dfrac{\partial \phi}{\partial t}\Bigg],
\label{chem_pot_evolve}
\end{align}
where $M$ is the atomic mobility that is explicitly related to 
the diffusivity $D$ as $D\dfrac{dc}{d\mu}$. For our calculations,
we assume a parabolic description $f\left(c\right)$ of the matrix
and precipitate phases with equal leading order terms 
$A = (1/2)\dfrac{\partial^{2}f}{\partial c^{2}}$, 
thereby the mobility becomes $\dfrac{D}{2A}$.
Finally, we compute the displacement field as a function of the 
spatial distribution of the order parameter, which is utilized 
to calculate the magnitude of the strain field across the domain.
Thus, we solve the damped wave equation iteratively which writes 
as, 
\begin{align}
  \rho\frac{d^2\mathbf{u}}{dt^2} + b\frac{d\mathbf{u}}{dt} &= \nabla 
  \cdot \boldsymbol{\sigma}. 
  \label{mech_equilibrium}
\end{align}
Eqn.~\eqref{mech_equilibrium} is solved until the equilibrium is reached, 
i.e., $\nabla \cdot \boldsymbol{\sigma}=\mathbf{0}$. 
The terms $\rho$ and $b$ are chosen 
such that the convergence is achieved in the fastest possible time. 

\section{Parameter initialization}
In this section, we list out the material parameters that will be 
used in the subsequent sections. We use a non-dimensionalization 
scheme where the energy density scale is derived from the magnitude 
of the shear modulus $1\times10^{9}\,J/m^{3}$, while the interfacial 
energy scale is given by $1\,J/m^{2}$. Dividing the interfacial energy  
scale with the  energy density scale sets the length 
scale of $l^{*}=1\,nm$. In this work, we report all the parameters
in terms of non-dimensional units. The anisotropy in the elastic 
energy is introduced through the Zener anisotropy parameter, i.e., 
$A_z$, which in turn modifies the magnitudes of the elastic constants,
that are $C_{11}=C_{1111}$, $C_{12}=C_{1122}$, and $C_{44}=C_{1212}$. 
These elastic constants can further be elaborated in terms of the 
shear modulus ($\mu$), Poisson's ratio ($\nu$), and Zener anisotropy 
parameter ($A_z$), which controls the evolution and orientation of 
the instability.
\begin{align}
  C_{44} = \mu, \quad
  C_{12} = 2\nu\left(\frac{\mu}{1-2\nu}\right), \quad
  C_{11} = C_{12} + \frac{2C_{44}}{A_z}.
\end{align}
In 2D, for values of $A_z$ greater than unity, 
$\langle10\rangle$ directions are elastically softer 
and $\langle11\rangle$ directions are elastically harder, 
whereas in 3D $\langle100\rangle$ directions 
are softer, while $\langle111\rangle$ directions are harder.
However, if $A_z$ is less than unity, elastically soft (hard) directions are 
$\langle11\rangle$ ($\langle10\rangle$) in 2D, and 
$\langle111\rangle(\langle100\rangle)$ 
in 3D. For all the cases, the precipitate and the matrix have 
the same magnitude of $A_z$. Unless otherwise specified, all results are 
produced with $\mu_{mat} =\mu_{ppt} =100$, $\nu_{mat}=\nu_{ppt}=0.3$, 
and $A_z$ varying between 2 and 3, typically observed in Ni-based superalloys. 
Here, the misfit strain or eigenstrain ($\epsilon^*$) 
is dilatational, i.e., the magnitude of the misfit strain is the same along 
all the principal directions while the off-diagonal terms are zero. 
The magnitude of misfit strain is varied in the range from 0.5 to 1\%.
The simulation is initialized with a precipitate of initial radius, $R_0=40$, 
that is placed at the center of the domain. The simulation domain obeys 
periodic boundary conditions. The size of the domain is chosen such that 
the diffusion fields of the neighboring precipitates do not interact with each 
other. The sizes are $3165d_0$ and 
$1099 d_0$ for 2D and 3D, respectively, where $d_0$ is the capillary length. 
Equilibrium compositions in the bulk precipitate 
$(c^{\alpha}_{eq})$ and the matrix $(c^{\beta}_{eq})$ are chosen to be 0.78125 and 0.5, 
respectively. 

In the following, we perform systematic simulation studies in both 2D and 3D 
in order to investigate the influence of supersaturation, misfit strain, and 
anisotropy strengths on the development of dendrite-like instabilities. 
The 2D simulations are performed using a finite-difference discretization on 
a regular grid and the code is written C and parallelized using MPI for running 
on several CPUs. Since the 3D simulations are more computationally intensive, 
we utilize the Fourier-spectral method that allows a quicker solution to the 
equations of mechanical equilibrium~\cite{SBhattacharyya2009}. 
The simulations are run on NVIDIA Tesla V100 GPUs using the 
optimized CUDA-based spectral solver for the governing 
equations~\eqref{phi_evolve_dend},~\eqref{chem_pot_evolve}, 
and~\eqref{mech_equilibrium}. 

\section{Evolution of the precipitate into dendrite-like shape}
\begin{figure}[!htbp]
 \centering
 \subfigure[]{\includegraphics[width=0.4\linewidth]{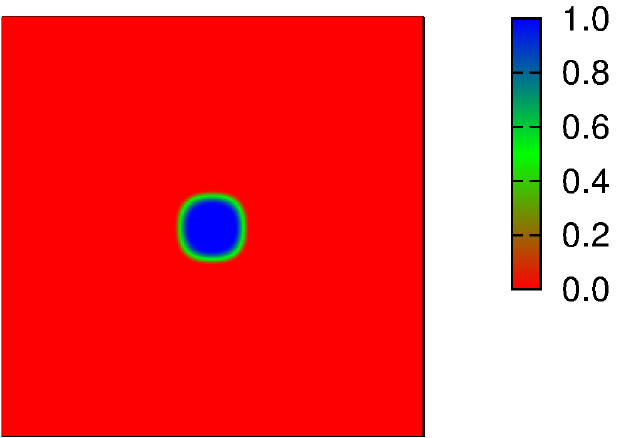}
 \label{phi_1k}
 }
 \centering
\subfigure[]{\includegraphics[width=0.4\linewidth]{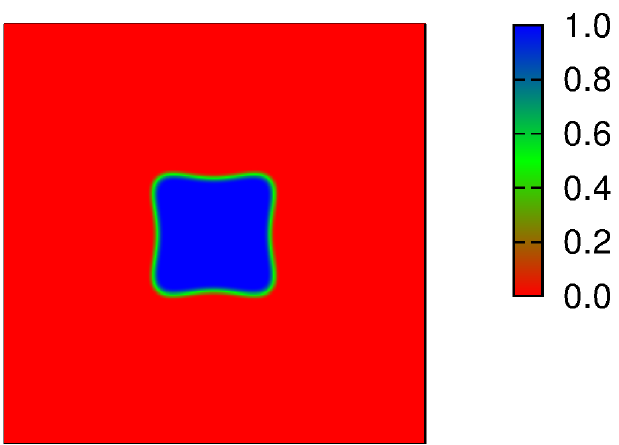}
\label{phi_10k}
 }
 \centering
 \subfigure[]{\includegraphics[width=0.4\linewidth]{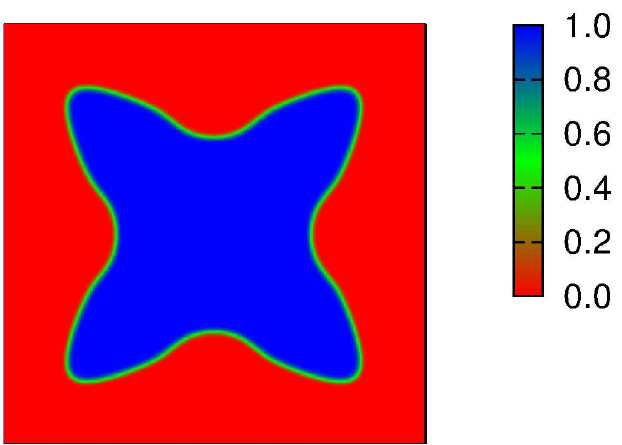}
 \label{phi_50k}
 }
 \centering
\subfigure[]{\includegraphics[width=0.4\linewidth]{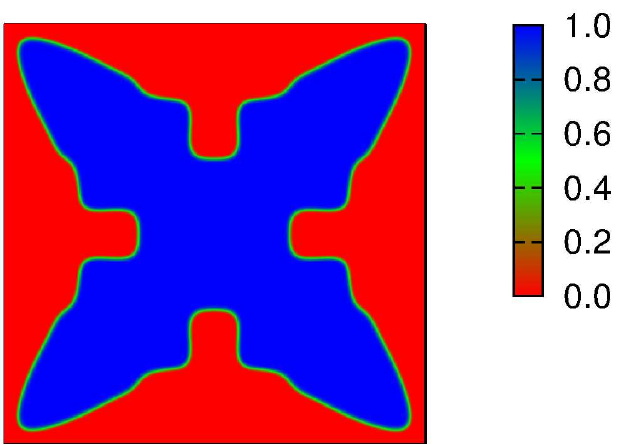}
\label{phi_150k}
 }
 \centering
\subfigure[]{\includegraphics[width=0.7\linewidth]{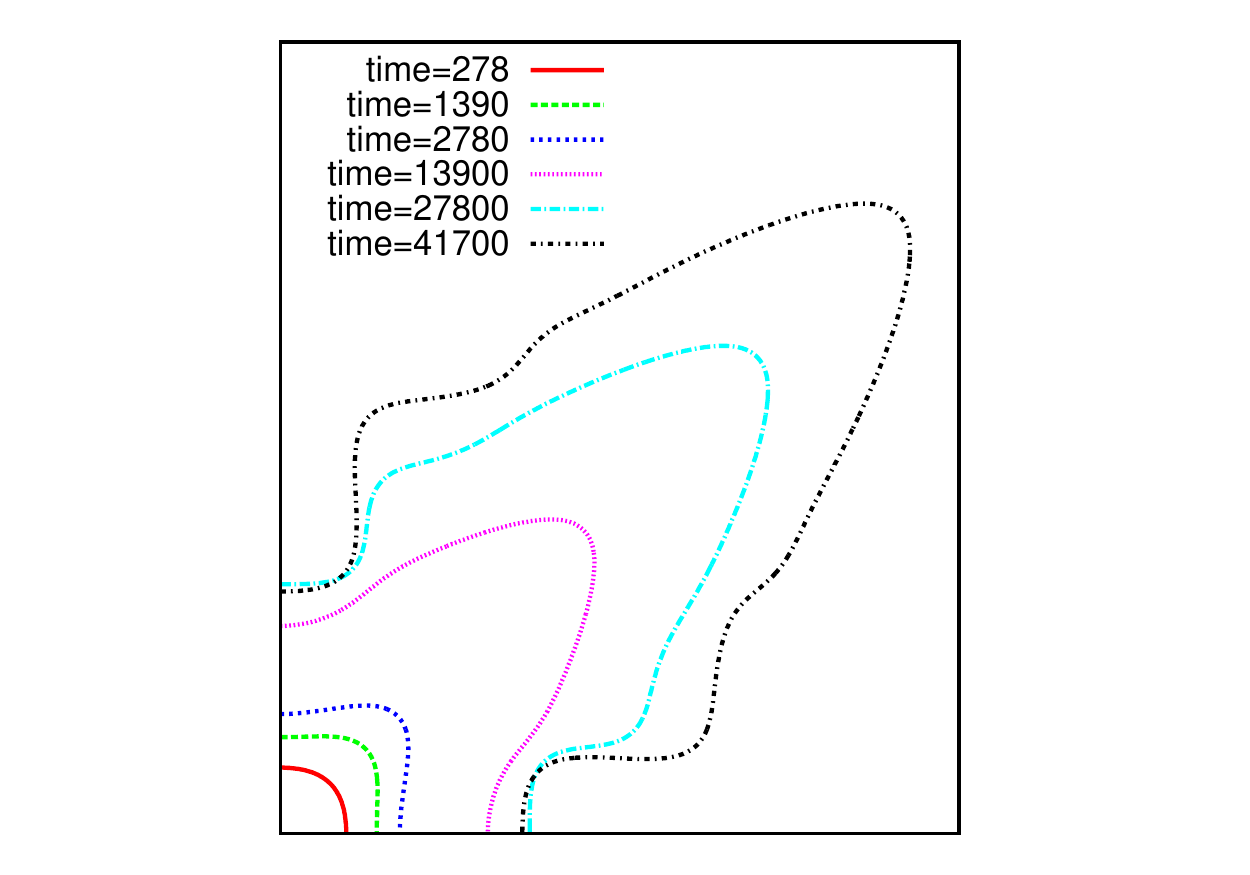}
\label{all_phi_profile}
 } 
\caption{Evolution of the precipitate into dendritic structure for 
$A_z=3.0$, $\epsilon^*=1\%$ shown as time snapshots at normalized times: (a)278,
(b)2780, (c)13900, (d)41700. (e) Contour plots of a one-fourth section of the 
evolving dendrite at different times.}
\label{evolution_dend_time}
\end{figure}

We begin by describing the shape evolution of the precipitate
giving rise to dendrite-like shapes.  
Fig.~\ref{evolution_dend_time} shows such an exemplary simulation 
in 2D where the matrix supersaturation ($\omega$) is 53\%, the misfit 
at the interface is $1\%$, Zener anisotropy parameter $3.0$, 
and the interfacial energy is isotropic. We note that experimental 
observations show dendritic structures usually at a lower supersaturation, 
which leads to larger inter-precipitate distances allowing 
free development of the instabilities without overlap of the 
composition and the elastic fields.

In the simulations, however, we are constrained
by having to resolve the large elastic and composition boundary layers.
Therefore, in order to derive results in a reasonable time,
we have performed simulations for higher supersaturations 
in comparison to experimentally observed situations. 
However, this should not alter the results that we 
derive from the simulations and should be generically 
applicable for interpreting experimentally observed 
microstructures and observed trends in the change of the 
morphologies upon alteration of the processing conditions or the material 
properties.

For 2D simulations, the precipitate with an initial circular shape 
grows into a square-like shape with rounded corners, where its faces are normal to 
$\langle10\rangle$ directions. In the absence of interfacial energy anisotropy, 
the precipitate shape and the eventual instabilities of the interface are determined 
by the anisotropy in the elastic energy. Further, with an increase in the size, the 
precipitate corners become sharper. Eventually, due to the point effect of diffusion, the
corners of the precipitate grow faster (along the $\langle11\rangle$ directions) 
compared to its faces, which further gives rise to concavities on 
the precipitate faces, as shown in Fig.~\ref{phi_50k}-~\ref{phi_150k}.
We note that this shape evolution is indeed due to a diffusive instability similar 
to the Mullins-Sekerka instability leading to the formation of dendrites during 
solidification for anisotropic interfacial energies, where the perturbations
of the composition boundary layer ahead of the interface provide an amplifying 
feedback to the interface velocity leading to the development of the instability.

\begin{figure}[!htbp]
 \centering
 \subfigure[]{\includegraphics[width=0.45\linewidth]{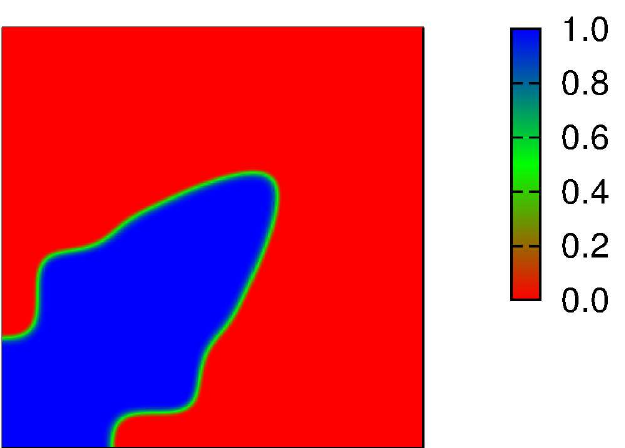}
 \label{pf_dend}
 }
  \centering
 \subfigure[]{\includegraphics[width=0.45\linewidth]{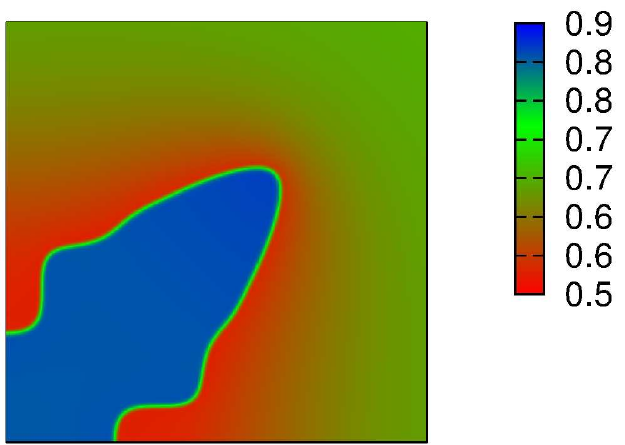}
 \label{comp_dend}
 }
 \caption{(a) Phase field and (b) composition field of a one-fourth section of 
 the precipitate growing into a dendritic structure. Here, supersaturation is 53\%, misfit 
 strain is 1\%, and Zener anisotropy parameter is 3.0.}
 \label{phase-field-comp-field}
\end{figure}

Among the different possible perturbation modes, the elastic energy anisotropy 
determines the eventual shape of the precipitate during growth giving rise to 
dendrite-like shapes. Fig.~\ref{all_phi_profile} shows the contour 
plot of phase-field profiles at $\phi=0.5$, i.e., the precipitate-matrix 
interface, which is plotted as a function of increasing time. 
The composition field around the precipitate also evolves 
as a function of time which is captured in Fig.~\ref{phase-field-comp-field}.


In 3D, the precipitate first develops into a cube-like shape with rounded 
corners where the faces are normal to $\langle100\rangle$ directions. 
Table~\ref{tab:prec_evolve} depicts the evolution of the precipitate 
into a dendrite-like morphology. The left panel shows the temporal evolution 
of the precipitate morphology (represented using isosurfaces drawn at 
$\phi = 0.5$) for normalized time 
$t = 4609, \, 13827,\,$ and $23045$. In addition, adjacent to the isosurface 
representation, we present the corresponding precipitate morphology in the $(110)$ 
plane passing through the center of the box.  
As a result of the point effect of diffusion, 
the precipitate starts developing ears along $\langle111\rangle$ 
directions (see $t = 4609$) i.e. protrusions along $\langle111\rangle$ 
directions and depressions along $\langle110\rangle$ and $\langle100\rangle$ 
directions. Further ahead in time, the ears develop into prominent primary 
dendrite arms, whereas concavities develop on the faces of the cube. 
The composition evolution during the development of the dendritic structure 
is revealed in the last column of Table~\ref{tab:prec_evolve}.

\begin{table}[htbp]
    \begin{tabular}{ m{2.0cm} m{2.75cm} m{4cm} m{4cm} }
         &  Isosurface view & \hspace{0.9cm} $\phi$ map 
         &  composition map \\
         $t = 4609$ &  
         \includegraphics[width=2.8cm]{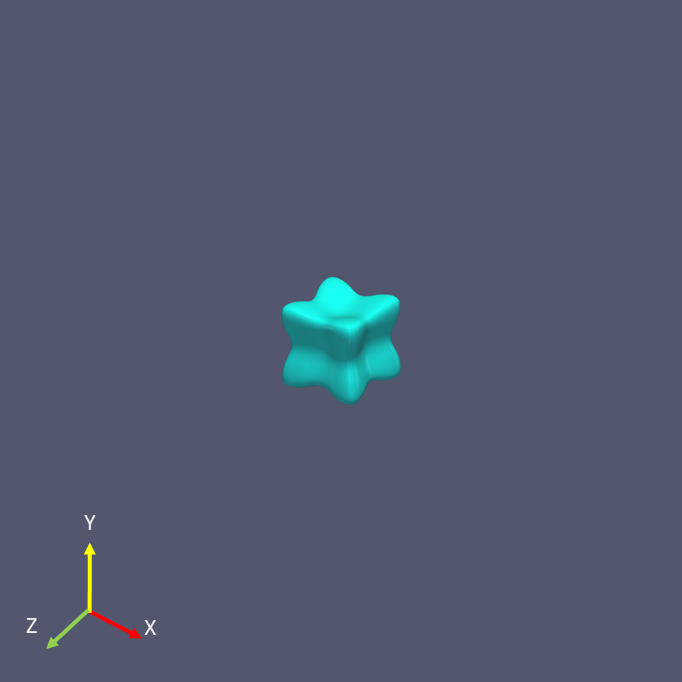}&
         \includegraphics[width=3.8cm]{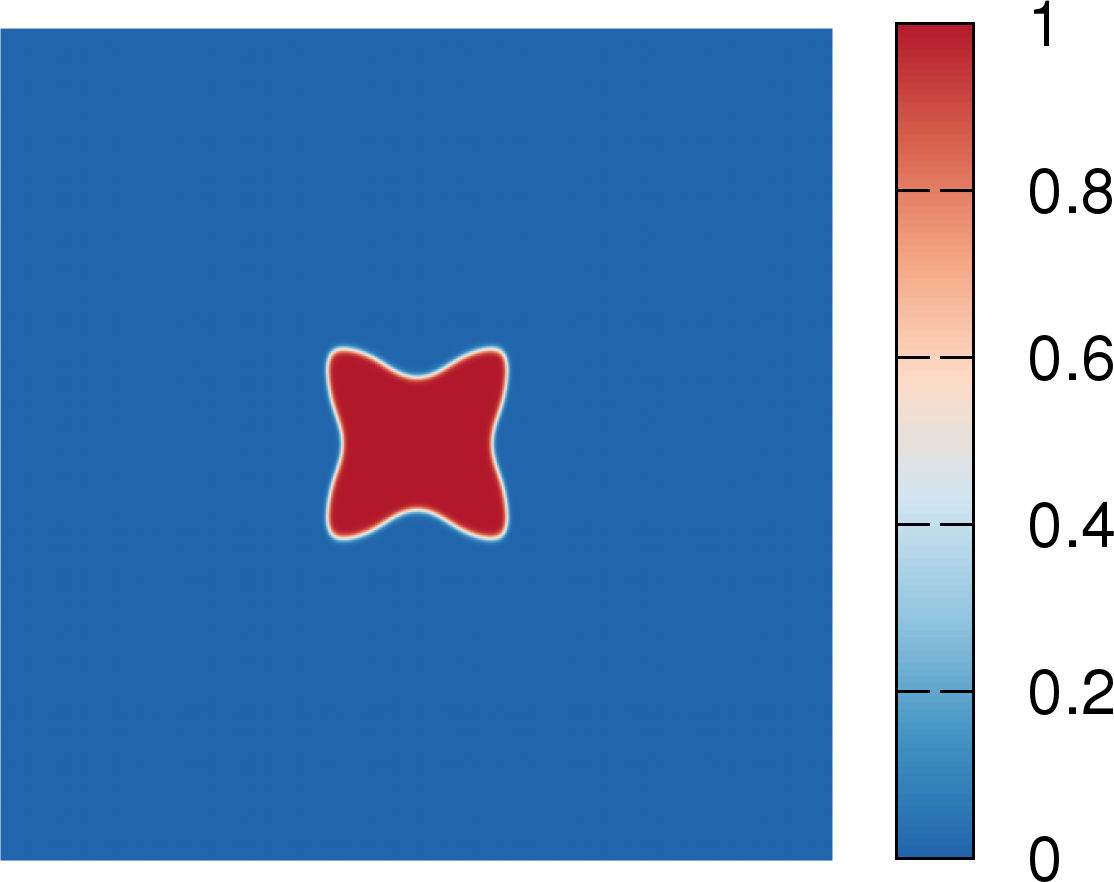} & 
         \includegraphics[width=3.8cm]{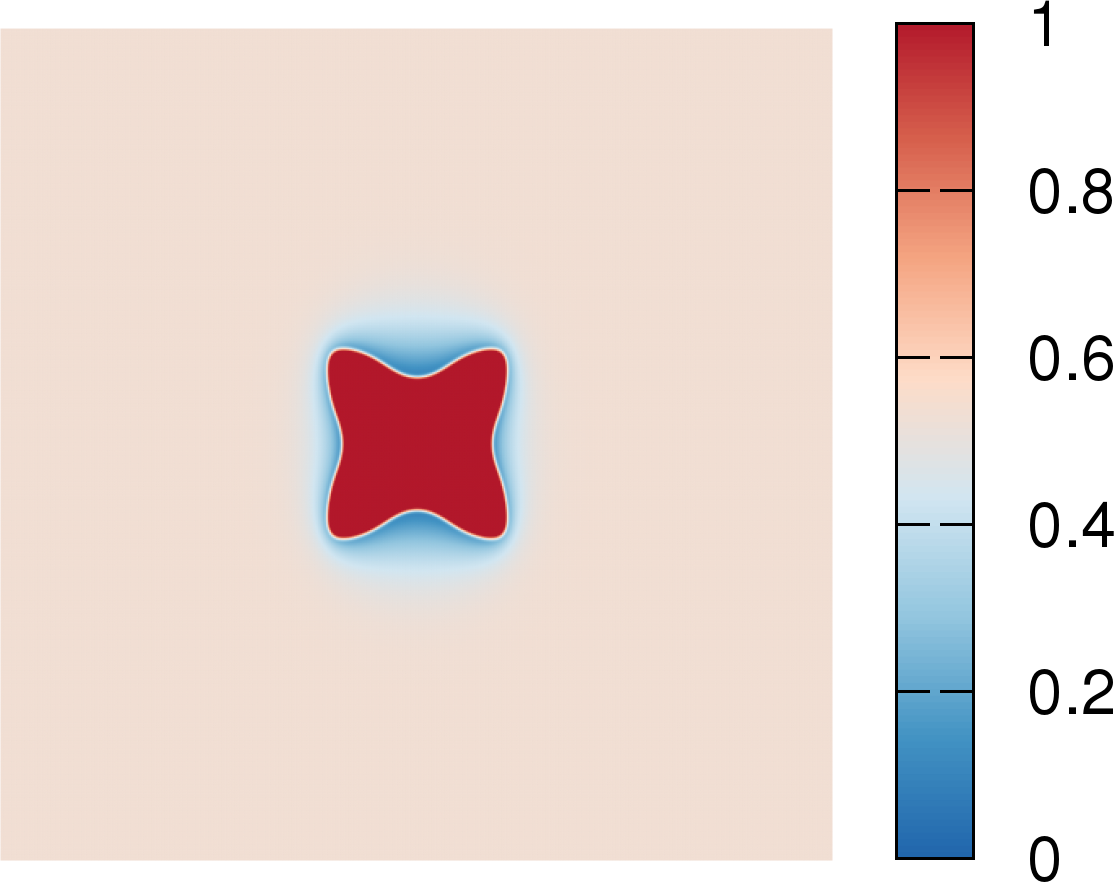}\\
         $t = 13827$ &  
         \includegraphics[width=2.8cm]{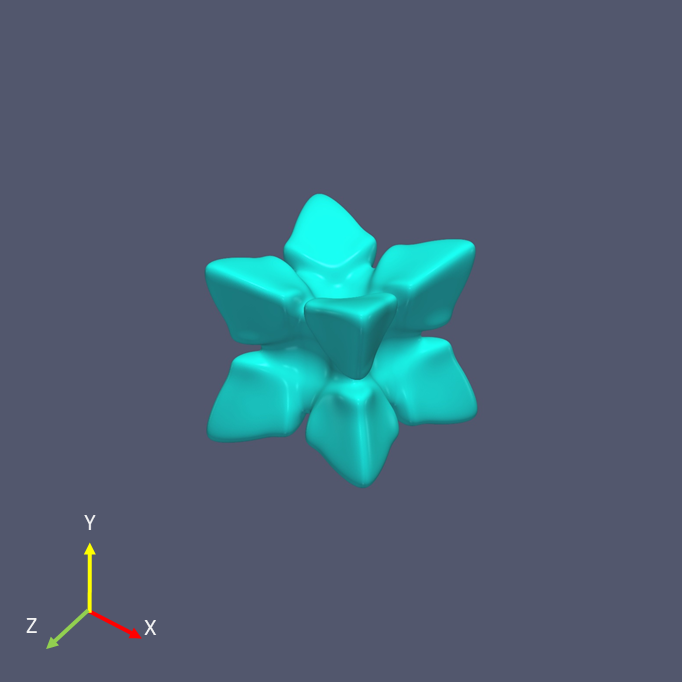}&
         \includegraphics[width=3.8cm]{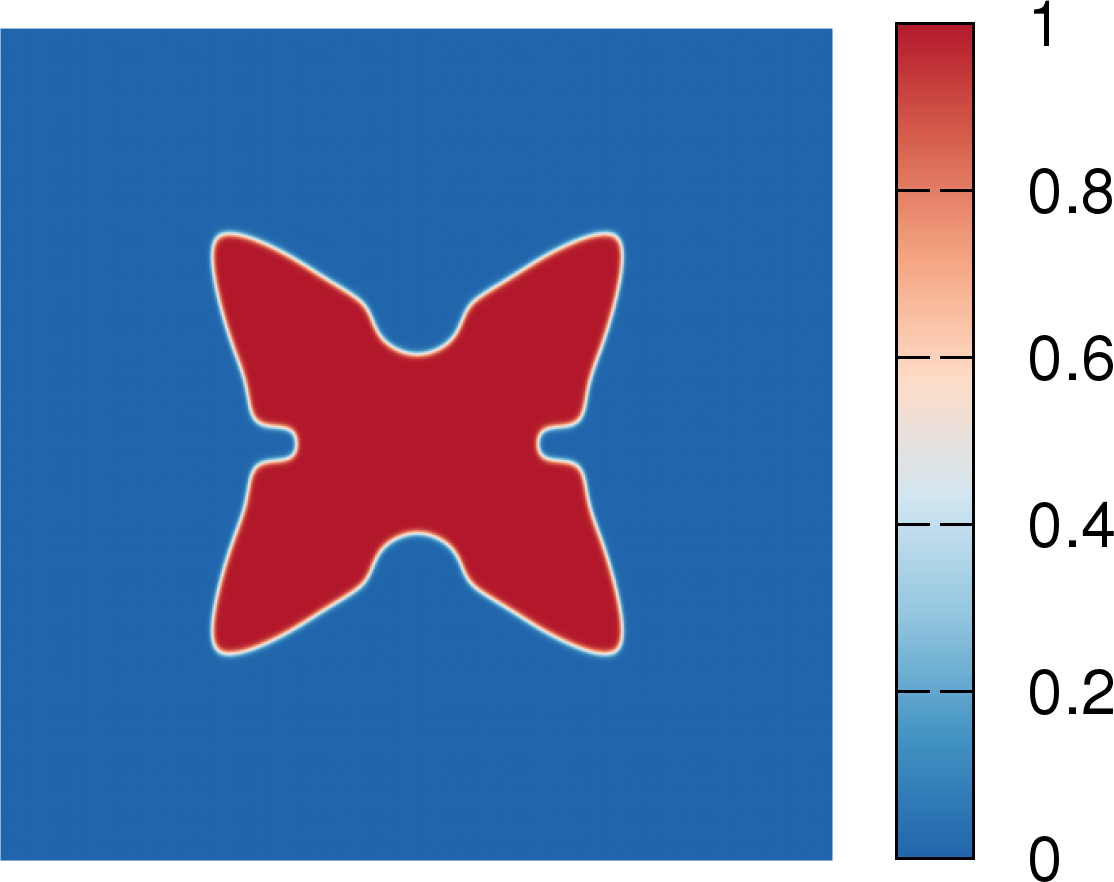} & 
         \includegraphics[width=3.8cm]{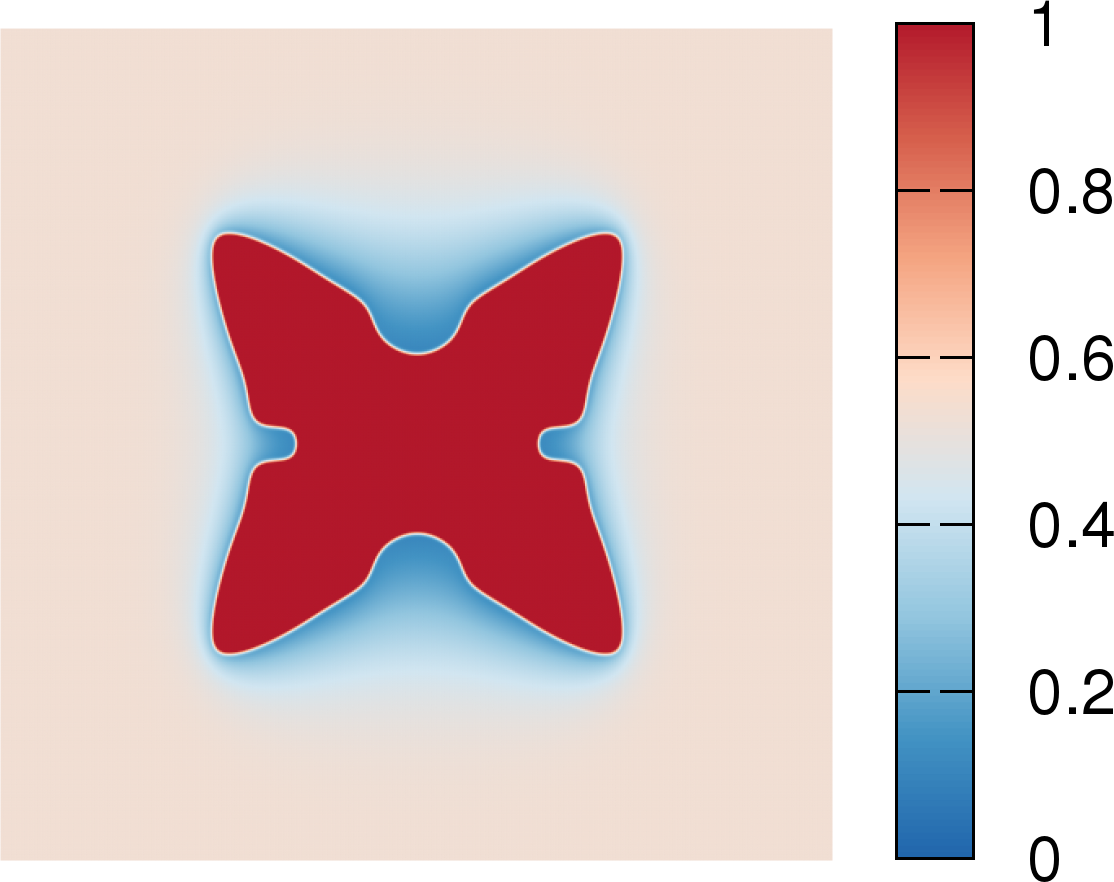}\\
         $t = 23045$ &  
         \includegraphics[width=2.8cm]{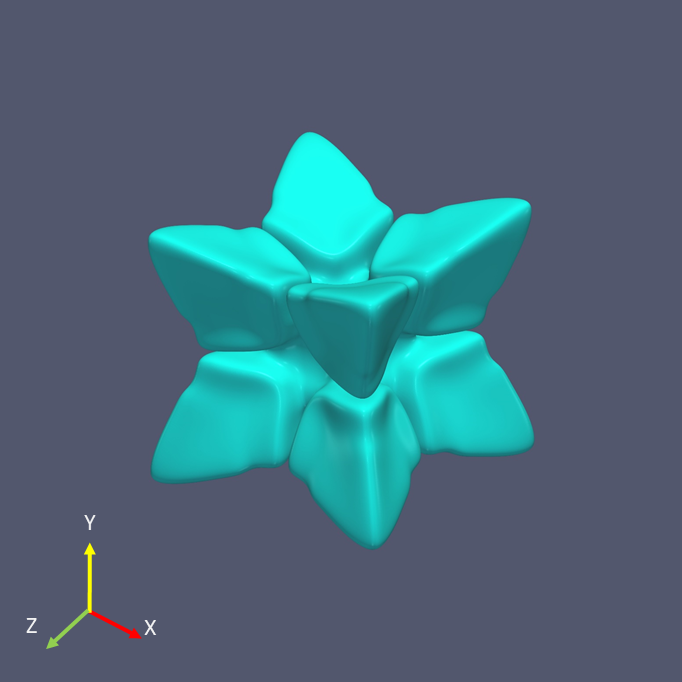}&
         \includegraphics[width=3.8cm]{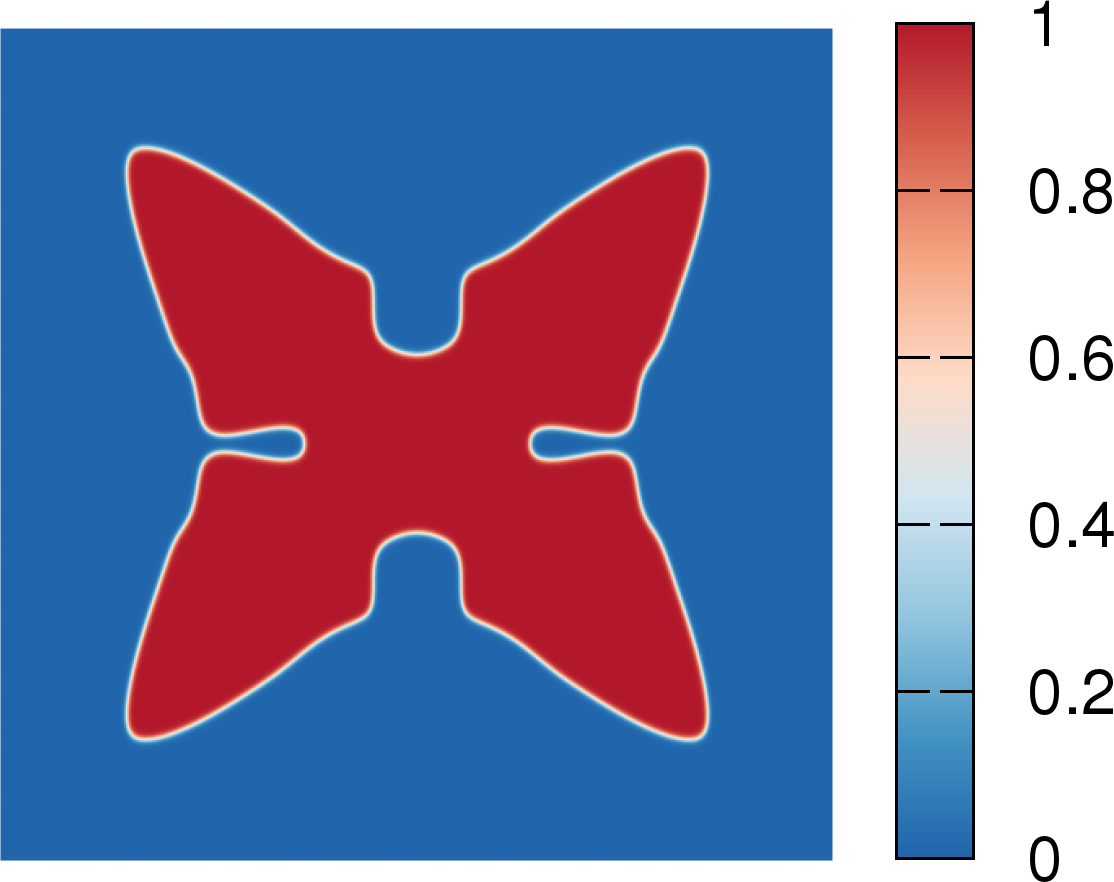} &
         \includegraphics[width=3.8cm]{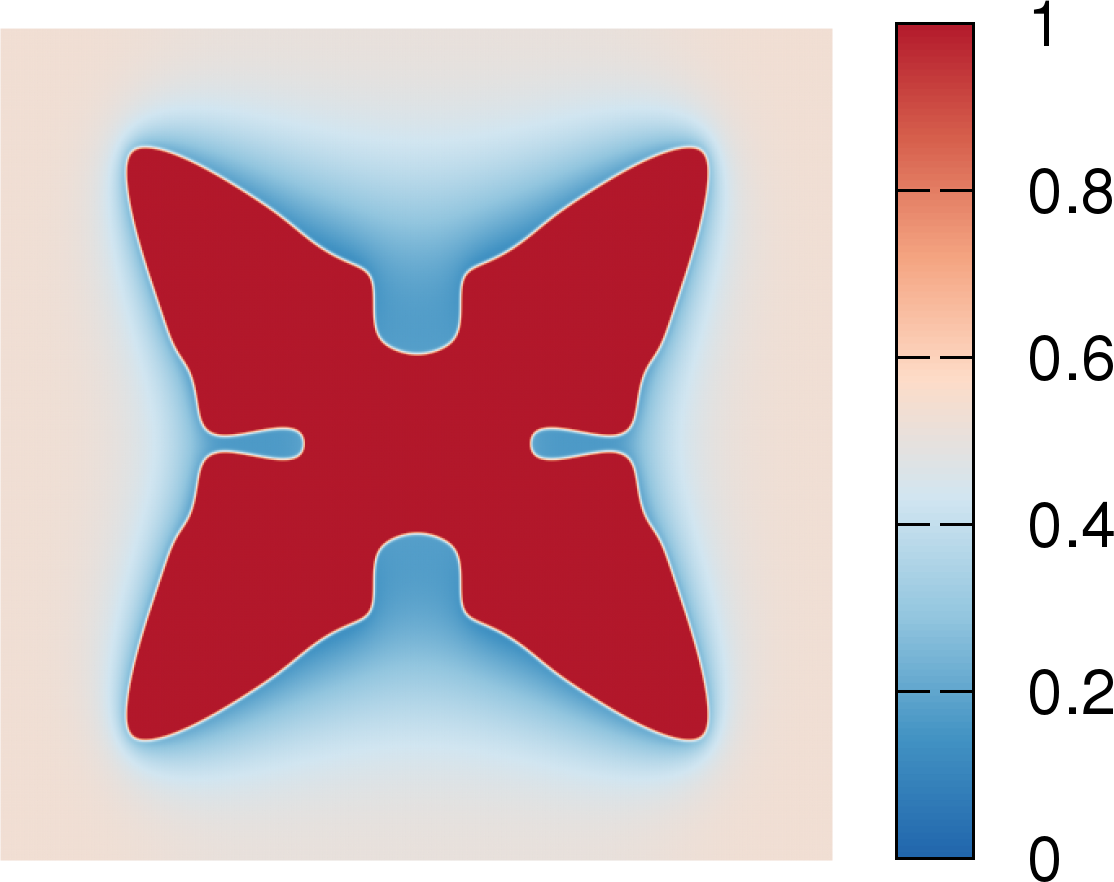}\\
    \end{tabular}
    \caption{Temporal evolution of a 3D dendritic morphology represented by time snapshots
    at $t = 4609$, $13827$ and $23045$. First column shows isosurfaces of the dendrites
    drawn at $\phi=0.5$. Second column shows the corresponding sections of the evolving dendrite
    on a $(110)$ plane passing through the center of the simulation box. 
    Last column shows the corresponding composition profiles on the same plane.  
    Here, we use misfit strain $\epsilon^{*} = 1\%$, supersaturation $\omega = 53\%$, and Zener anisotropy parameter $A_z = 3$.}
    \label{tab:prec_evolve}
\end{table}

\begin{figure}[!htbp]
 \centering
\includegraphics[width=0.5\linewidth]{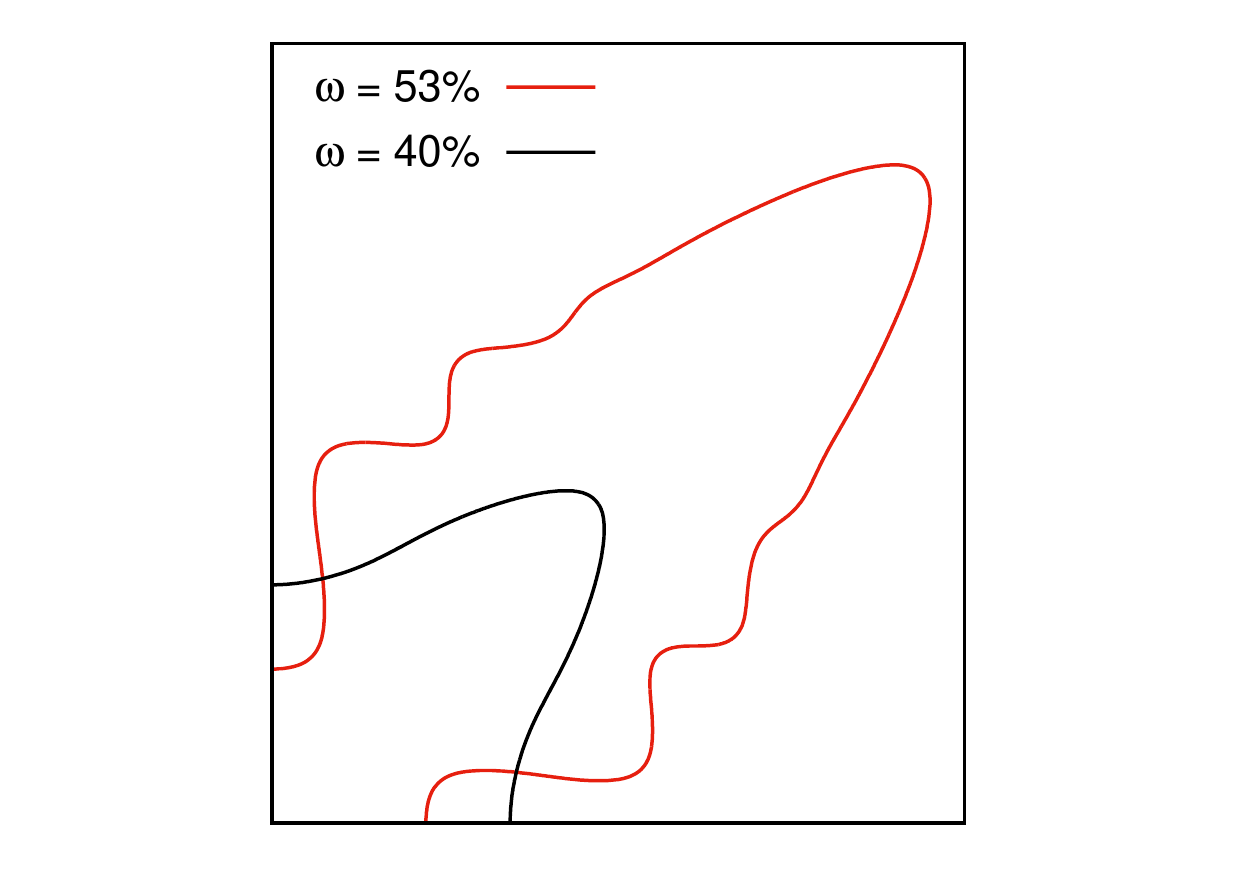}
 \caption{Evolution of precipitate into dendritic structure for different magnitudes 
	  of the supersaturation in the matrix phase captured at the same time. At a higher
	  supersaturation, prominent primary dendritic arms develop. Here, misfit strain is 
	  1\% and Zener anisotropy parameter is 3.0.}
 \label{dend_supersaturation_compare}	  
\end{figure}

With an increase in the amount of the supersaturation, 
the precipitate grows faster into a dendritic shape. This can be seen in 
Fig.~\ref{dend_supersaturation_compare}, where the precipitate with supersaturation 
$\omega=53\%$, grows its arms faster along the $\langle11\rangle$ 
directions giving rise to a dendrite-like morphology 
as compared to the one with supersaturation $\omega=40\%$.
Typical solidification dendrites are characterized by a unique tip radius 
and velocity and a dendritic shape 
that is a function of the strength of the anisotropy. 
In the following, we perform a similar characterization of our morphologies.
A usual characteristic of dendrites during solidification is that apart from
the selection of a Peclet number, the product of the square of the dendrite 
tip radius with the velocity is also a constant. 
This constant is typically 
known as the microsolvability constant $\sigma^{*}=2d_0D/R^2_{tip}V_{tip}$,
where $d_0$ is the capillary length, D is the diffusivity, and $R_{tip}$ 
and $V_{tip}$ are the radius and the velocity  
of the dendritic tip respectively, that is either estimated
using the linearized microsolvability theory as 
presented in~\cite{barbieri1989} or using dynamical phase-field 
simulations~\cite{karma1996}. We follow the latter 
route, where we derive the value of the selection constant $\sigma^{*}$
as a function of the material parameters and supersaturation from the 
measured values of the dendrite-tip radius and velocity from our phase-field 
simulations.

\begin{figure}[!htbp]
 \centering
 \subfigure[]{\includegraphics[width=0.45\linewidth]{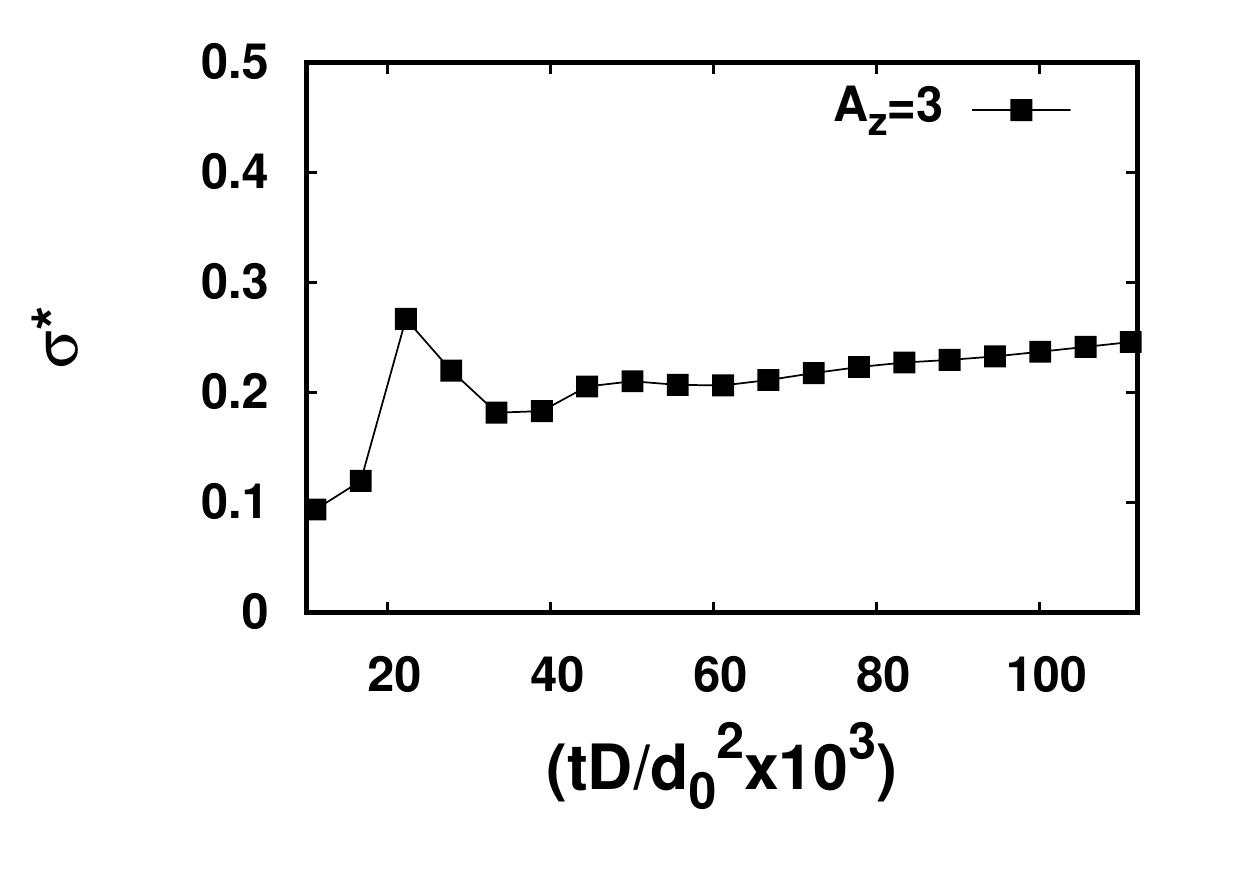}
  \label{sigma_star_Az3}
 }
 \centering
 \subfigure[]{\includegraphics[width=0.45\linewidth]{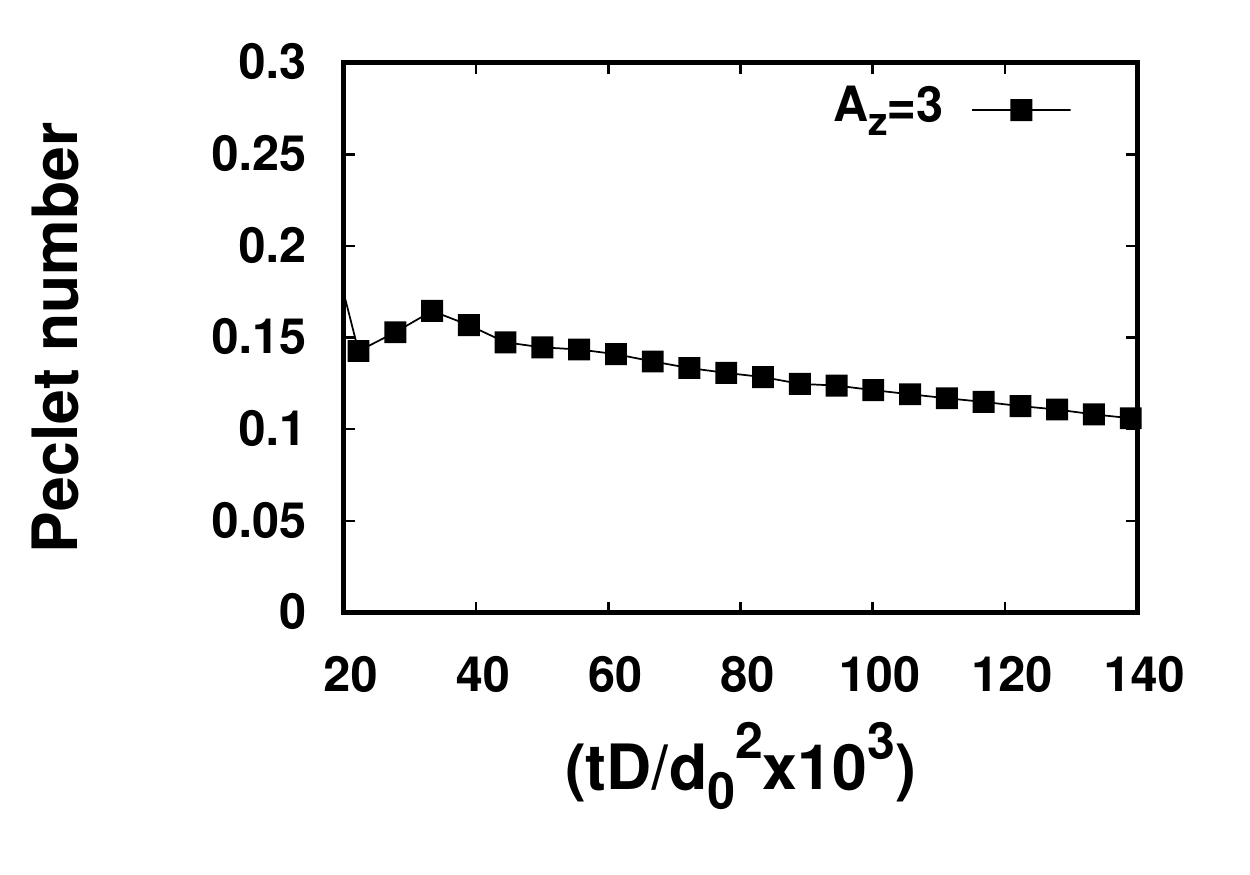}
  \label{Pe_estrain} 
 }
 \centering
 \subfigure[]{\includegraphics[width=0.45\linewidth]{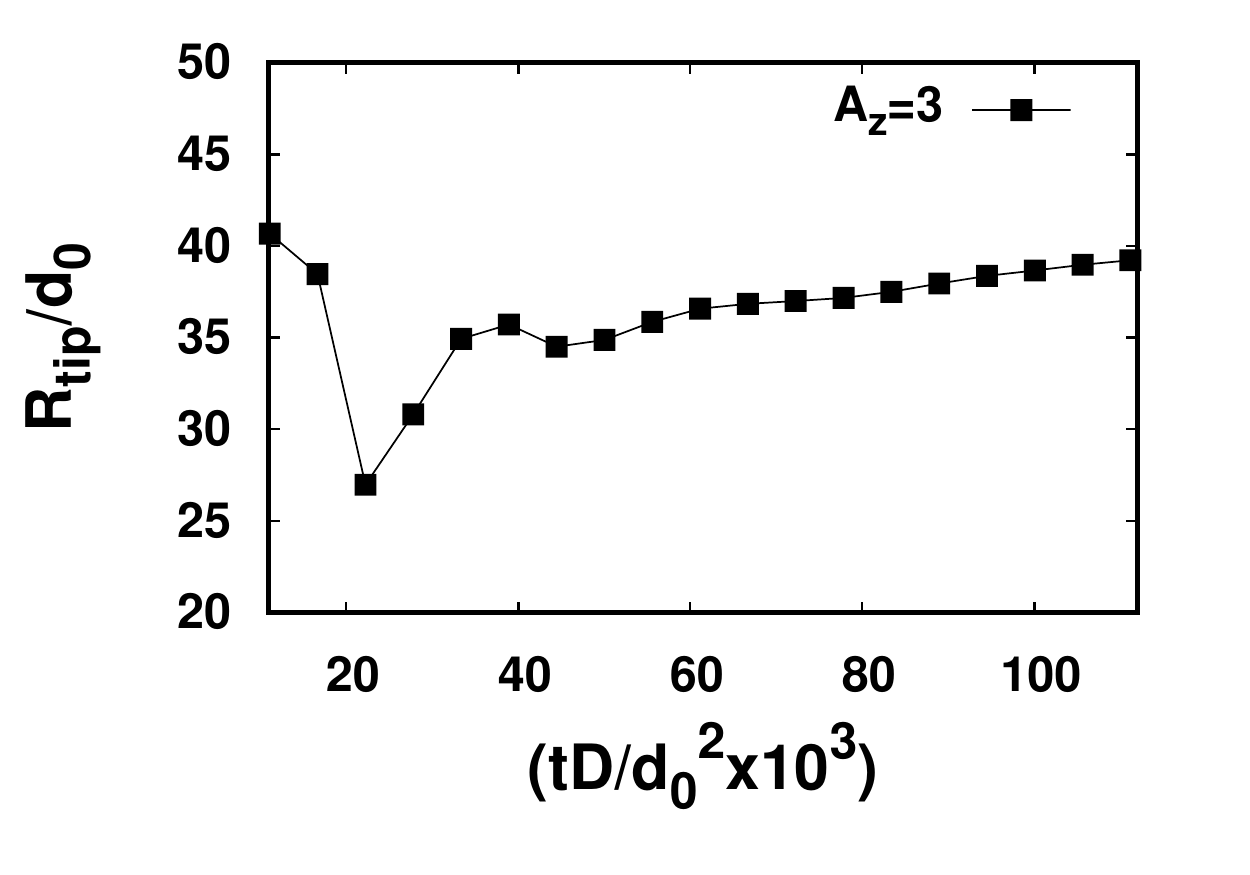}
  \label{Rtip_az3}
 }
 \centering
 \subfigure[]{\includegraphics[width=0.45\linewidth]{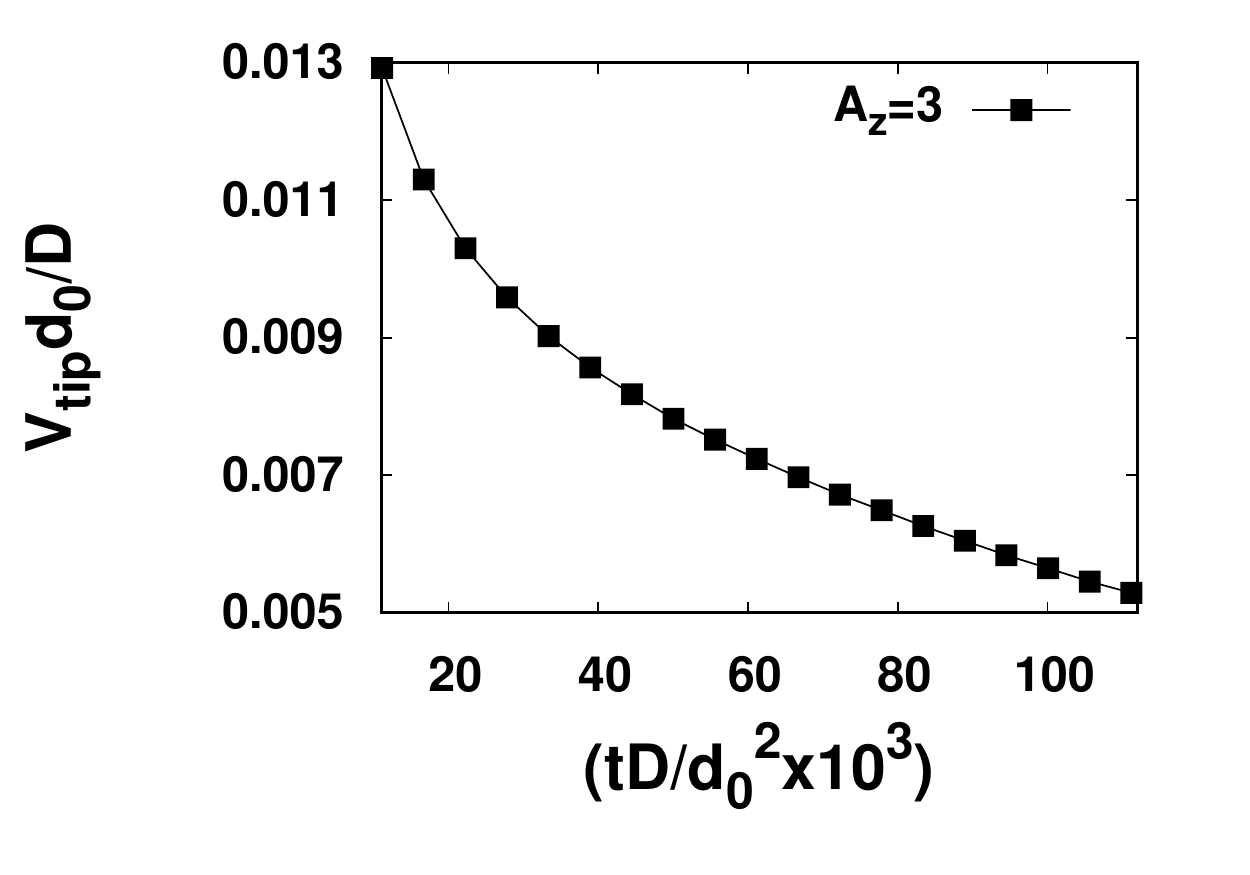}
  \label{Vtip_az3}
 }
 \caption{(a) Plot of the variation of $\sigma^{*}$ as a function of time in the
 simulations  (b)Variation of the Peclet number as a function of scaled time (c)
 and (d) show the variation of the  dendrite tip radius and velocity as a 
 function of time.}
 \label{Approach_to_steady_state}
\end{figure}

\begin{figure}[!htbp]

 \centering
 \subfigure[]{\includegraphics[width=0.45\linewidth]{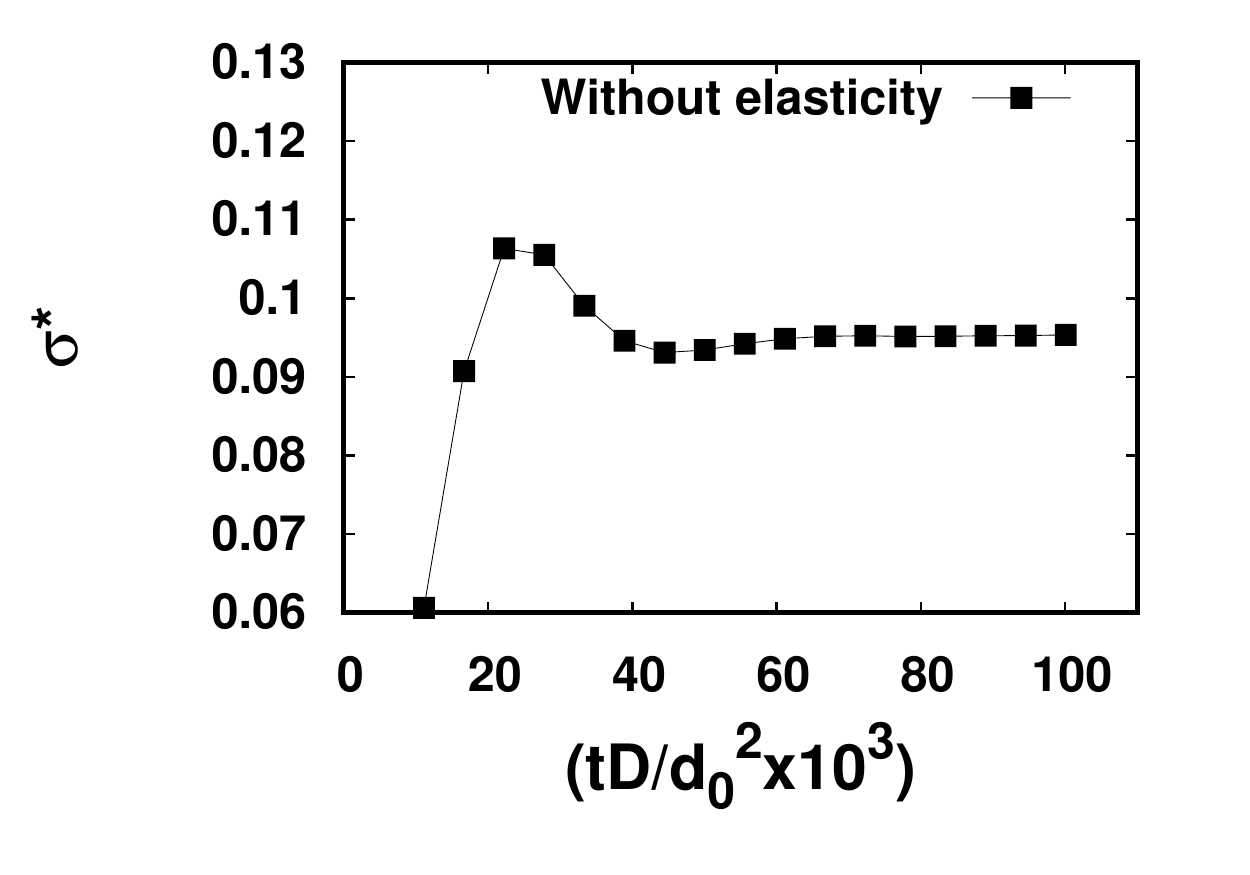}
 \label{sigma_vary_no_elasticity}
 }
 \centering
 \subfigure[]{\includegraphics[width=0.45\linewidth]{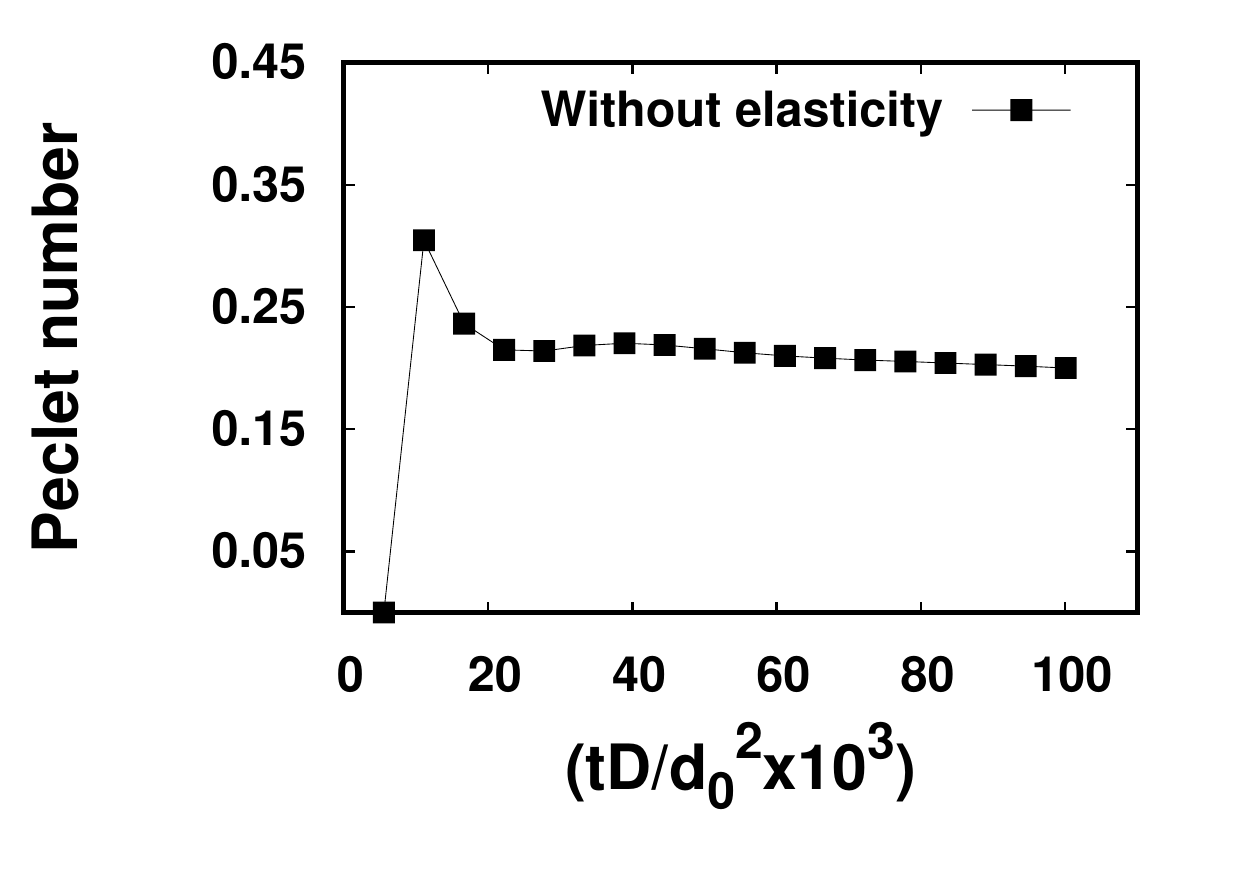}
 \label{Pe_vary_no_elasticity}
 }
 \centering
 \subfigure[]{\includegraphics[width=0.45\linewidth]{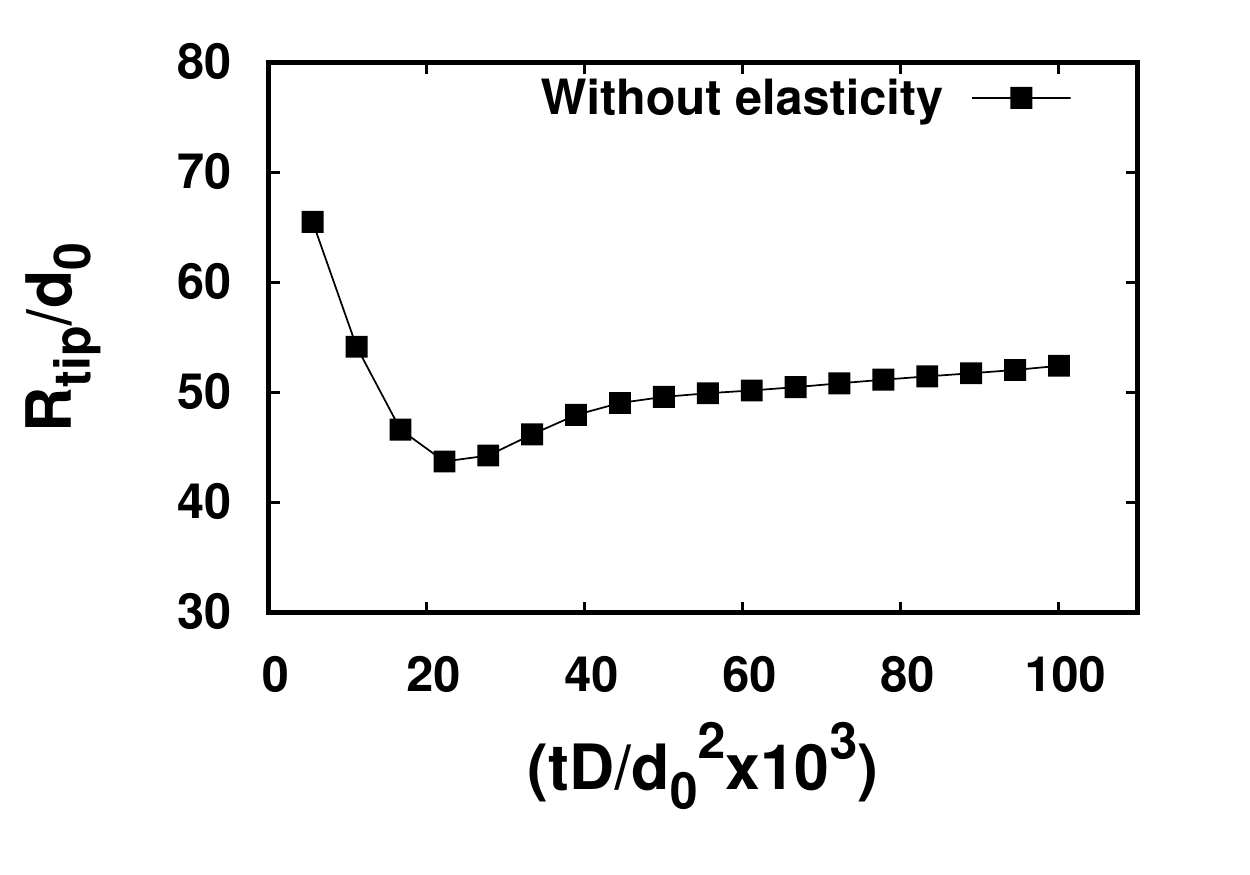}
 \label{radius_vary_no_elasticity}
 }
  \centering
 \subfigure[]{\includegraphics[width=0.45\linewidth]{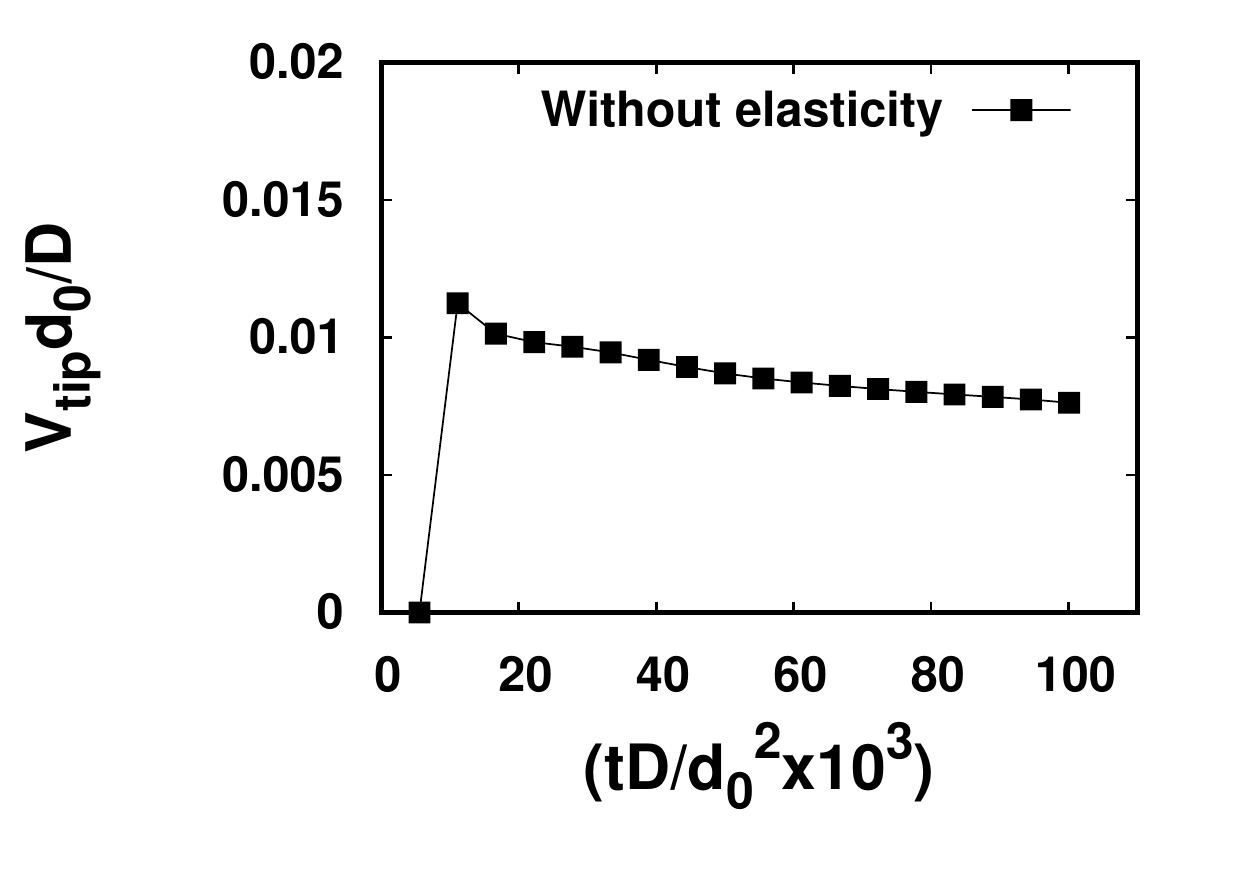}
 \label{velocity_vary_no_elasticity}
 }  
\caption{ Variation of (a) $\sigma^*$, (b) Peclet number, (c) $R_{tip}$, and (d) $V_{tip}$  
as a function of normalized time without elasticity. Here, we choose 
supersatuartion of $53\%$. The selection constant $\sigma^*$ achieves
a relatively steady state. In addition, the Peclet number
nearly attains a steady state value. }
\label{rtip_vtip_vary_no_elasticity}
\end{figure}

Fig.~\ref{sigma_star_Az3} shows the variation of the selection constant 
$\sigma^{*}$ as a function of time, scaled with the characteristic diffusion 
time. We find that the value of the selection constant has a transient where 
there is both an increase and decrease in the magnitude before settling into 
a regime where the values continue to change approximately linearly with 
simulation time.
This linear regime initiates approximately when the primary arms have emerged
as a result of the instability. This is quite different to dendrites in the 
presence of just interfacial energy anisotropy, where the value of $\sigma^{*}$
becomes relatively constant (see Fig.~\ref{sigma_vary_no_elasticity}) quite early 
in the simulation (just as the primary 
arms appear), even though the $R_{tip}$ (Fig.~\ref{radius_vary_no_elasticity}) and 
$V_{tip}$ (Fig.~\ref{velocity_vary_no_elasticity}) themselves have not yet 
attained their steady state values. A similar variation is also seen in the Peclet
number $\dfrac{R_{tip}V_{tip}}{2D}$, where it continues to decrease approximately 
linearly with time, as depicted in Fig.~\ref{Pe_estrain} in the presence of 
elasticity, while a nearly constant value is attained for the dendrite with just 
interfacial energy anisotropy (see Fig.~\ref{Pe_vary_no_elasticity}). 
Therefore, we also do not derive any steady state with respect to the shape of the 
tip as well as the velocity as highlighted in Figs.~\ref{Rtip_az3} and~\ref{Vtip_az3} 
respectively in the presence of elasticity.


\begin{figure}[htbp]
 \centering
 \subfigure[]{\includegraphics[width=0.48\linewidth]{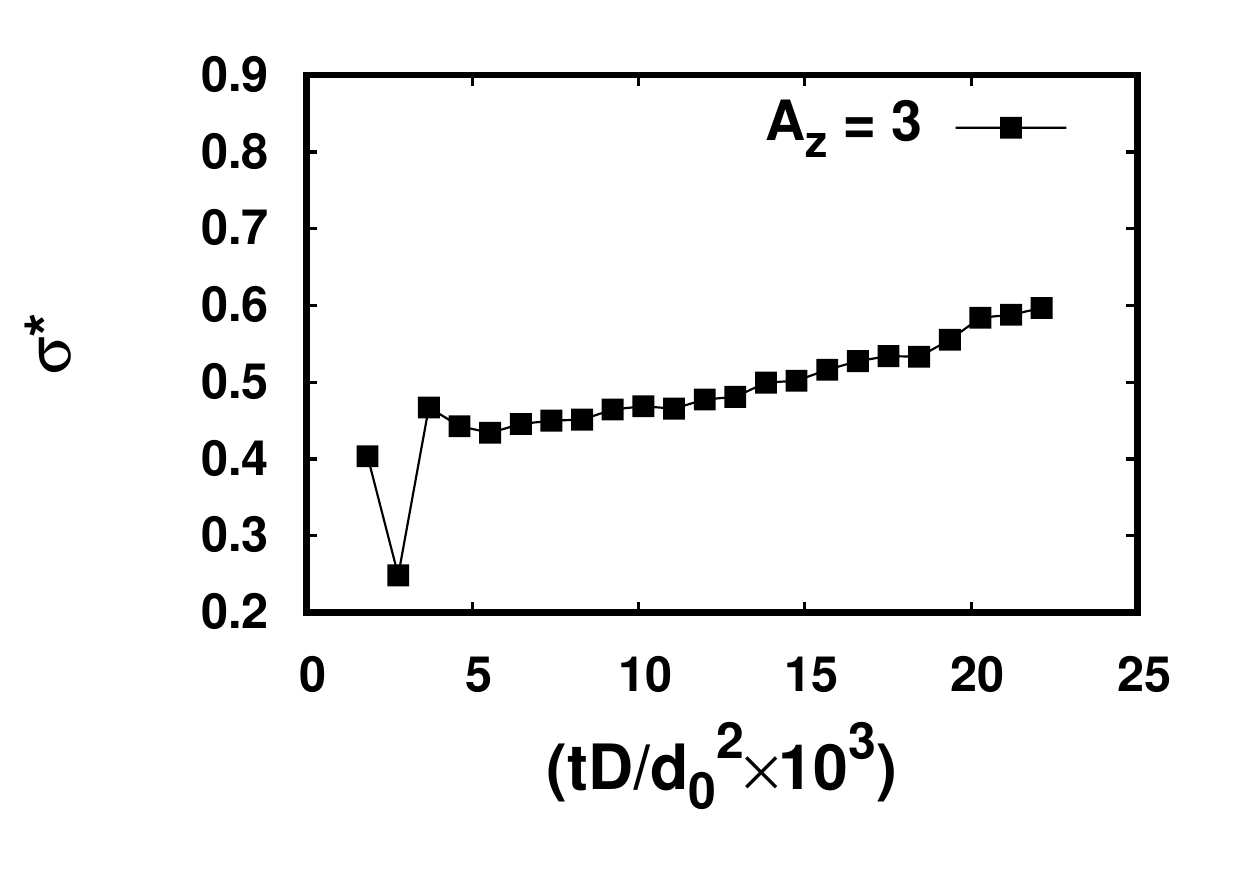}
  \label{sigma_star_Az3_3d}
 }
 \centering
 \subfigure[]{\includegraphics[width=0.48\linewidth]{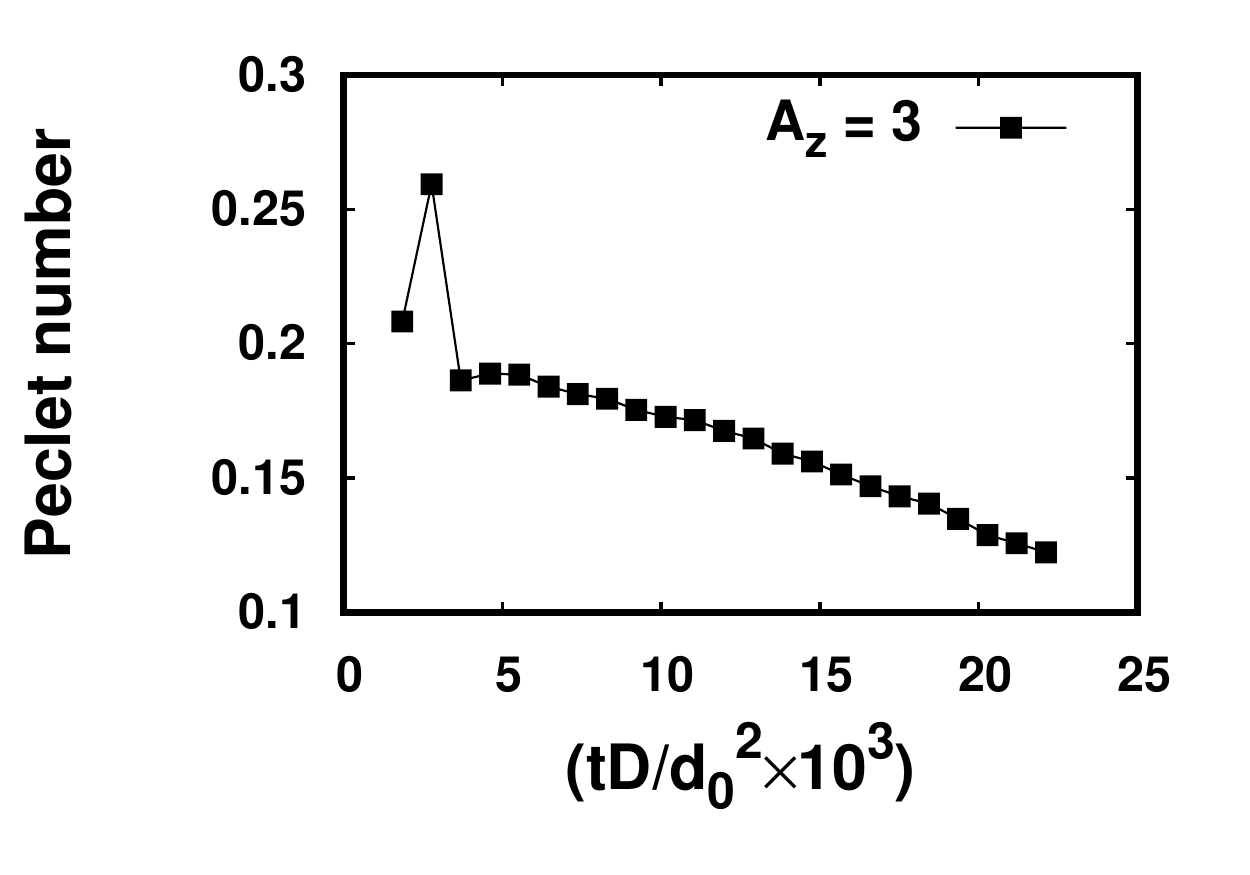}
  \label{Peclet_num_3d} 
 }
 \centering
 \subfigure[]{\includegraphics[width=0.48\linewidth]{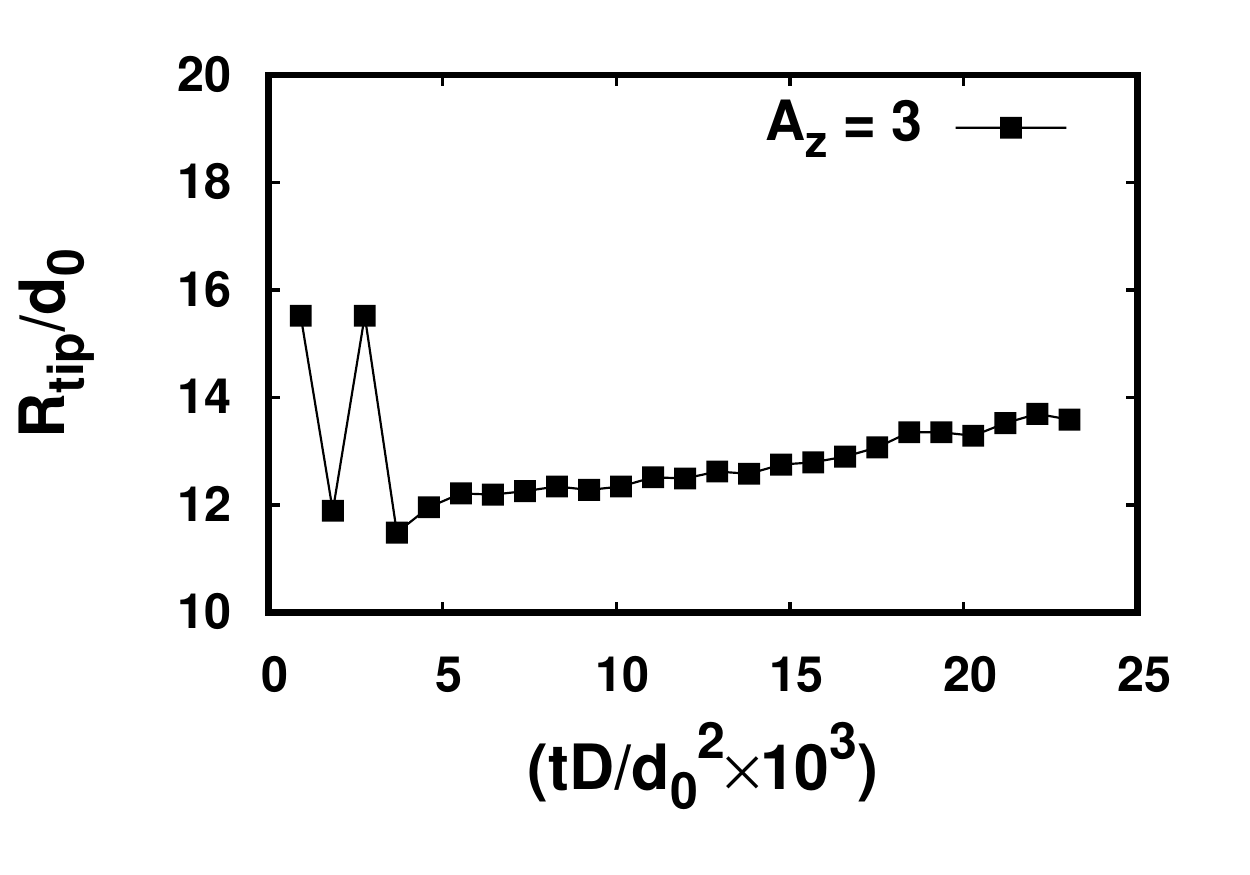}
  \label{Rtip_az3_3d}
 }
 \centering
 \subfigure[]{\includegraphics[width=0.48\linewidth]{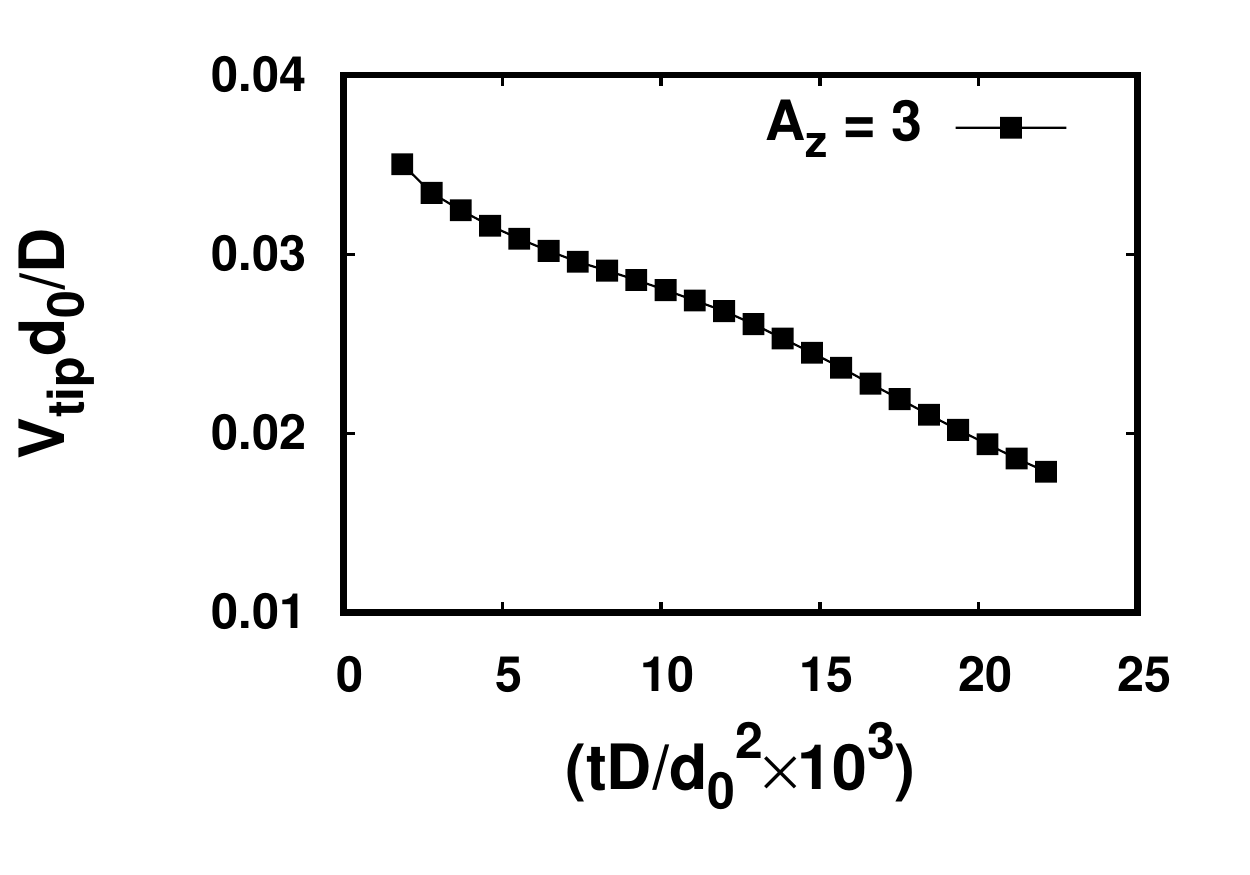}
  \label{Vtip_az3_3d}
 }
 \caption{Temporal evolution of the dendritic characteristics: 
 (a) selection constant ($\sigma^{*}$) (b) Peclet number (c) dendritic tip 
 radius ($R_{tip}$) (d) dendritic tip velocity ($V_{tip}$). Here, supersaturation is
 $53\%$, misfit strain is $1\%$, and Zener anisotropy parameter is
 $3$. All of these dendritic characteristics do not achieve steady state values}
 \label{Approach_to_steady_state_3d}
\end{figure}

Similar trends are also noticed for the 3D simulations 
as well. In 3D, we measure the dendritic tip radius ($R_{tip}$) at different 
times by fitting the dendritic tip surface to a paraboloid of 
revolution. The tip position at different times is used to calculate 
the velocity ($V_{tip}$) along the 
$\langle 111 \rangle$ direction as this is the growth direction 
for $A_z>1$. Similar to the 2D simulations, we calculate the 
microsolvability constant ($\sigma^*$) and Peclet number at different times. 
Fig.~\ref{sigma_star_Az3_3d} shows the temporal variation of $\sigma^*$ and 
Fig.~\ref{Peclet_num_3d} depicts the variation of Peclet number as a function of
scaled time. The values of $\sigma^*$ and Peclet number continue to vary 
with time without the attainment of a steady state. Figs.~\ref{Rtip_az3_3d} 
and~\ref{Vtip_az3_3d} represent the variation of $R_{tip}$ and $V_{tip}$ as a
function of scaled time that reveals that both of these quantities 
also do not achieve a steady state. The $R_{tip}$ continues to increase, 
whereas $V_{tip}$ continues to decrease with time, and the variation is 
similar in nature to that observed in the 2D simulations.

\begin{figure}[!htbp]
 \centering
 \subfigure[]{\includegraphics[width=0.48\linewidth]{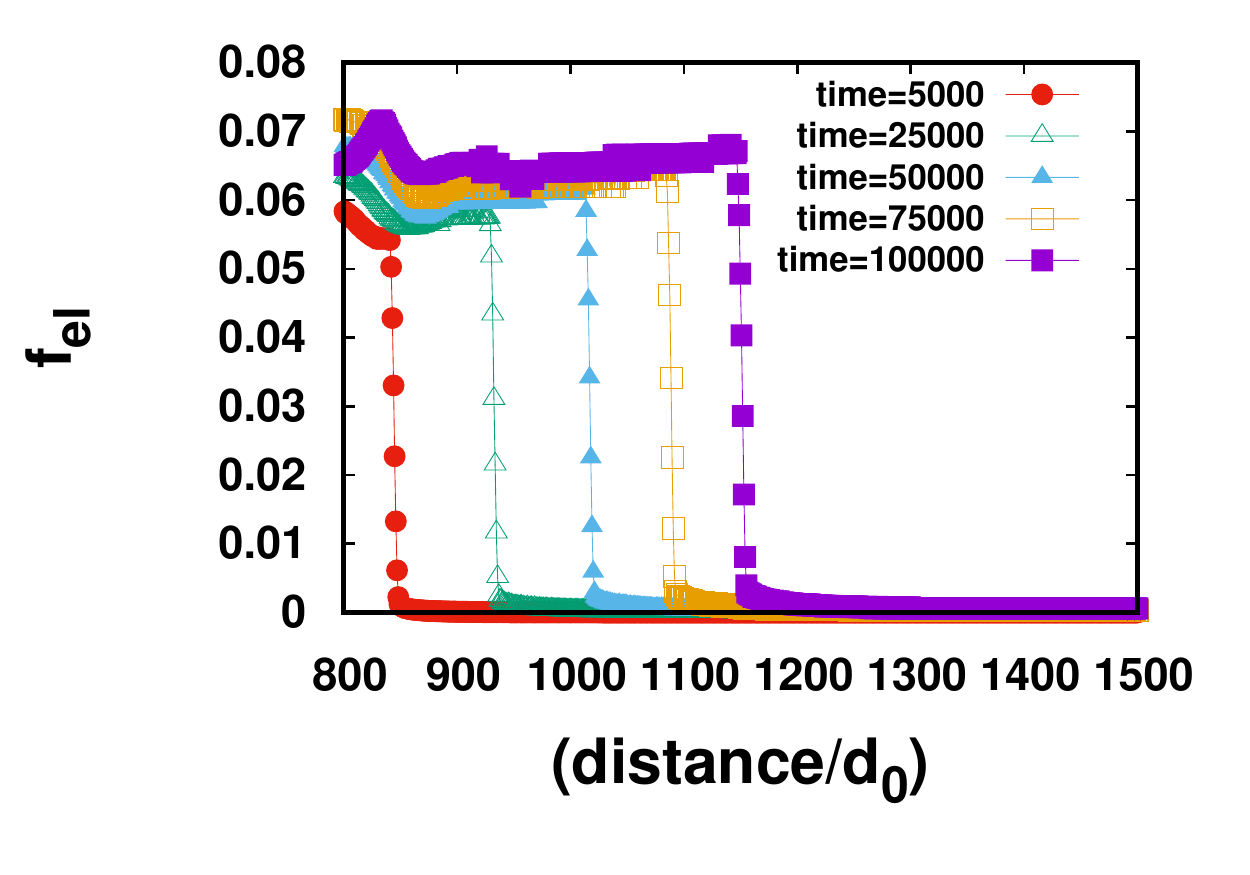}
 \label{elast_energy_11}}
 \centering
 \subfigure[]{\includegraphics[width=0.48\linewidth]{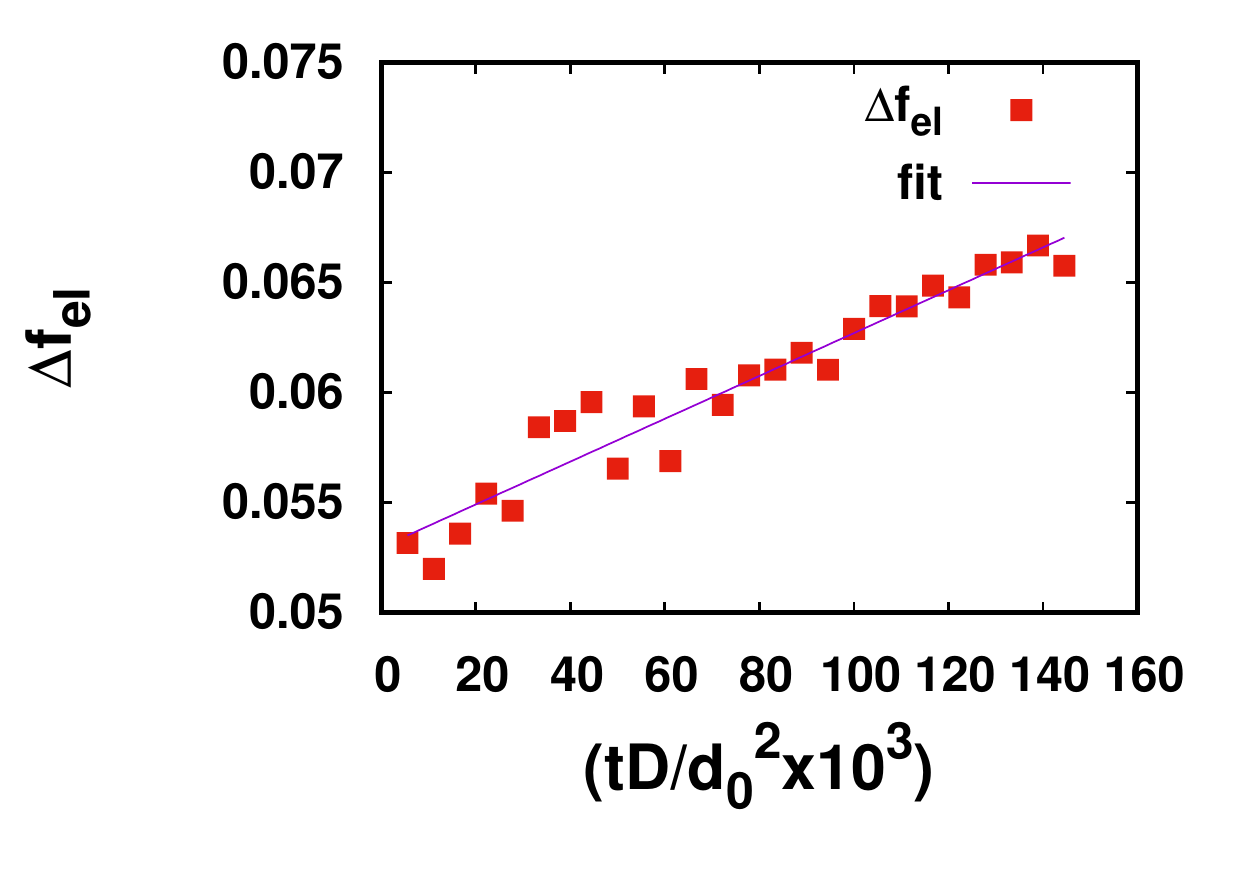}
 \label{delta_fel}}
 \caption{(a) Variation of elastic energy density along 
 $[11]$ direction, i.e., along the tip of the precipitate at 
 simulation times of $5000$, $25000$, $50000$, $75000$, $100000$. The bulk
 elastic energy in the precipitate increases with time. (b) Temporal evolution of 
 $\Delta f_{el}$ calculated along $[11]$ direction. The line fit to the jump in elastic energy 
 data suggests linear increment of $\Delta f_{el}$ with time. Here, Zener anisotropy parameter 
 is $3$, supersaturation is $53\%$, and misfit strain is $1\%$.}
\end{figure}

The reason for the non-attainment of a steady state is possibly linked to the 
variation of the jump in the value of the elastic energy at the interface.
We have extracted the elastic energy along the direction normal to the 
dendrite-tip for different simulation times. From this, we calculate the 
jump in the elastic energy ($\Delta f_{el}$) by computing the difference in the 
values of the elastic energy on the precipitate side 
with the matrix side. For the 2D simulations, 
Fig.~\ref{elast_energy_11} highlights
the variation of the elastic energy density as a function
of time, while the jump in the elastic energy density at the 
interface is depicted in Fig.~\ref{delta_fel} that clearly 
shows an increase with simulation time.

\begin{figure}[htbp]
    \centering
    \subfigure[]{\includegraphics[width=0.48\linewidth]{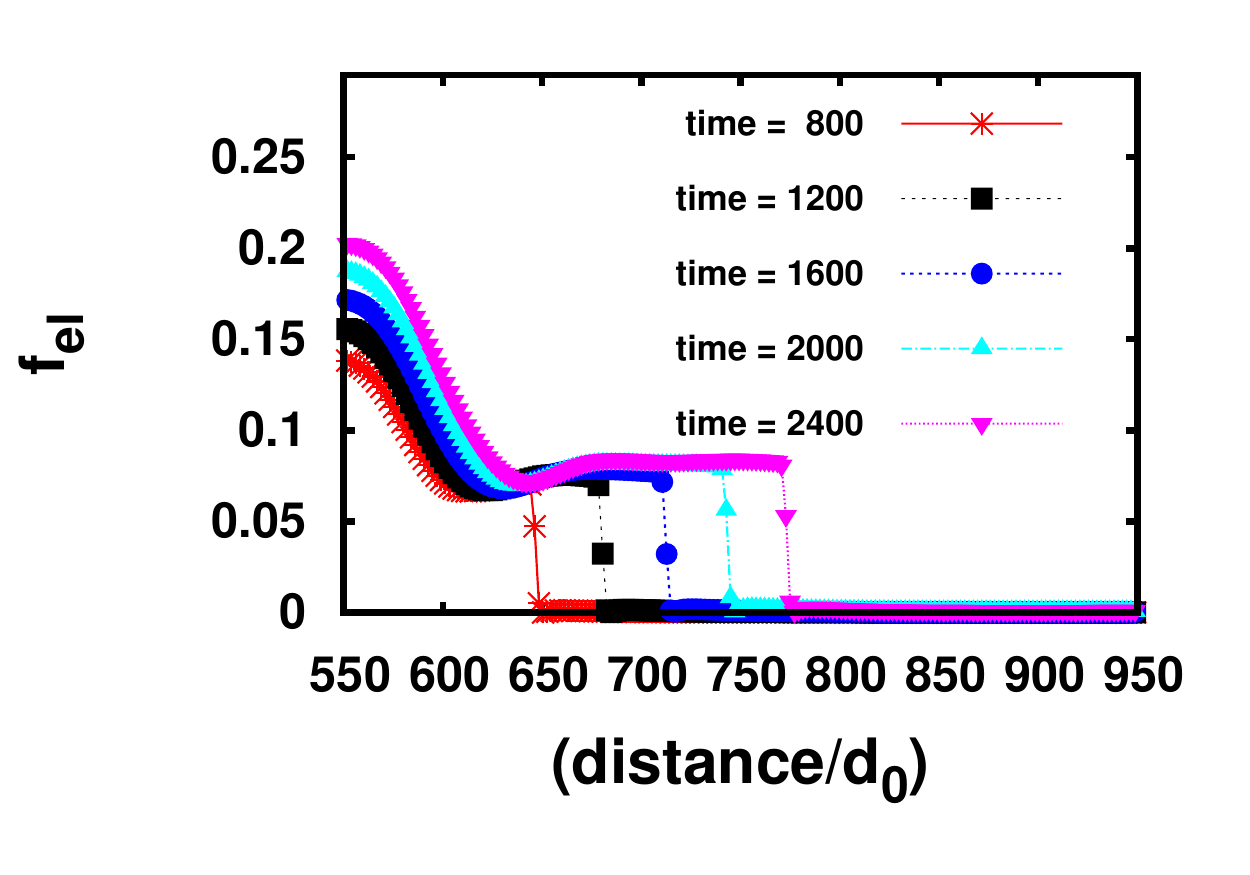}
    \label{fig:elast_energy_prof}}
    \centering
    \subfigure[]{\includegraphics[width=0.48\linewidth]{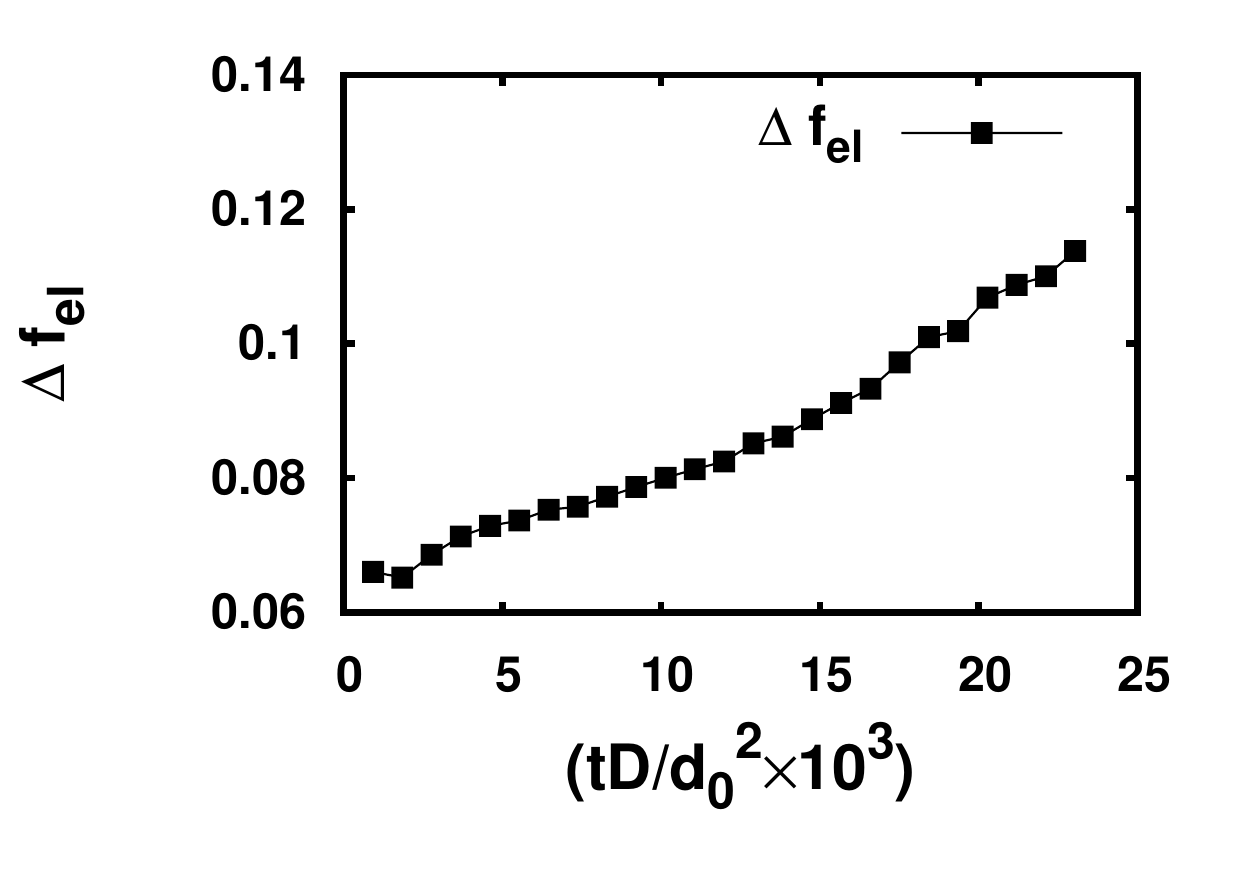}
    \label{fig:elast_energy_jump}}
    \caption{(a) Variation of elastic energy density across the interface along 
    $[111]$ at different simulation times of $t = 800$, $1200$, $1600$, $2000$, and $2400$. 
    Elastic energy inside precipitate increases with time. (b) Temporal evolution of 
    $\Delta f_{el}$ along $[111]$ direction. The jump in the elastic energy along $[111]$  
    increases with time indicating absence of steady state. Here, supersaturation is $53\%$, 
    misfit strain is $1\%$, and Zener anisotropy parameter is $3$. Elastic energy inside 
    precipitate increases with time.}
\end{figure}

Similarly, for 3D, Fig.~\ref{fig:elast_energy_prof}
depicts the variation of elastic energy density across the interface 
along $[111]$ direction at time $t = 800$, $1200$, $1600$, 
$2000$, and $2400$, while the jump of the elastic energy 
density at the interface in the $[111]$ direction 
is plotted in Fig~\ref{fig:elast_energy_jump}.

We note that although the variation is small, this will, in turn, lead
to the change in the interfacial compositions as a function 
of time. The reason for this is as follows: as the jump in elastic energy increases with time,
the dendritic tip experiences varying elastic fields ahead of interface 
indicating an increasing contribution of elastic energy to the interfacial 
equilibrium conditions during the growth of precipitate. The increasing elastic contribution
to the interfacial equilibrium conditions is countered by a decreasing curvature contribution.
As a result, there is continuous increase in $R_{tip}$ with increasing time. 
The decreasing curvature contribution possibly leads to a reduction in the 
point effect of diffusion, leading to slower interface dynamics that result in 
a continuous decrease of $V_{tip}$ with time. Therefore, the Peclet number as well
as the selection constant $\sigma^{*}$ do not attain saturated values. This also
means that the dendrite tip radius as well as the tip velocity never reach a 
steady state in contrast to the situation of dendritic growth in the 
presence of interfacial energy anisotropy with no elastic contribution. 
(e.g., dendritic growth during solidification). 
Therefore, in relation to the classical dendritic structures observed during solidification, 
the simulated structures may only be referred to as dendrite-like. 

\begin{figure}[!htbp]
 \centering
 \includegraphics[width=0.5\linewidth]{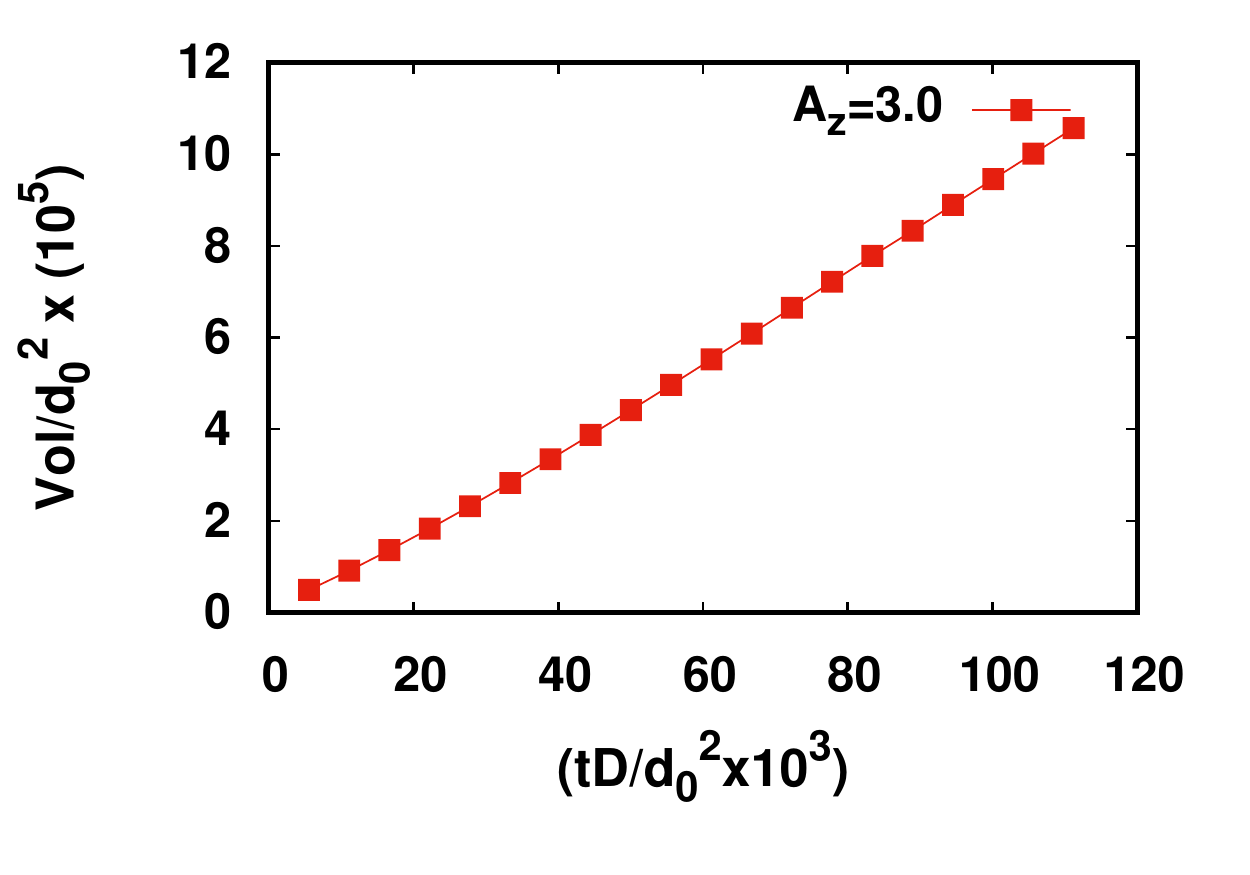}
 \caption{Evolution of scaled precipitate volume as a function of scaled 
 time $tD/d_0^2$. Here, Zener anisotropy parameter id $3.0$, misfit strain is $1\%$, 
 and supersaturation is $53\%$. The linear variation of precipitate volume suggests 
 the diffusion-controlled growth of precipitate.}
 \label{ppt_volume}  
\end{figure}

Although the variation of the $\Delta f_{el}$ 
at the dendrite tip will lead to a change in the interfacial 
compositions and thereby the Peclet number, 
the overall volume (area for 2D) still varies linearly with time (see 
Fig.~\ref{ppt_volume}).
The linear temporal variation of the volume of precipitate suggests 
the parabolic law of diffusion-controlled growth of the precipitate.

\begin{figure}[!htbp]
 \centering
\includegraphics[width=0.5\linewidth]{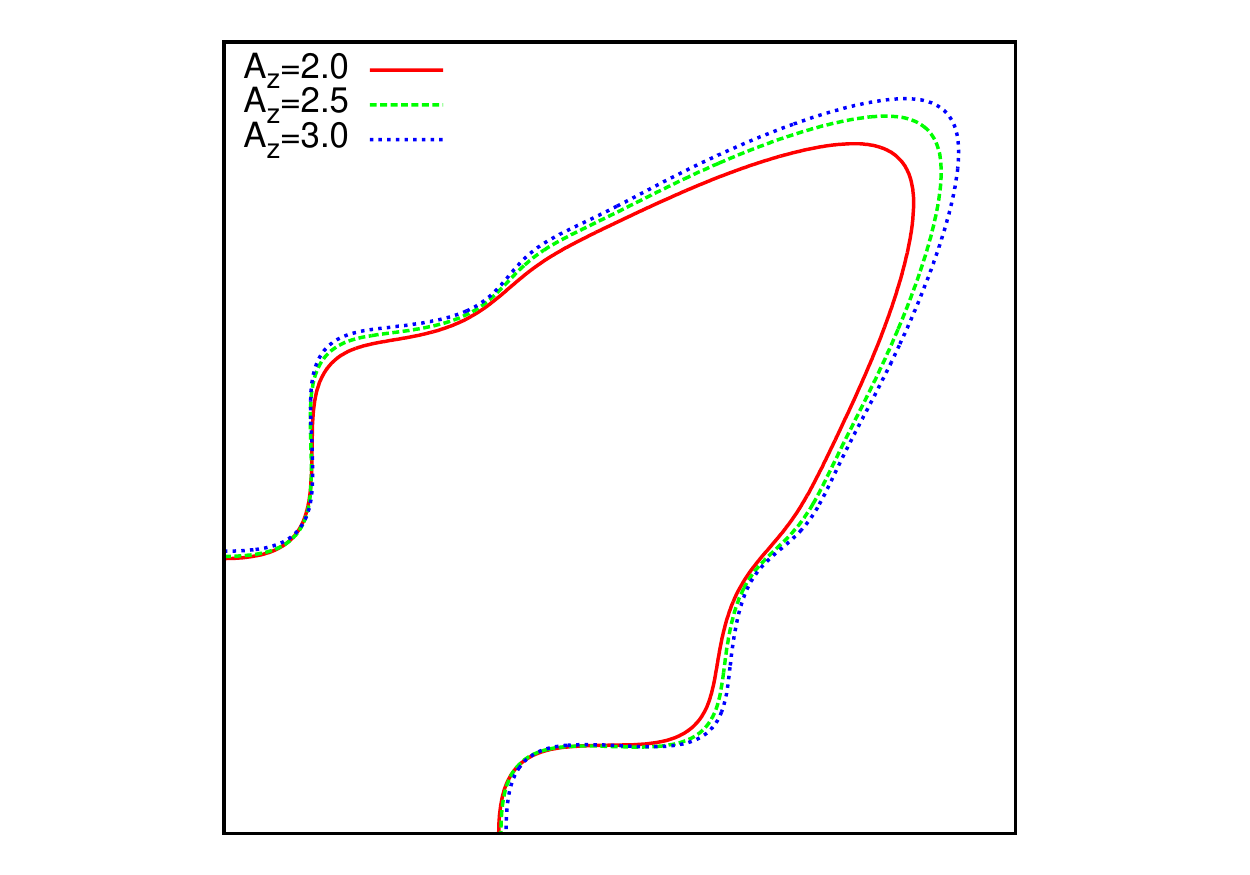}
 \caption{Contours of $\phi=0.5$ showing one-fourth section of the dendritic 
 structure at a normalized time of 41700 for different strengths of anisotropy 
 in elastic energy ($A_z = 2.0$, $2.5$, and $3.0$). The higher Zener anisotropy parameter
 gives rise to faster growth of dendritic tip.}
 \label{dend_az} 
\end{figure}

Further, we determine the effect of elastic anisotropy on the behavior of 
precipitate morphologies. For 2D situations, Fig.~\ref{dend_az} shows the 
difference in the tip shapes as a function of $A_z$ at a normalized time=41700, 
where we plot only one of the symmetric quadrants for simplicity.

\begin{figure}[!htbp]
 \centering
 \subfigure[]{\includegraphics[width=0.48\linewidth]{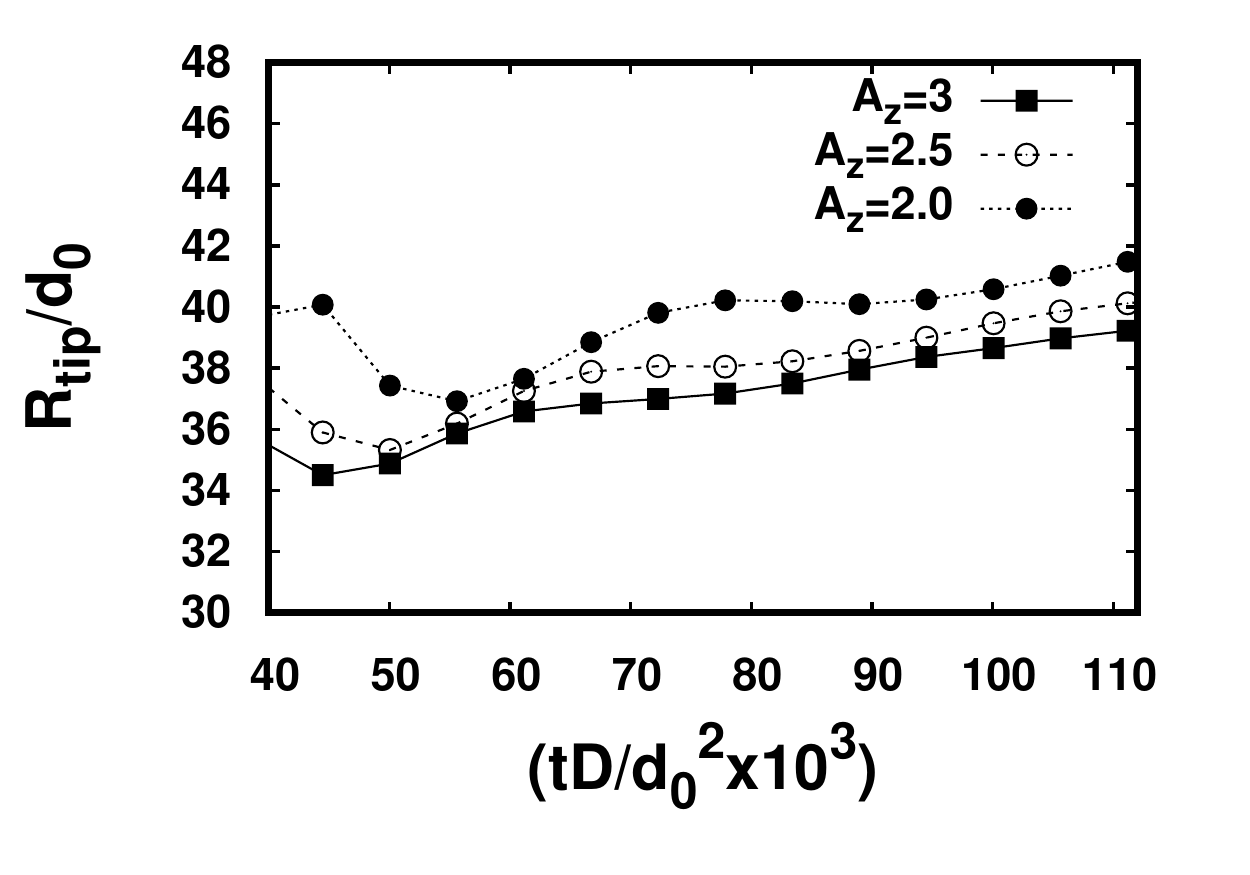}
 \label{radius_vary_az_es1}
 }%
  \centering
 \subfigure[]{\includegraphics[width=0.48\linewidth]{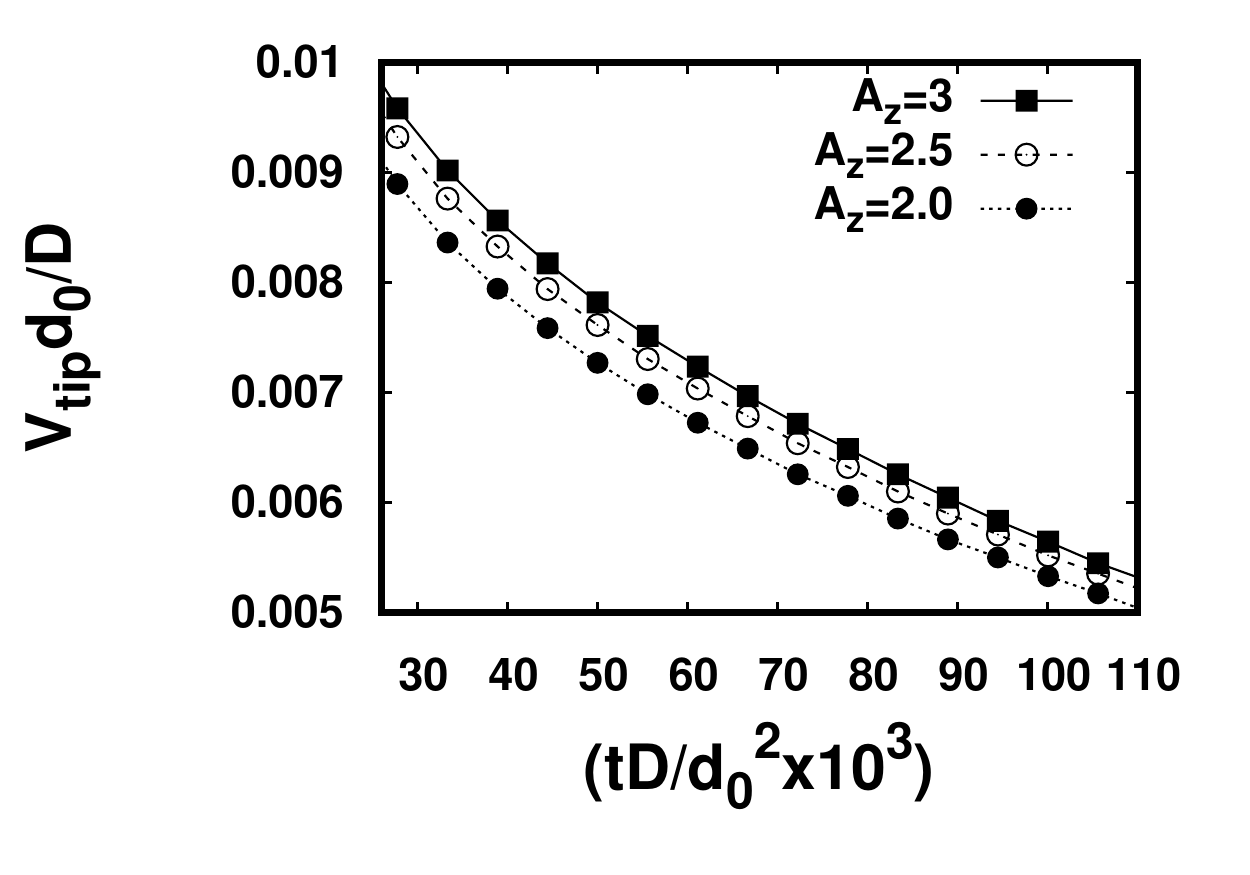}
 \label{velocity_vary_az_es1}
 } 
 \subfigure[]{\includegraphics[width=0.48\linewidth]{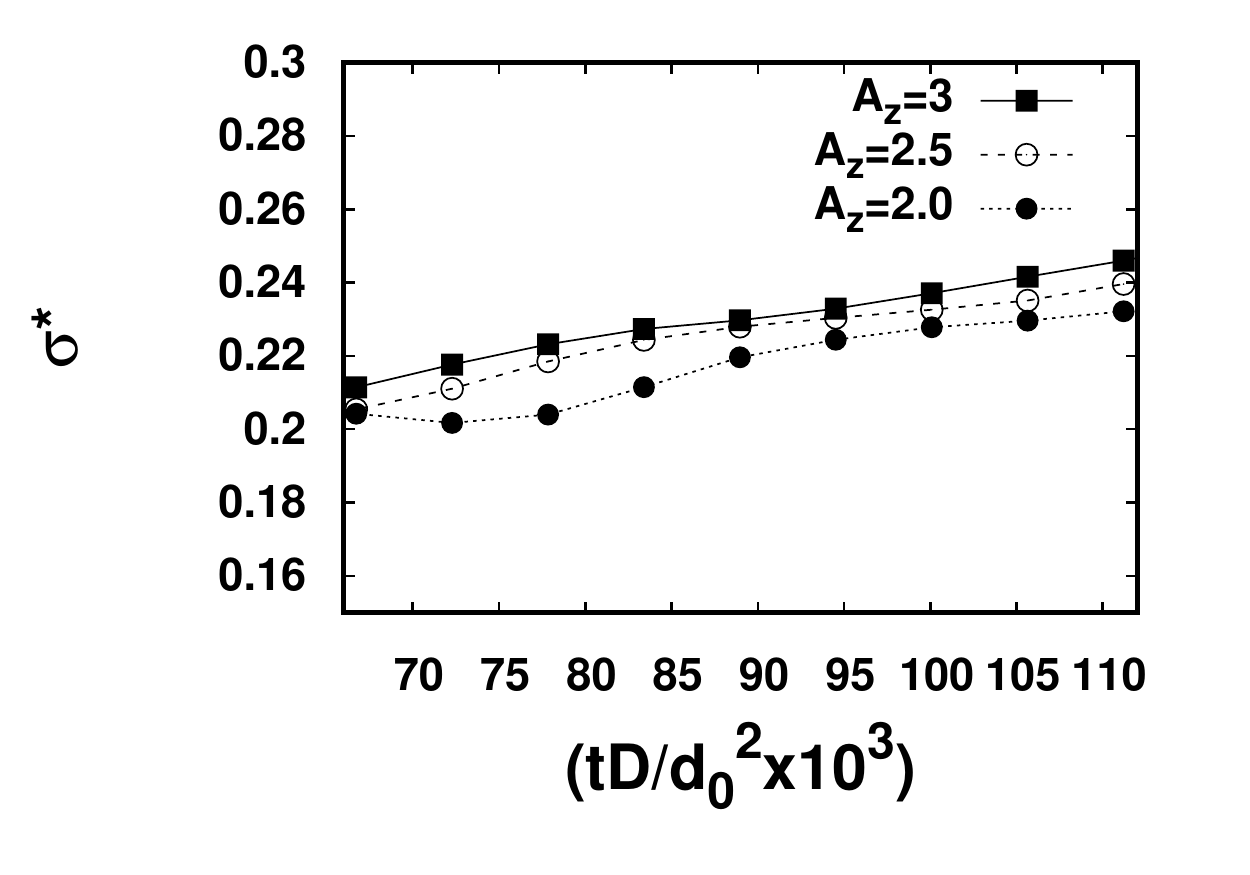}
 \label{sigma_vary_az} 
 }
\caption{ Effect of strength of anisotropy in elastic energy 
on the temporal variation of (a) $R_{tip}$ and (b) $V_{tip}$, and (c) $\sigma^*$.
Here, the supersaturation is 53\% and the misfit strain 
is 1\%. With the increase in $A_z$, tip becomes more sharper,
$V_{tip}$ increases, and $\sigma^*$ increases.}
\label{rtip_vtip_vary_az_es1}
\end{figure}

Fig.~\ref{dend_az} shows that with 
increase in the strength of anisotropy in the elastic energy ($A_z>1.0$), the radius 
of the tip of the precipitate reduces, i.e., the tip morphologies become sharper and 
elongated along $\langle 11 \rangle$ directions, that is also revealed in the plots 
showing the dendrite tip radius in Fig.~\ref{radius_vary_az_es1}. 
Correspondingly, the velocities at the tip are also higher for larger values of $A_z$
as highlighted in Fig.~\ref{velocity_vary_az_es1}.

\begin{figure}[!htbp]
 \centering
 \subfigure[]{\includegraphics[width=0.48\linewidth]{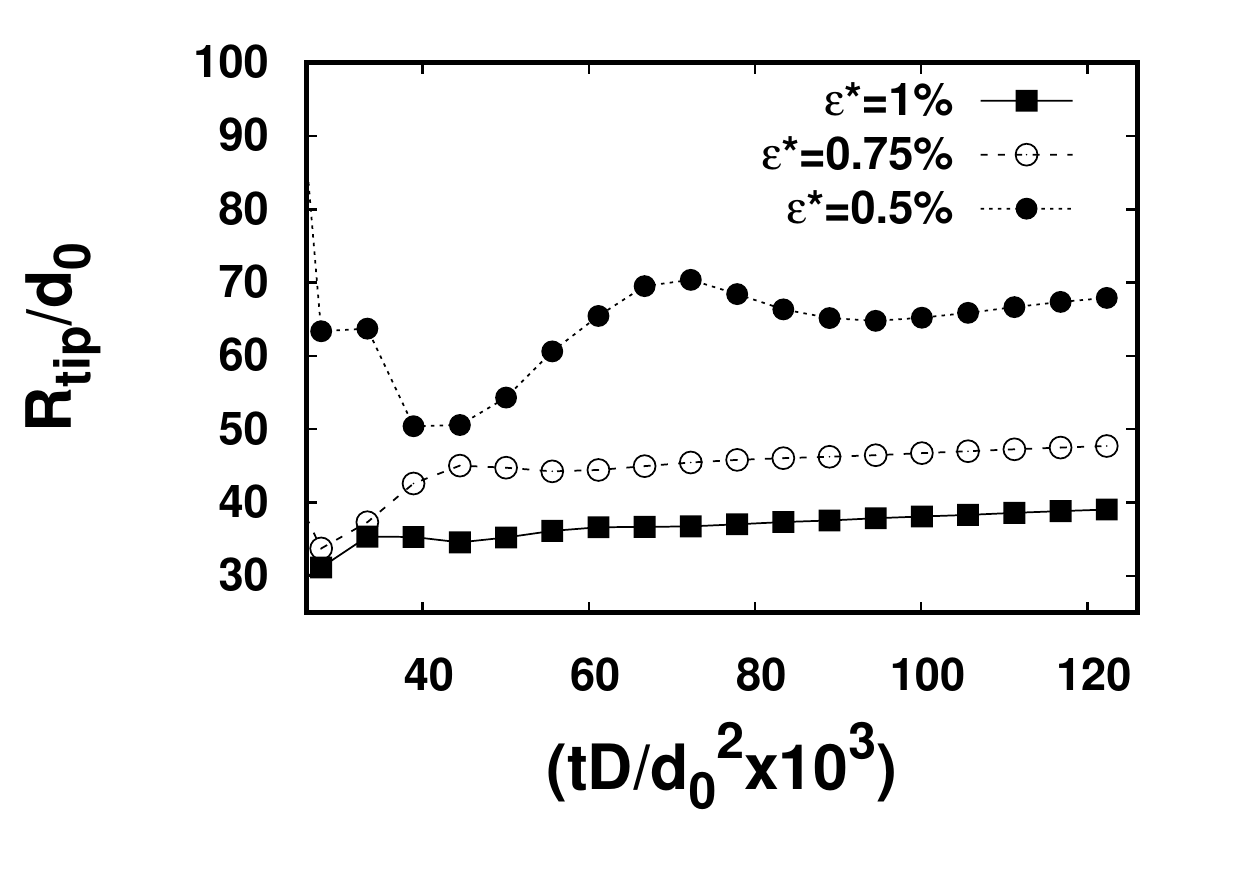}
 \label{radius_vary_es_az3}
 }%
  \centering
 \subfigure[]{\includegraphics[width=0.48\linewidth]{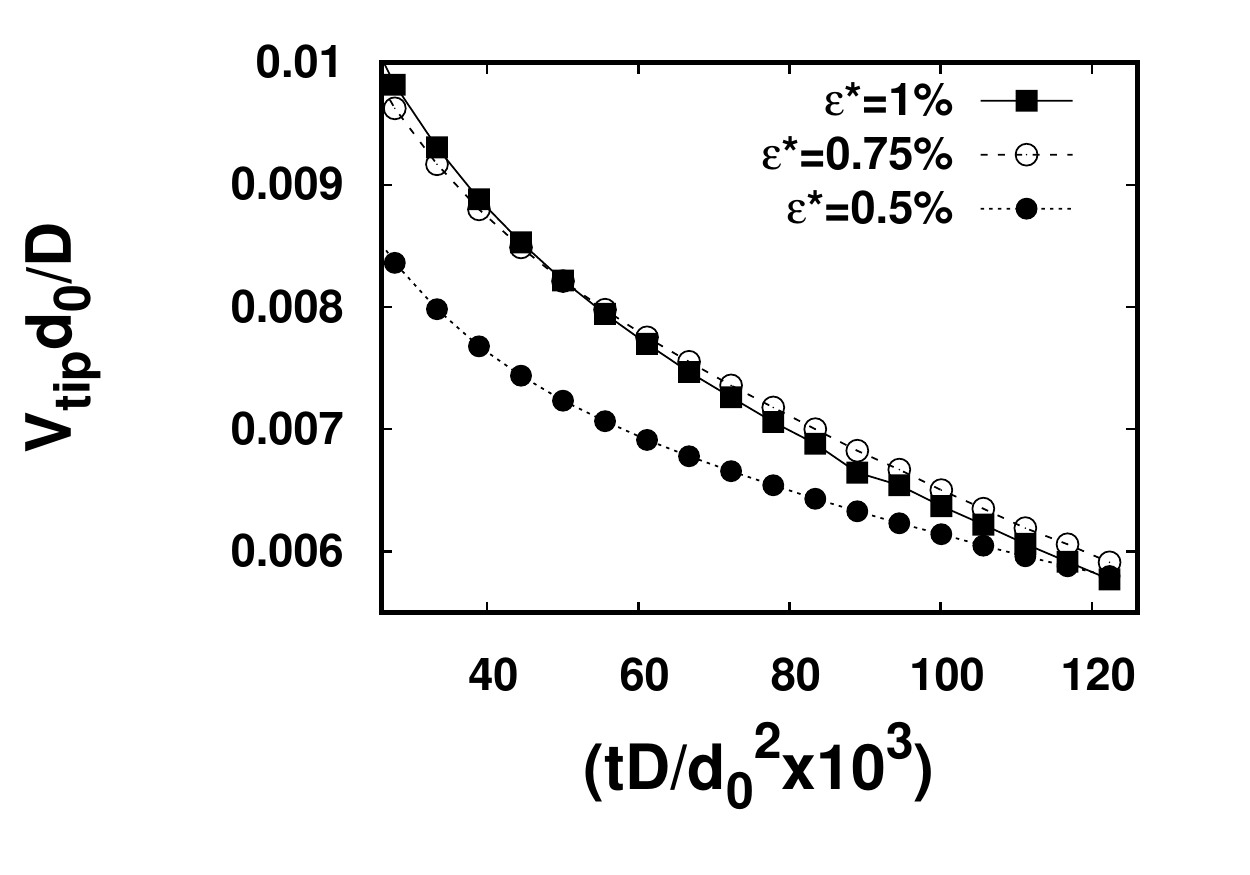}
 \label{velocity_vary_es_az3}
 } 
 \centering
 \subfigure[]{\includegraphics[width=0.48\linewidth]{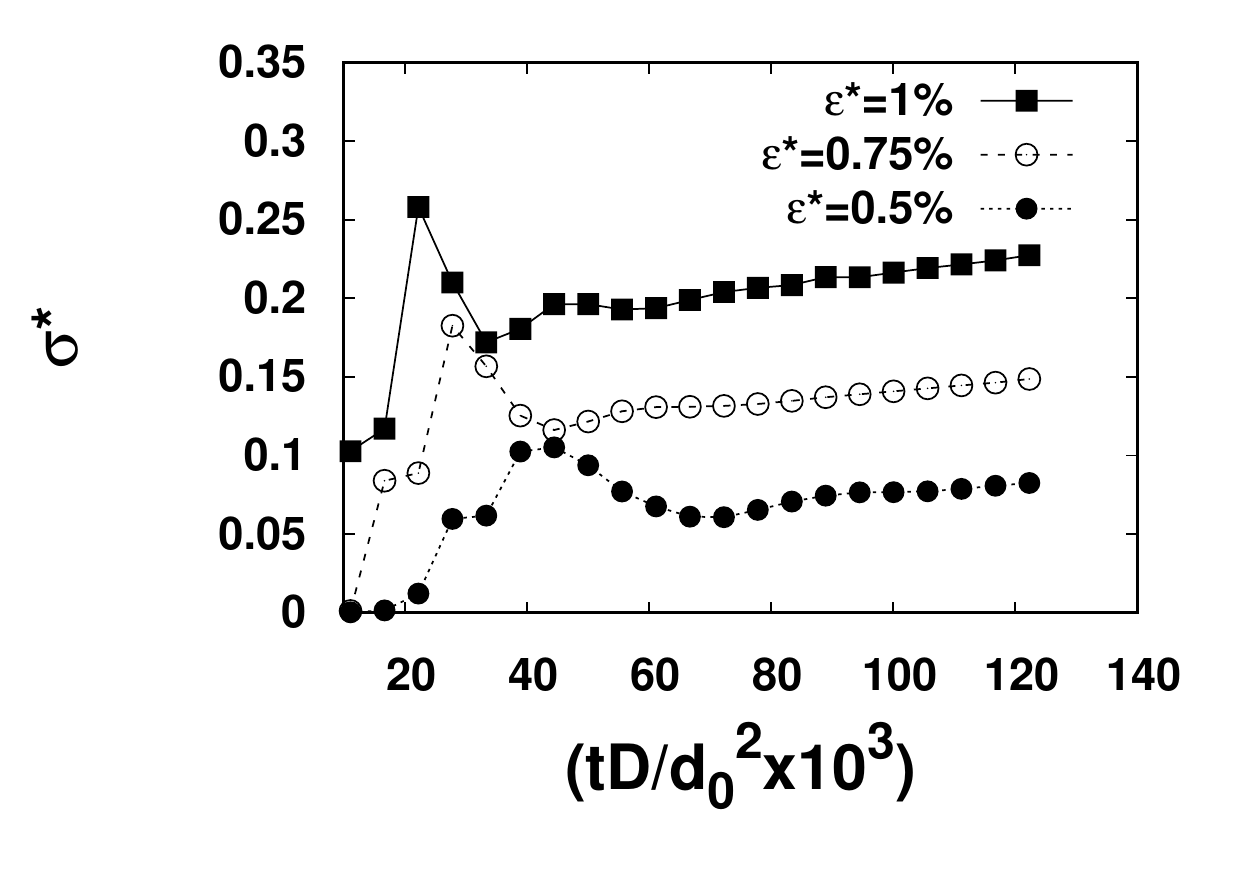}}
 \label{sigma_estrain}
\caption{ Effect of misfit strain on the temporal 
variation of (a) $R_{tip}$, (b) $V_{tip}$, and 
$\sigma^*$. Here, Zener anisotropy parameter is 3 and 
supersaturation is 53\%. With the increase in misfit 
strain, tip becomes sharper, $V_{tip}$ increases, and 
$\sigma^*$ increases.}
\label{rtip_vtip_vary_es_az3}
\end{figure}

The plot in Fig.~\ref{sigma_vary_az} shows that the magnitude of 
$\sigma^*$ consistently increases with normalized time $(tD/d_0^2)$ 
after the initial transient, whereas the magnitude of $\sigma^*$ increases with the 
strength of anisotropy in the elastic energy, i.e., $A_z$ 
at a given time. Similarly, Fig.~\ref{sigma_estrain} shows that the magnitude of 
$\sigma^*$ increases with time for a given misfit strain. The plot also shows that, 
as the magnitude of the misfit strain increases, the value of the selection constant 
also becomes larger. There is no clear trend observable with the change in the 
velocities as seen in Fig.~\ref{velocity_vary_es_az3}, while the tip radius reduces 
with increase in the value of the misfit strain as depicted in 
Fig.~\ref{radius_vary_es_az3}. Thus, there is no unique value of the selection 
constant $(\sigma^*)$ at a given anisotropy strength of elastic energy and the 
magnitude of the misfit strain. While the variation of $\sigma^{*}$ in the linear 
regime at a given time, is approximately linear with the variation of $A_z$, it 
changes approximately as $\epsilon^{*1.5}$ with the misfit 
strain (comparing values for a given time).

\begin{figure}[!htbp]
    \centering
    \includegraphics[width=0.5\linewidth]{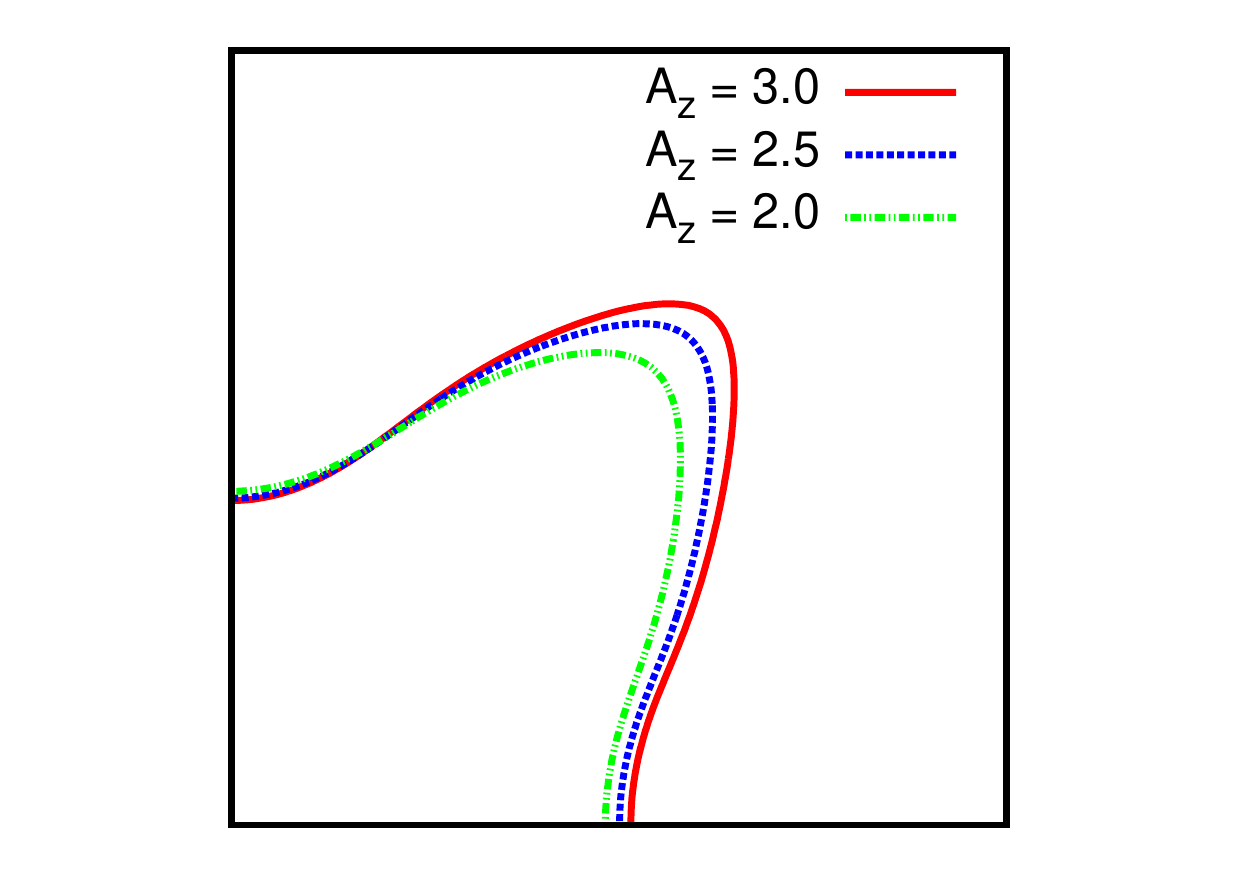}
    \caption{Contours of $\phi = 0.5$ at a 
    normalized time of $18436$ in the plane $(110)$ passing through 
    the center of simulation box for different magnitudes of $A_z$ 
    ($A_z = 2.0$, $2.5$, and $3.0$). Higher $A_z$ shows
    faster growth of the dendritic tip.}
    \label{fig:contour_t_4000}
\end{figure}

\begin{figure}[htbp]
    \centering
    \subfigure[]{\includegraphics[width=0.49\linewidth]{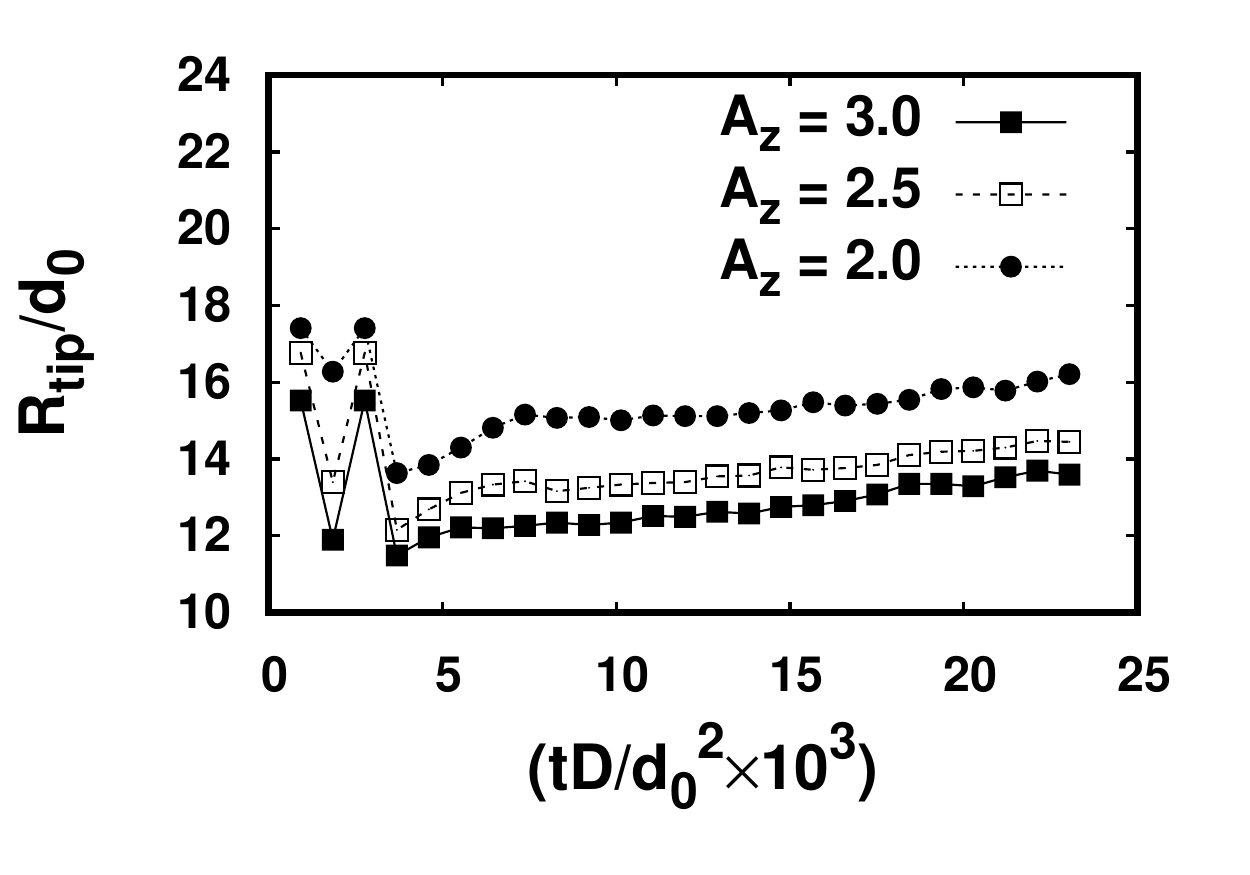}
    \label{fig:rtip_Az_effect}}%
    \centering
    \subfigure[]{\includegraphics[width=0.49\linewidth]{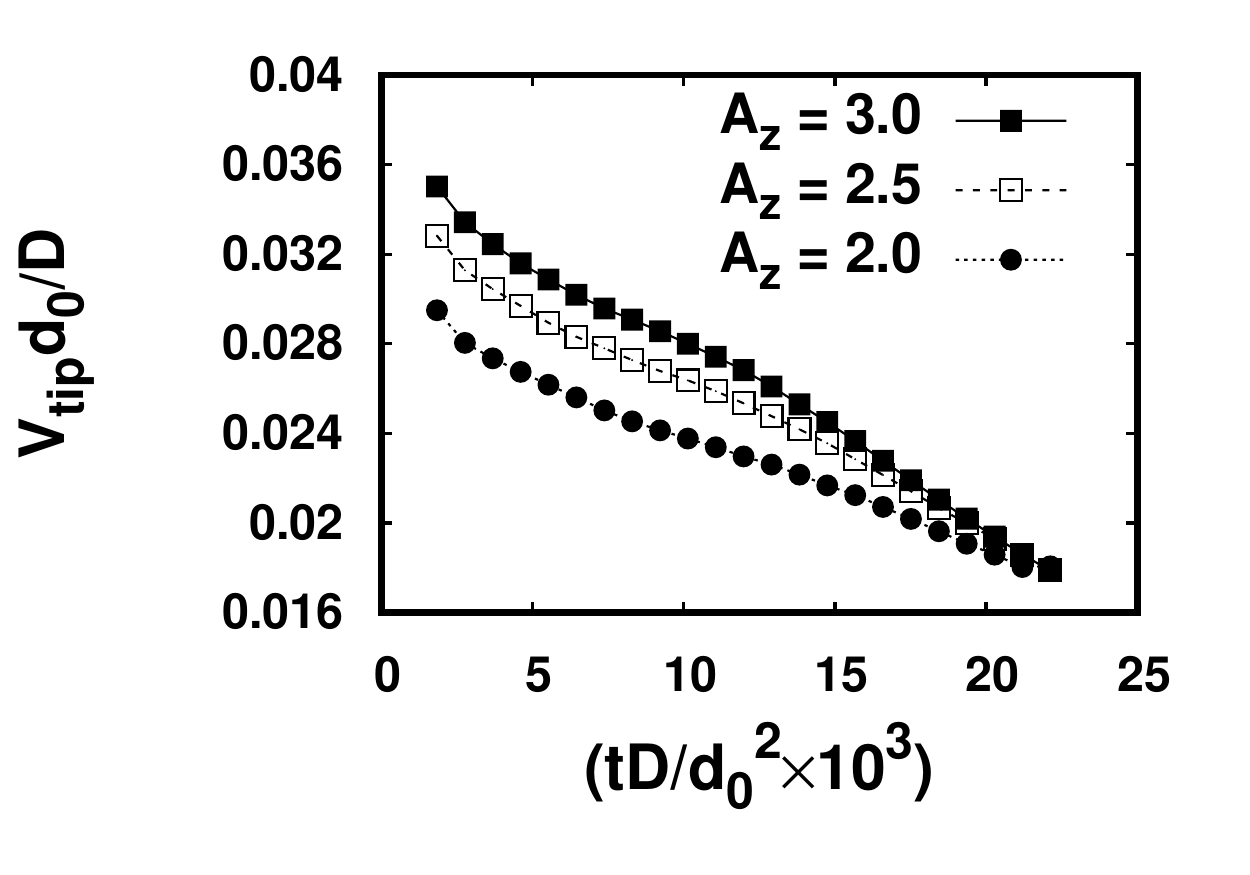}
    \label{fig:vtip_Az_effect}}
    \centering
    \subfigure[]{\includegraphics[width=0.49\linewidth]{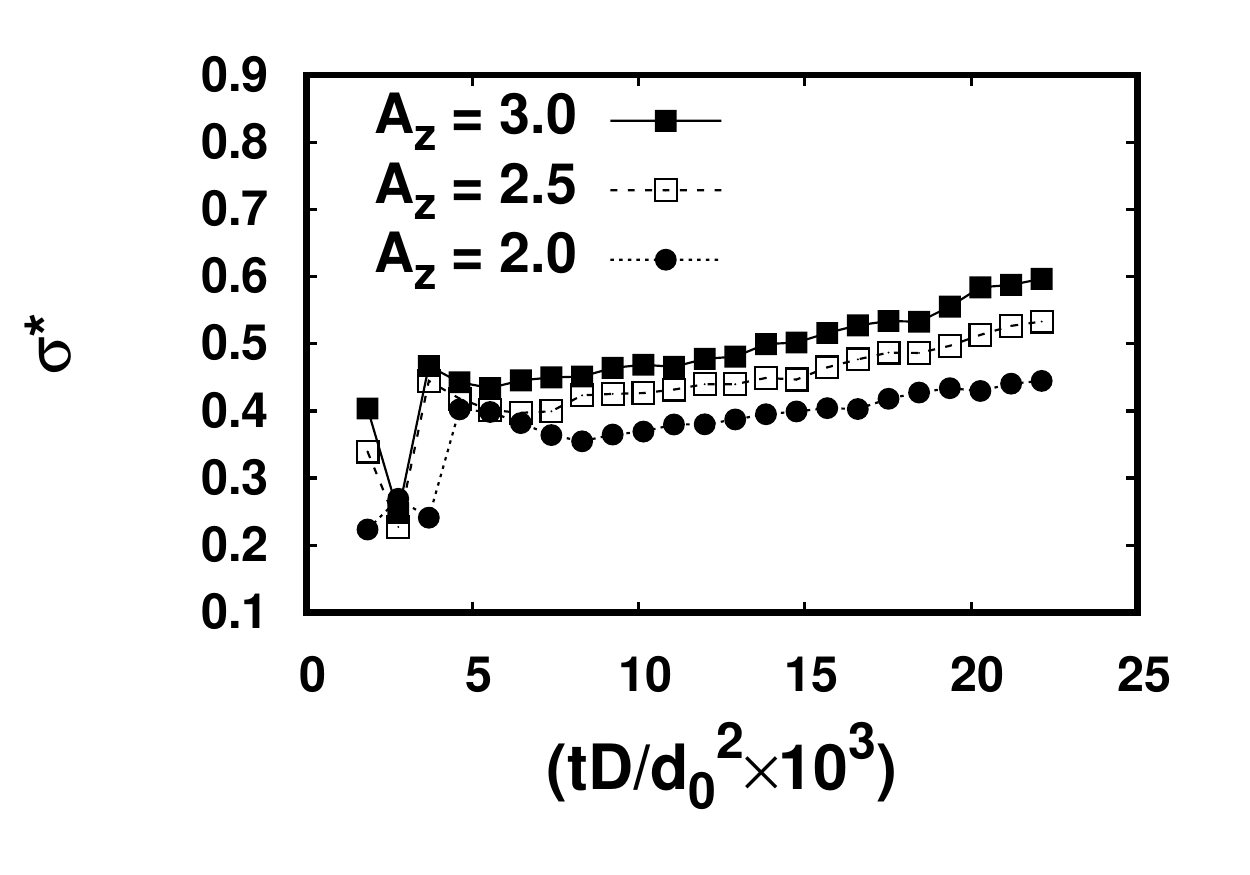}
    \label{fig:sigma_star_Az_effect}}
    \caption{Effect of strength of anisotropy in elastic energy on
     the variation of (a) $R_{tip}$ (b) $V_{tip}$, and (c) 
     $\sigma^*$ as a function of scaled time. 
     Here, misfit strain is $1\%$ and supersaturation is $53\%$.
     As $A_z$ increases, the dendritic tip becomes
     sharper, $V_{tip}$ increases, and $\sigma^*$ tends to 
     increase.}
    \label{fig:Az_effect}
\end{figure}

\begin{figure}[htbp]
    \centering
    \subfigure[]{\includegraphics[width=0.49\linewidth]{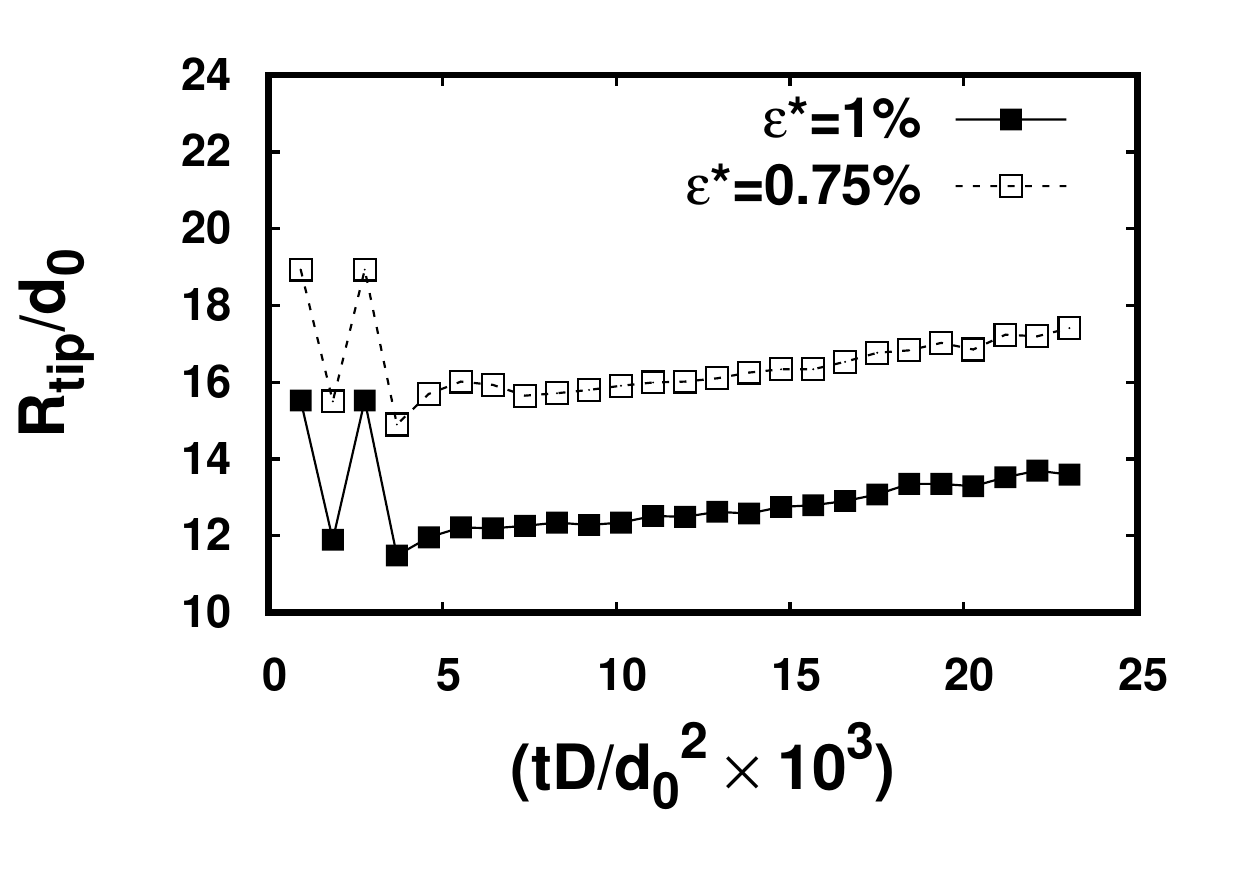}
    \label{fig:rtip_evolve_misfit_effect}}%
    \centering
    \subfigure[]{\includegraphics[width=0.49\linewidth]{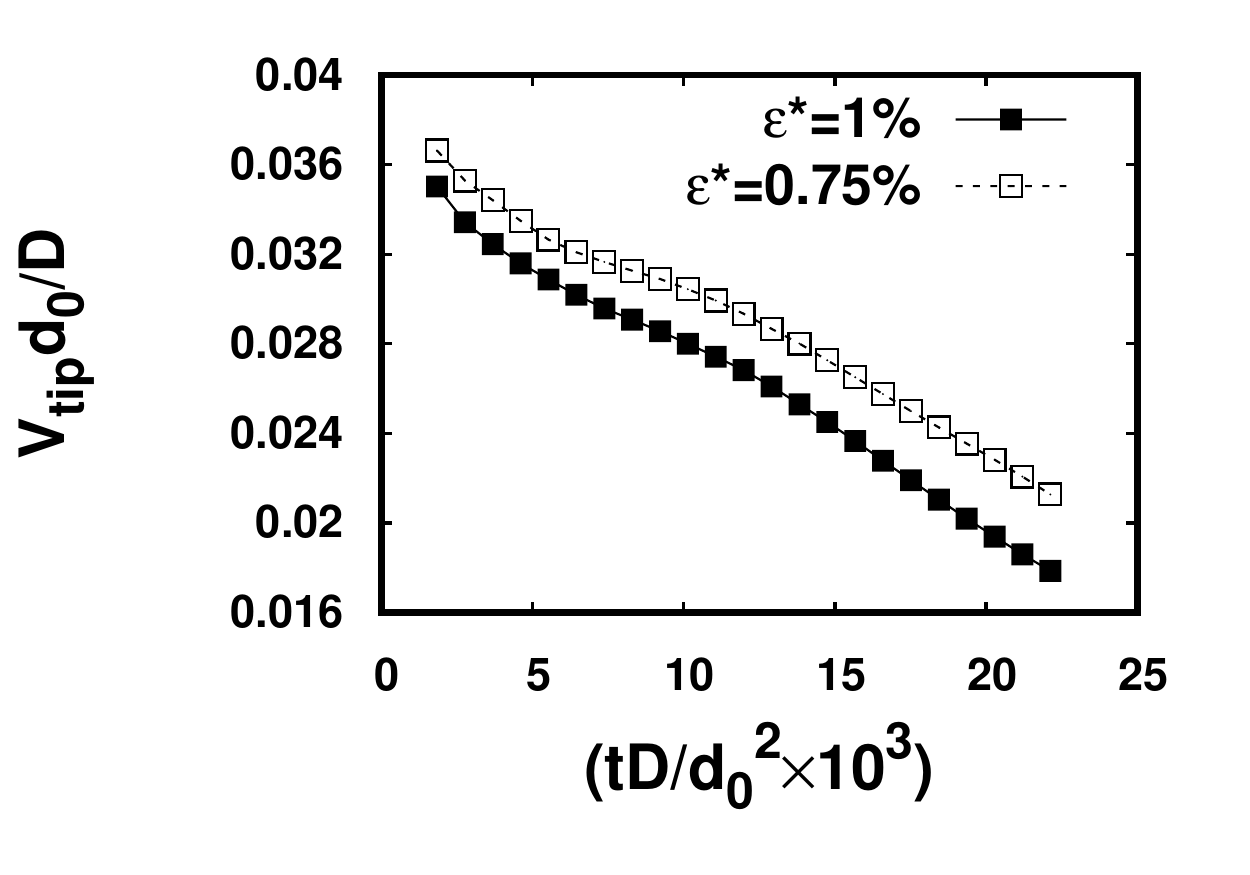}
    \label{fig:vtip_evolve_misfit_effect}}
    \centering
    \subfigure[]{\includegraphics[width=0.49\linewidth]{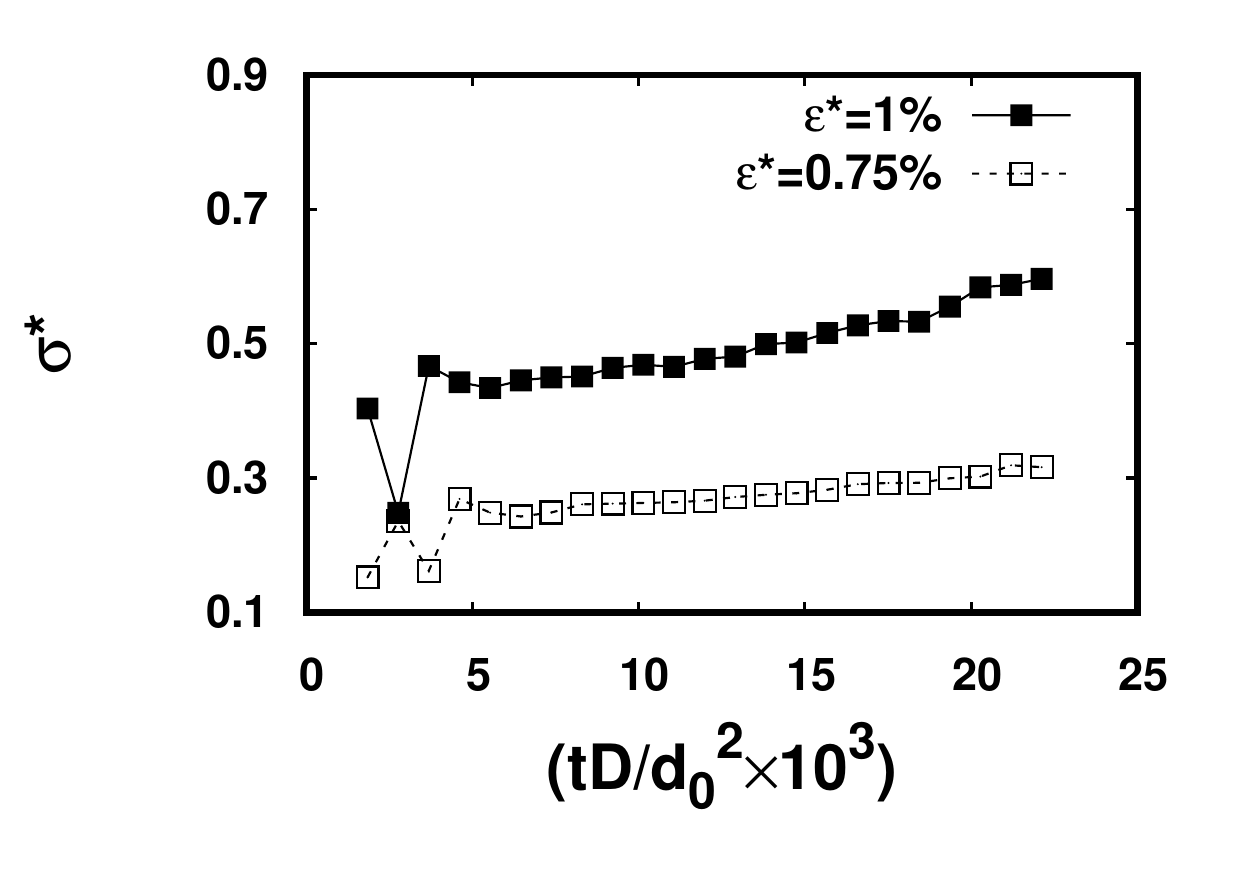}
    \label{fig:sigma_star_evolve_misfit_effect}}
    \caption{Effect of misfit strain on the variation of (a) $R_{tip}$ 
    (b) $V_{tip}$, and (c) $\sigma^*$ as a function of scaled 
    time $tD/d_0^2$. Here, the supersaturation is 53\% and Zener 
    anisotropy parameter is 3. With the decrease in misfit 
    strain, dendritic tip becomes more blunt, tip grows faster, 
    and $\sigma^*$ decreases.}
    \label{fig:misfit_effect}
\end{figure}

3D simulations were also performed for different values of 
$A_z$ for a supersaturation $\omega=45\%$. 
Fig~\ref{fig:contour_t_4000} represents the 
contour plot of $\phi = 0.5$ in the $(110)$ plane at different
levels of elastic anisotropy. The figure shows that as the 
anisotropy in elastic energy increases, the precipitate grows 
faster with a sharper dendrite tip. 
Figs.~\ref{fig:rtip_Az_effect} and~\ref{fig:vtip_Az_effect} 
reflects the same effect of decrease in $R_{tip}$ and increase
in $V_{tip}$ with increase in $A_z$, respectively. 
Here, too, we observe that $R_{tip}$ and $V_{tip}$ do 
not achieve a steady state value at all 
levels of $A_z$. Fig.~\ref{fig:sigma_star_Az_effect} shows the
variation $\sigma^*$ with the scaled time at different levels of 
$A_z$. The microsolvability constant $\sigma^*$ also does not 
achieve a steady state value over time at all 
levels of $A_z$. The influence of the misfit-strain on the 
observed microstructures is depicted in 
Fig.~\ref{fig:rtip_evolve_misfit_effect} where the
radius of the tip becomes sharper with an increase in the 
value of the misfit strain. Along with this, the velocity 
shows an increasing trend with higher values of the misfit 
strain as highlighted in 
Fig.~\ref{fig:vtip_evolve_misfit_effect}.
The selection constant on the other hand also increases with 
larger misfit, however no steady state is 
achieved as shown in Fig.~\ref{fig:sigma_star_evolve_misfit_effect}.

\begin{figure}[!htbp]
 \centering
 \subfigure[]{\includegraphics[width=0.48\linewidth]{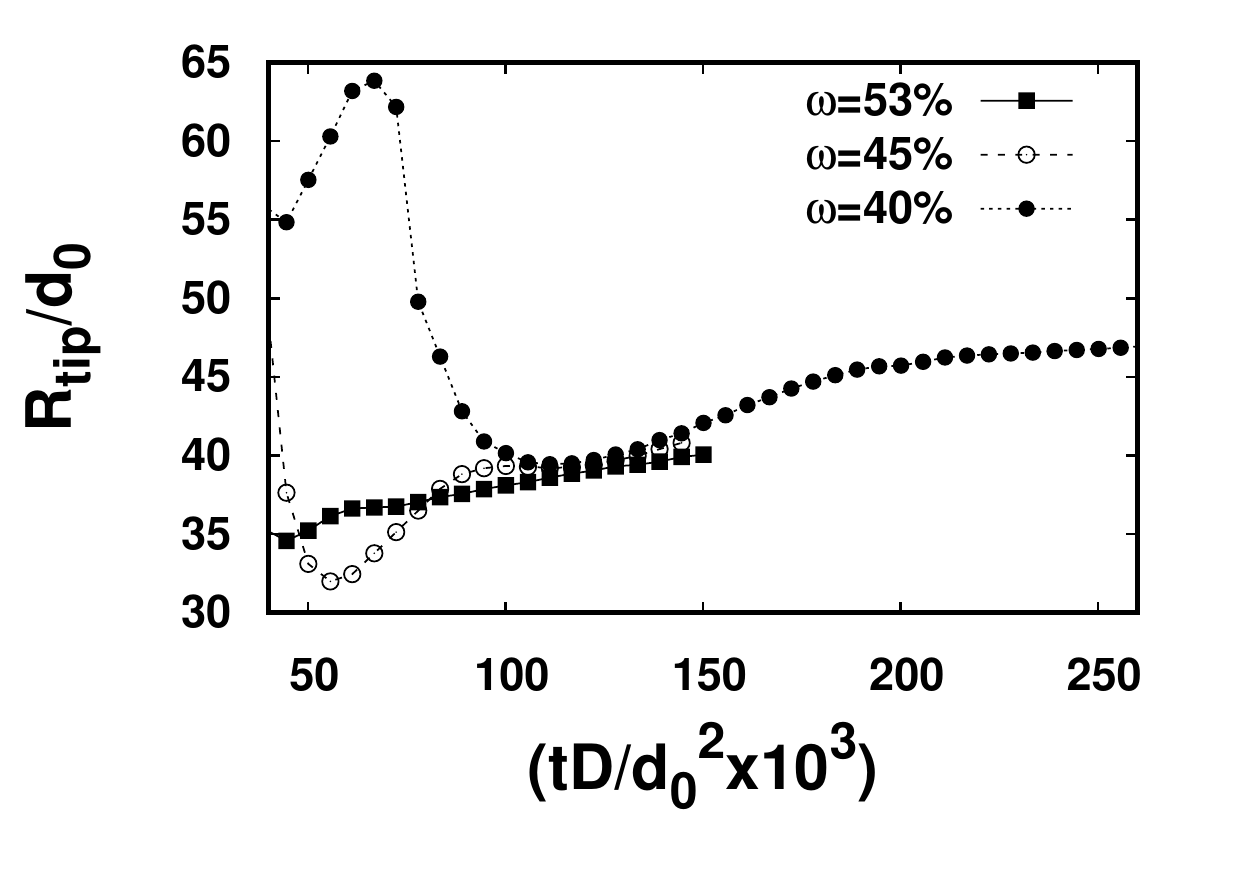}
 \label{radius_super}
 }
  \centering
 \subfigure[]{\includegraphics[width=0.48\linewidth]{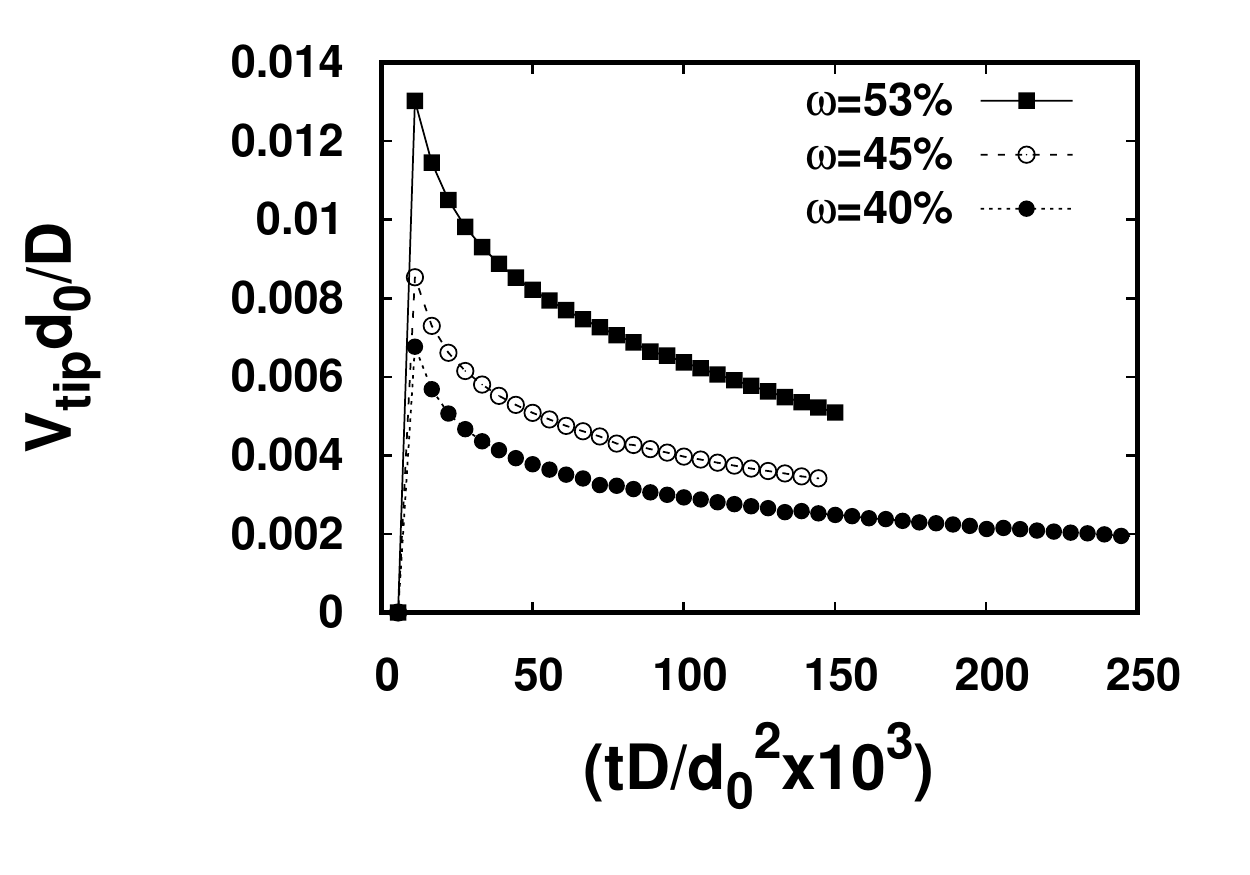}
 \label{velocity_super}
 }
 \centering
 \subfigure[]{\includegraphics[width=0.48\linewidth]{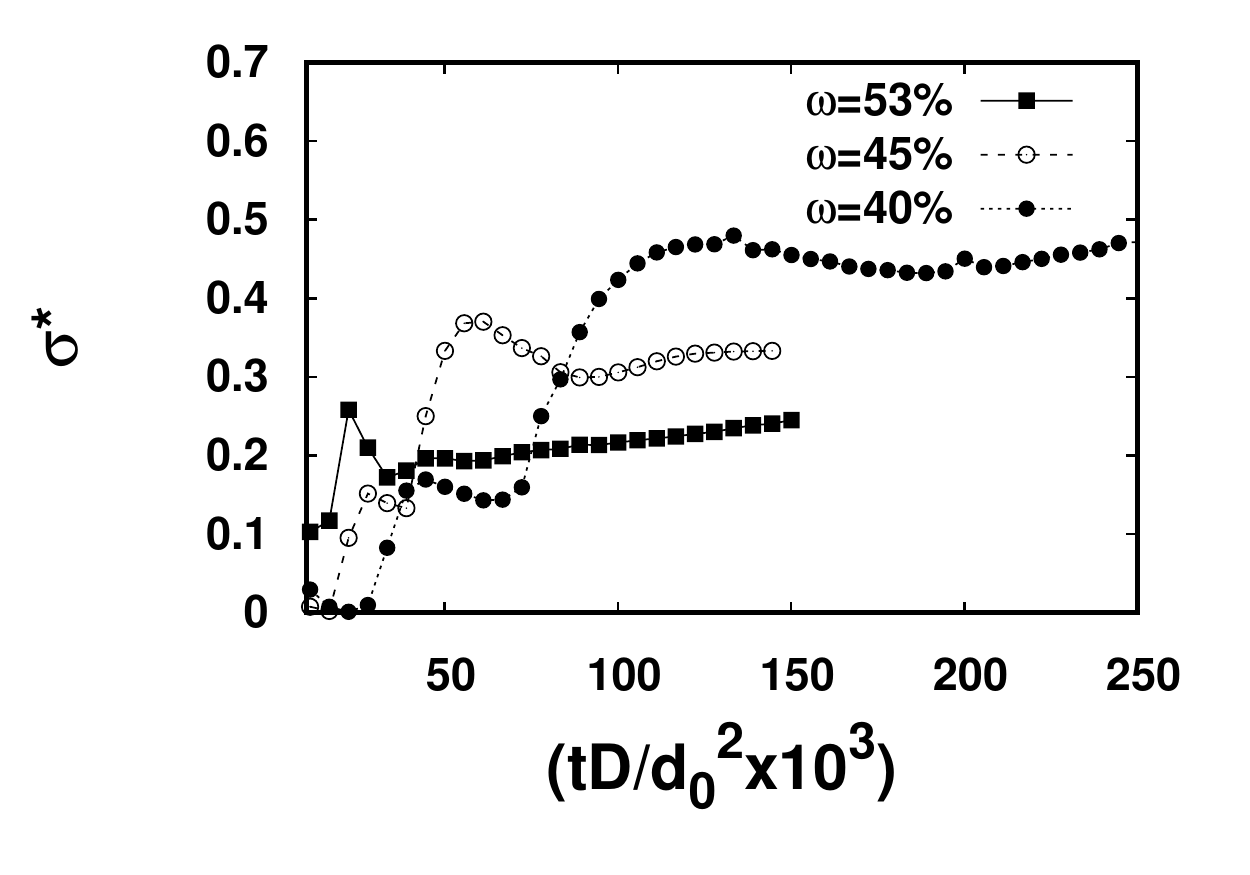}
 \label{sigma_super}}
\caption{ Effect of supersaturation on the temporal evolution 
of (a) $R_{tip}$ and (b) $V_{tip}$, (c) $\sigma^*$. 
Here, Zener anisotropy parameter is $3.0$ and supersaturation is 
53\%. With the increase in supersaturation, tip becomes more 
sharper, $V_{tip}$ increases, and $\sigma^*$ decreases.}
\label{rtip_vtip_vary_super}
\end{figure}

Fig.~\ref{rtip_vtip_vary_super} depicts the effect of change in 
the degree of the supersaturation in the matrix, while the 
anisotropy in the elastic energy as well as the magnitude of the 
misfit strain are kept constant, i.e., $A_z=3.0$ and 
$\epsilon^*=1\%$. The plots show that the radius of the 
dendritic tip becomes sharper (see Fig.~\ref{radius_super}) with 
increase in the magnitude of supersaturation whereas
the velocity increases (see Fig.~\ref{velocity_super}), but does 
not saturate. Similarly, the magnitude of the selection constant 
increases with time as shown in Fig.~\ref{sigma_super}.%

\begin{figure}[htbp]
    \centering
    \subfigure[]{\includegraphics[width=0.49\linewidth]{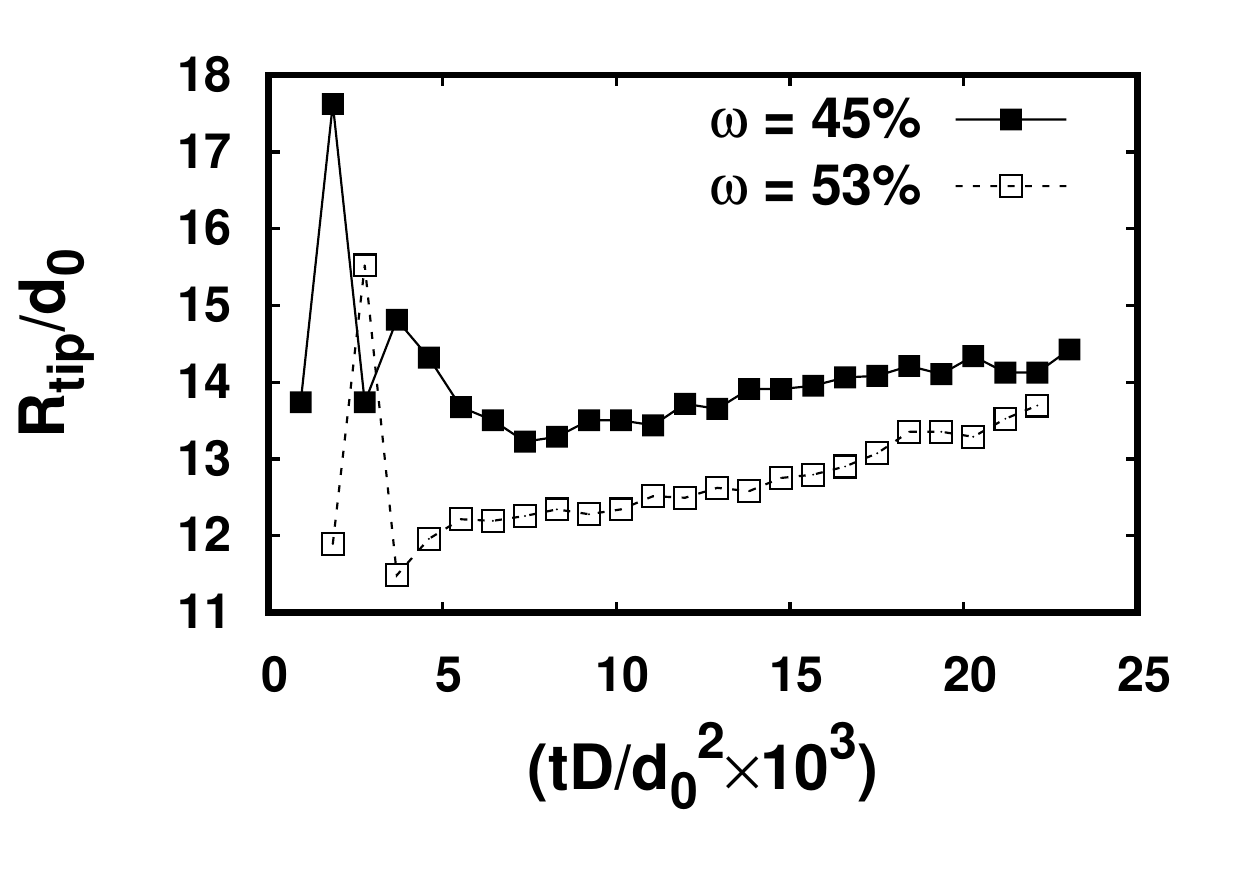}
    \label{fig:rtip_evolve_c0_effect}}%
    \centering
    \subfigure[]{\includegraphics[width=0.49\linewidth]{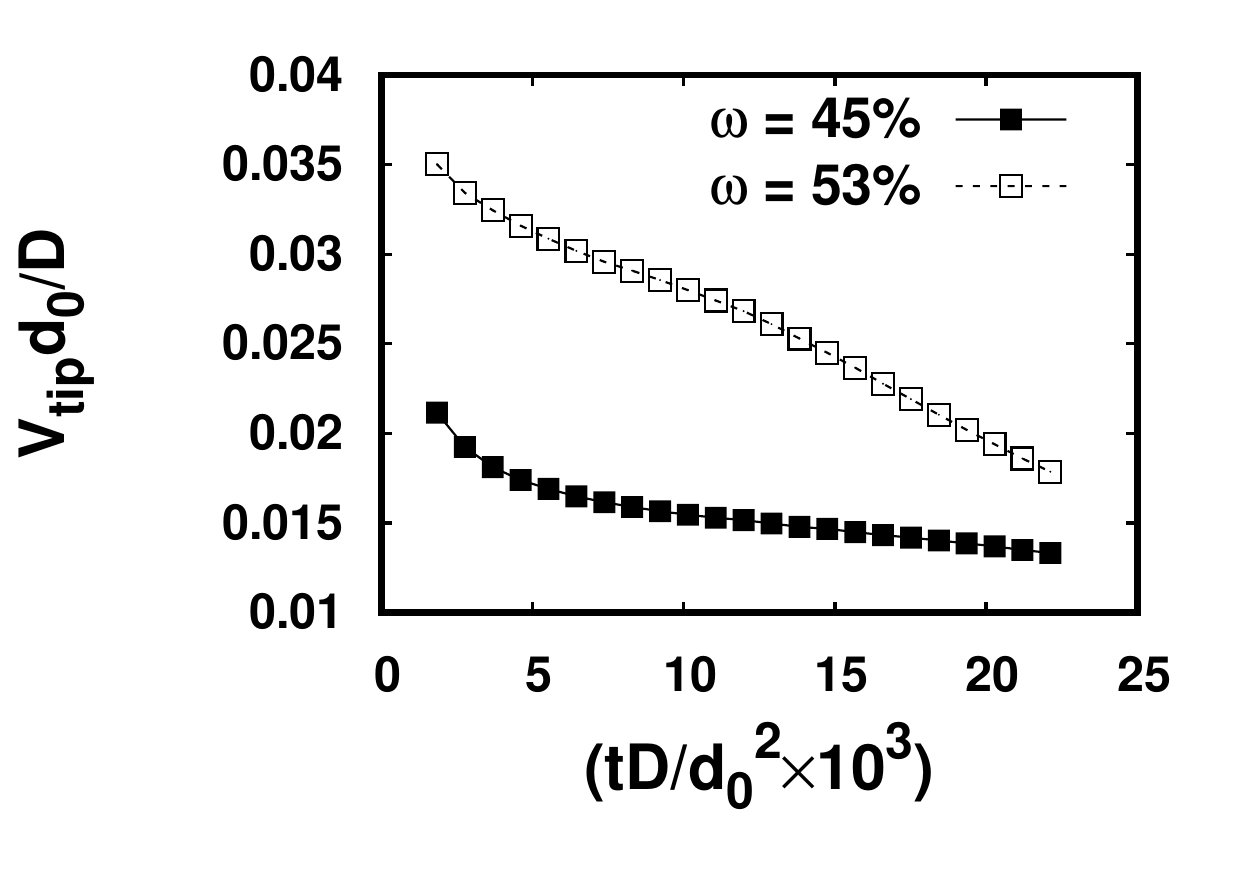}
    \label{fig:vtip_evolve_c0_effect}}
    \centering
    \subfigure[]{\includegraphics[width=0.49\linewidth]{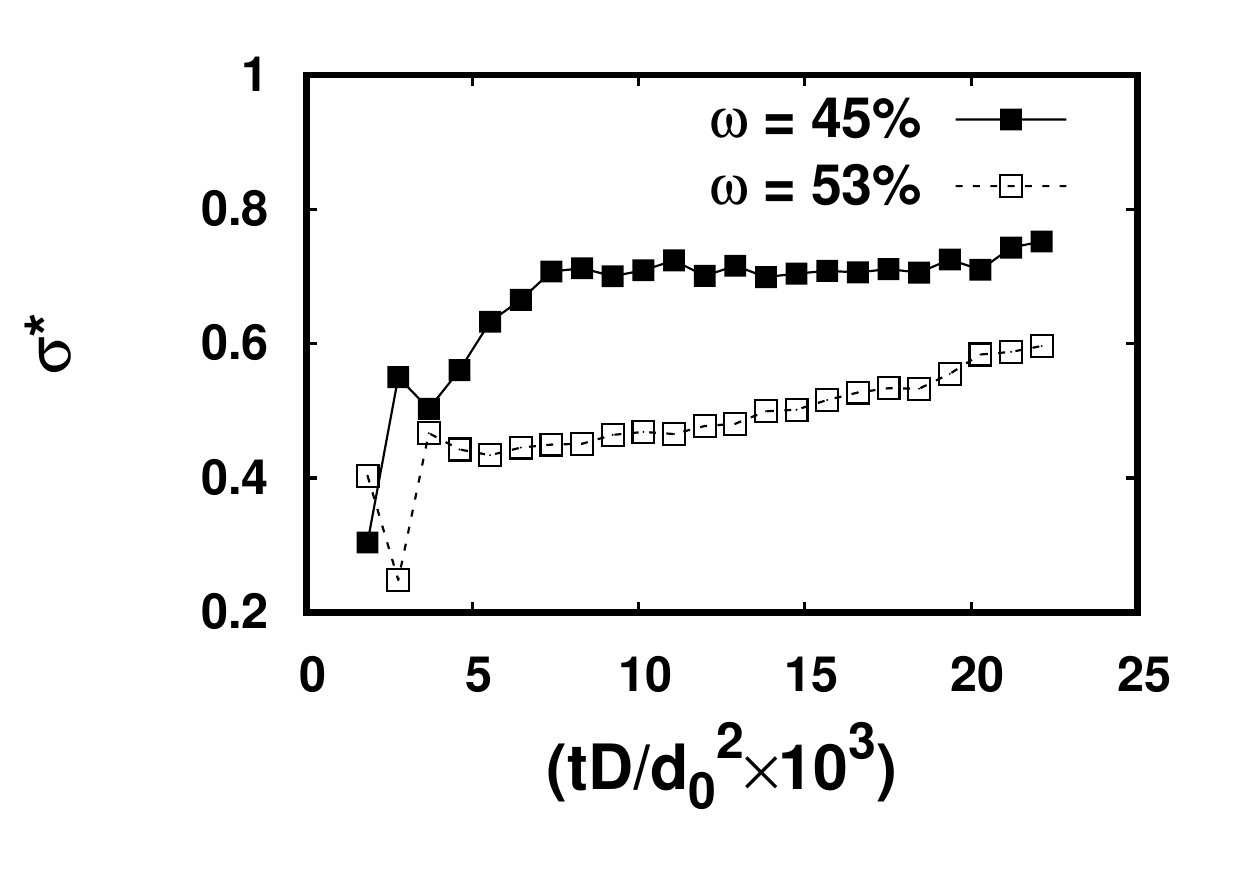}
    \label{fig:sigma_star_evolve_c0_effect}}
    \caption{Effect of supersaturation on the variation 
    of (a) $R_{tip}$ (b) $V_{tip}$, and (c) $\sigma^*$ 
    as a function of scaled time $tD/d_0^2$. Here, the 
    misfit strain is 1\% and strength of elastic energy 
    anisotropy is 3.0. As the supersaturation increases, 
    tip becomes more sharper, tip grows faster, and 
    $\sigma^*$ decreases.}
    \label{fig:c0_effect}
\end{figure}

3D simulations with $A_z=3.0$ and $\epsilon^*=1\%$ and 
different supersaturations were also performed with similar
conclusions. Fig.~\ref{fig:rtip_evolve_c0_effect}
reveal a decrease in the radius of the tip and a corresponding
increase in the velocity Fig.~\ref{fig:vtip_evolve_c0_effect} 
with increasing supersaturation. The 3D simulations were limited 
by domain size and we could not access supersaturations that are lower. 
The selection constant also shows a decreasing trend with higher 
supersaturations as revealed in Fig.~\ref{fig:sigma_star_evolve_c0_effect}.

\section{Competition between anisotropies in the interfacial energy and elastic energy}
In the previous section, we have seen the effect of different 
variables such as the variation of supersaturation, misfit strain and 
anisotropy in elastic energy on the evolution of the 
dendrite-like morphologies. In this section, we investigate the 
competition between the influence of anisotropies 
in the elastic and the interfacial energies. 
For brevity, we will only utilize 2D simulations in this section for
investigating the competition between the two forms of anisotropy 
and the influence on the dynamics of the instability. The anisotropy
in the interfacial energy is modeled using the 
function~\cite{karma1998}, Eqn.~\ref{Eqn_anisotropy}, which leads 
to the formation of dendrites aligned along the $\langle 10 \rangle$
directions.
 
Here, we keep the anisotropy in elastic energy $(A_z)$ constant 
and allow the strength in anisotropy in interfacial energy to vary 
in the range of 0.0 to 0.03. Also, here we have considered two broad 
categories with different magnitudes of elastic anisotropies, i.e., 
$A_z=3.0$ and $A_z=0.5$. In the first case, as the magnitude of 
$A_z$ is above one (similar to the case studied before),
the interfacial energy and the elastic energy anisotropies 
lead to the formation of dendrites in 
$\langle 11 \rangle$ and $\langle 10 \rangle$ directions 
respectively, whereas, in the second case, for values of $A_z$ less 
than unity, the anisotropies in the elastic energy and the 
interfacial energy superimpose.\\ 
\subsection{Case A: Strength of elastic anisotropy $(A_z)>1.0$}
\begin{figure}[!htbp]
 \centering
 \includegraphics[width=0.3\linewidth]{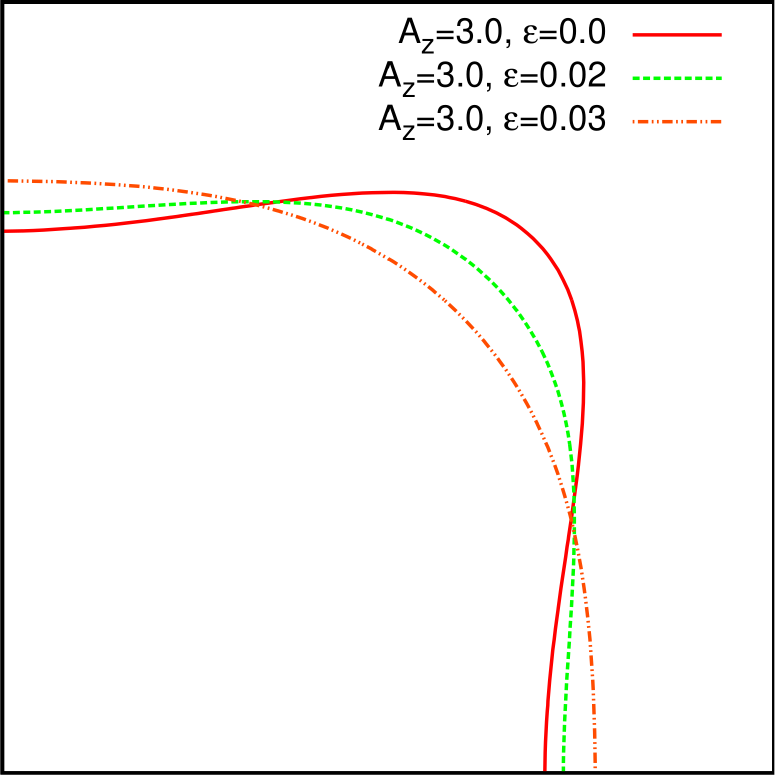}
 \caption{Early stage precipitate growth under combined effect of 
 anisotropies in the elastic energy and the interfacial energy at  
 normalized time of 2780 for  Zener anisotropy parameter of $3.0$.
 Here, at $A_z=3.0$ and $\varepsilon=0.03$, the precipitate acquires
 nearly circular shape. }
  \label{early_stage_dendrits}
\end{figure}

\begin{figure}[!htbp]
 \centering
 \includegraphics[width=0.3\linewidth]{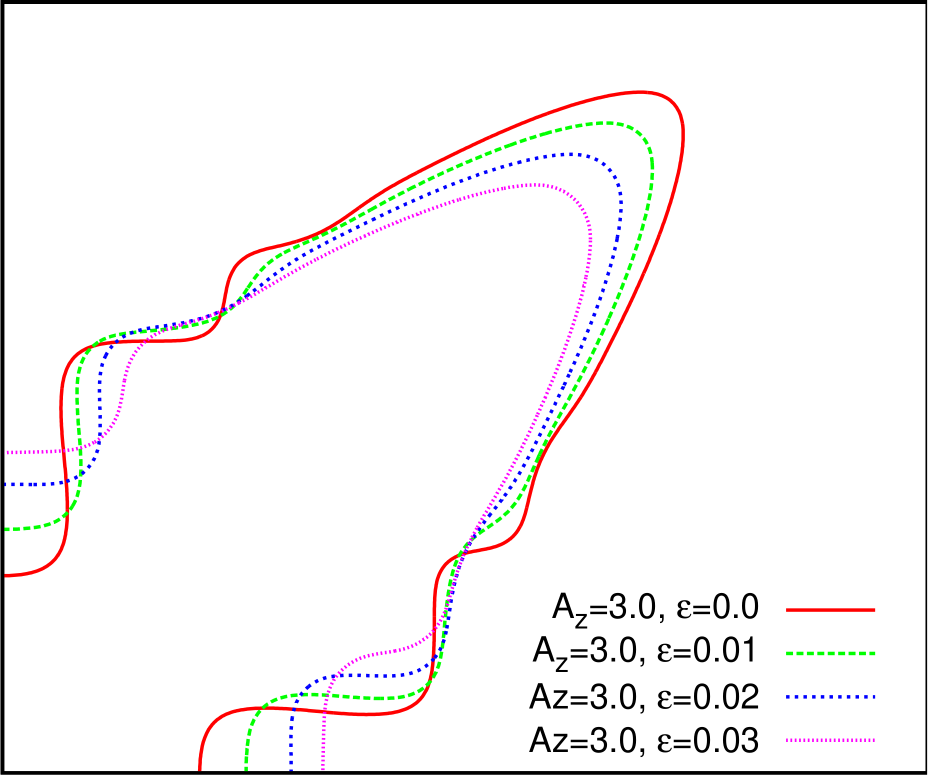}
 \caption{Contours of $\phi = 0.3$ at a normalized time  $t=55600$ 
 showing one-fourth section of the symmetric dendritic structure under the 
 combined effect of anisotropies in the elastic energy and the 
 interfacial energy. Here, Zener anisotropy parameter is 3 and 
 strength of anisotropy in interfacial energy varies from 
 $\varepsilon=0.0-0.03$. The increase in $\varepsilon$ slows down the dendritic growth along $\langle 11\rangle$.}
  \label{dendrits_sfaniso}
\end{figure}

\begin{figure}[!htbp]
 \centering
 \subfigure[]{\centering 
 \includegraphics[width=0.48\linewidth]{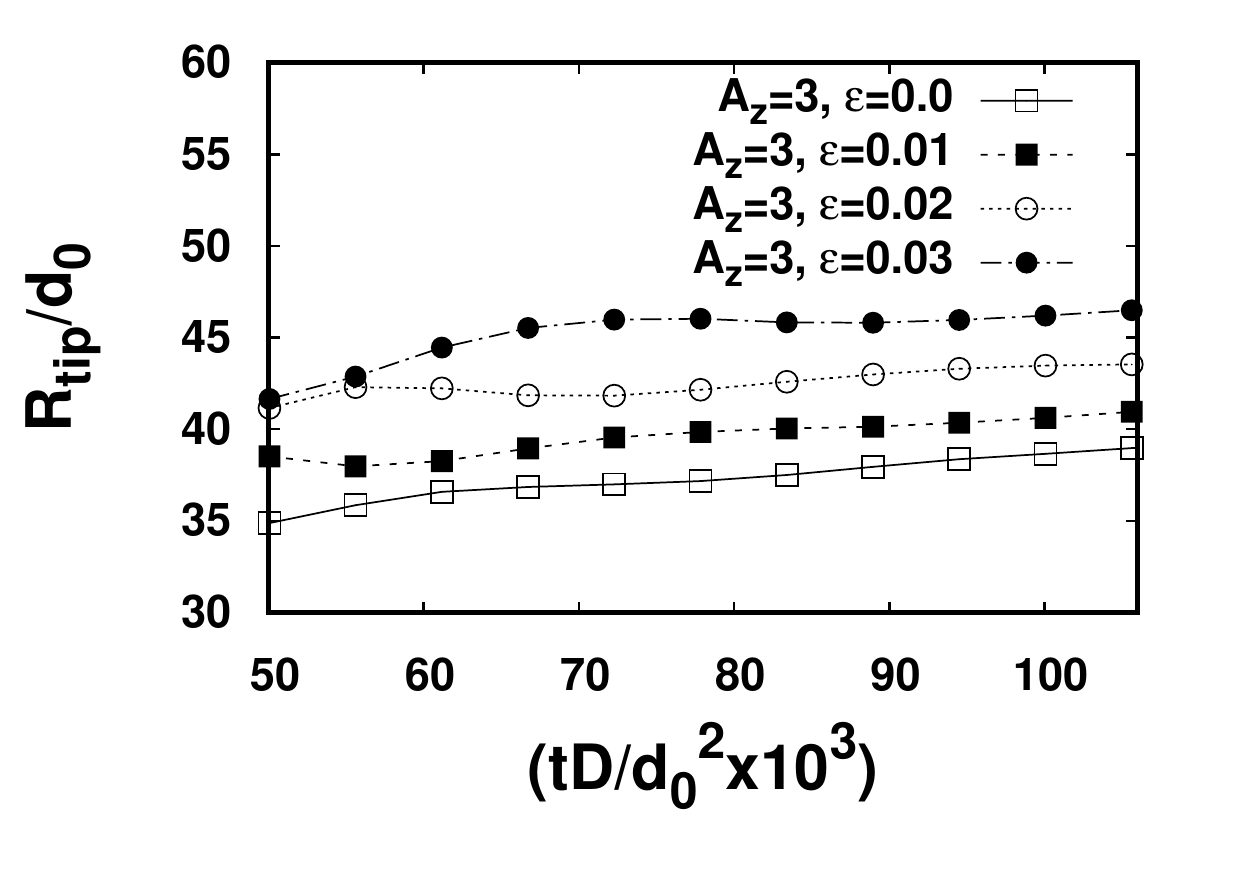}
 \label{radius_vary_inter_aniso_es1_az3}
 }%
 \subfigure[]{\centering 
 \includegraphics[width=0.48\linewidth]{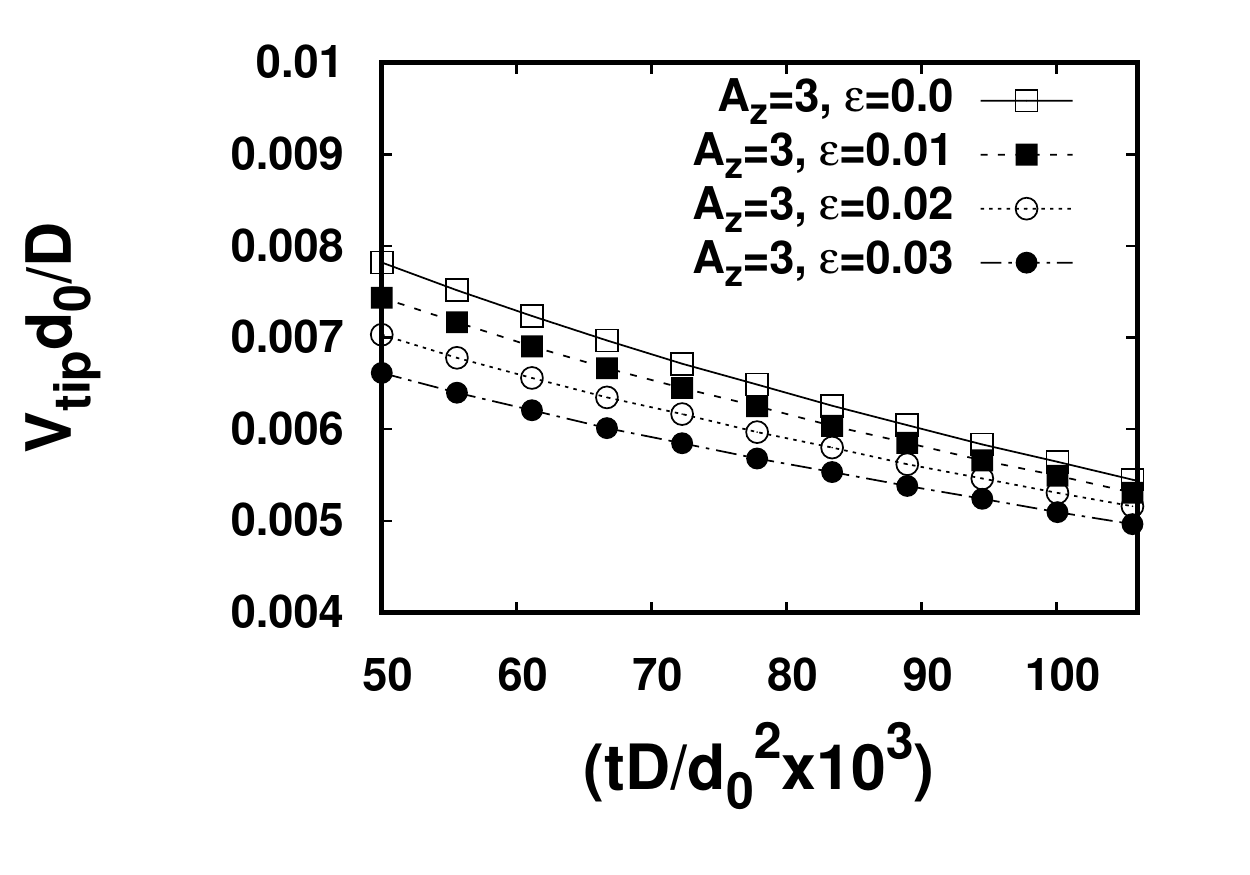}
 \label{velocity_vary_inter_aniso_es1_az3}
 } 
 \subfigure[]{\includegraphics[width=0.48\linewidth]{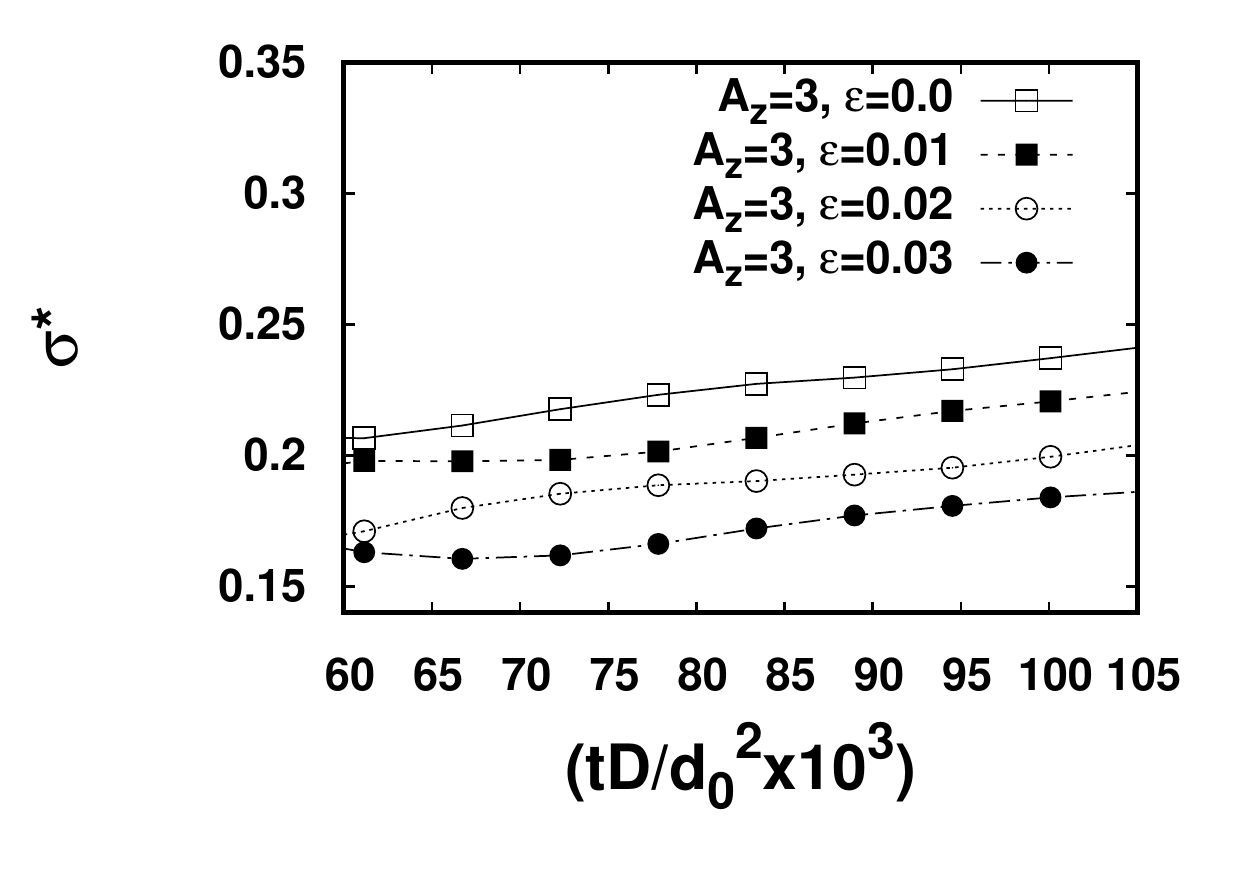}
 \label{sigma_diff_az} 
 }
\caption{ Effect of anisotropy in interfacial energy on the 
variation of (a) $R_{tip}$, (b) $V_{tip}$ and (c) $\sigma^*$ at $A_z = 3.0$. 
Here, the misfit strain is 1\% and supersaturation is 53\%. 
The increase in $\varepsilon$ leads to wider dendritic tip with slower 
tip velocity and decrease in $\sigma^*$.}
\label{rtip_vtip_vary_inter_aniso_es1_az3}
\end{figure}

For this case where the anisotropies in the interfacial energy and 
elasticity lead to dendrite-like structures in different directions, 
the combined influence of anisotropies in both the 
energies gives rise to a precipitate shape which is nearly circular in 
the early stage of growth as shown in Fig.~\ref{early_stage_dendrits}, at $A_z=3.0$ and 
$\varepsilon=0.03$.
But, with decrease in the strength of anisotropy in the interfacial 
energy from 0.03 to 0.0, the precipitate not only develops sharp 
corners but also shows strong alignment along $\langle 11 \rangle$ 
directions, that is the elastically preferred direction. 
Fig.~\ref{dendrits_sfaniso} compares the morphologies for different
strengths of anisotropies in interfacial energy
$(\varepsilon=0.0-0.03)$ for a given value of $A_z=3.0$ at the 
normalized simulation time of $t=55600$.
Fig.~\ref{radius_vary_inter_aniso_es1_az3} depicts that as the 
interfacial energy anisotropy increases, the dendrite tip radius 
becomes larger while the velocity of the dendrite tip reduces as 
highlighted in Fig.~\ref{velocity_vary_inter_aniso_es1_az3}.

Similarly, while the magnitude of $\sigma^*$ again has a linearly 
increasing trend with simulation time (after the initial transient), 
the competition between the elastic energy anisotropy and the 
interfacial energy anisotropy leads to a decrease in the magnitude 
of $\sigma^*$ at a given time of the evolution of
the precipitate, which is expected as this is similar to the 
effective reduction of anisotropy in the system (see 
Fig.~\ref{sigma_diff_az}).  

\subsection{Case B: Strength of elastic anisotropy $(A_z)<1.0$}
\begin{figure}[!htbp]
 \centering
 \includegraphics[width=0.5\linewidth]{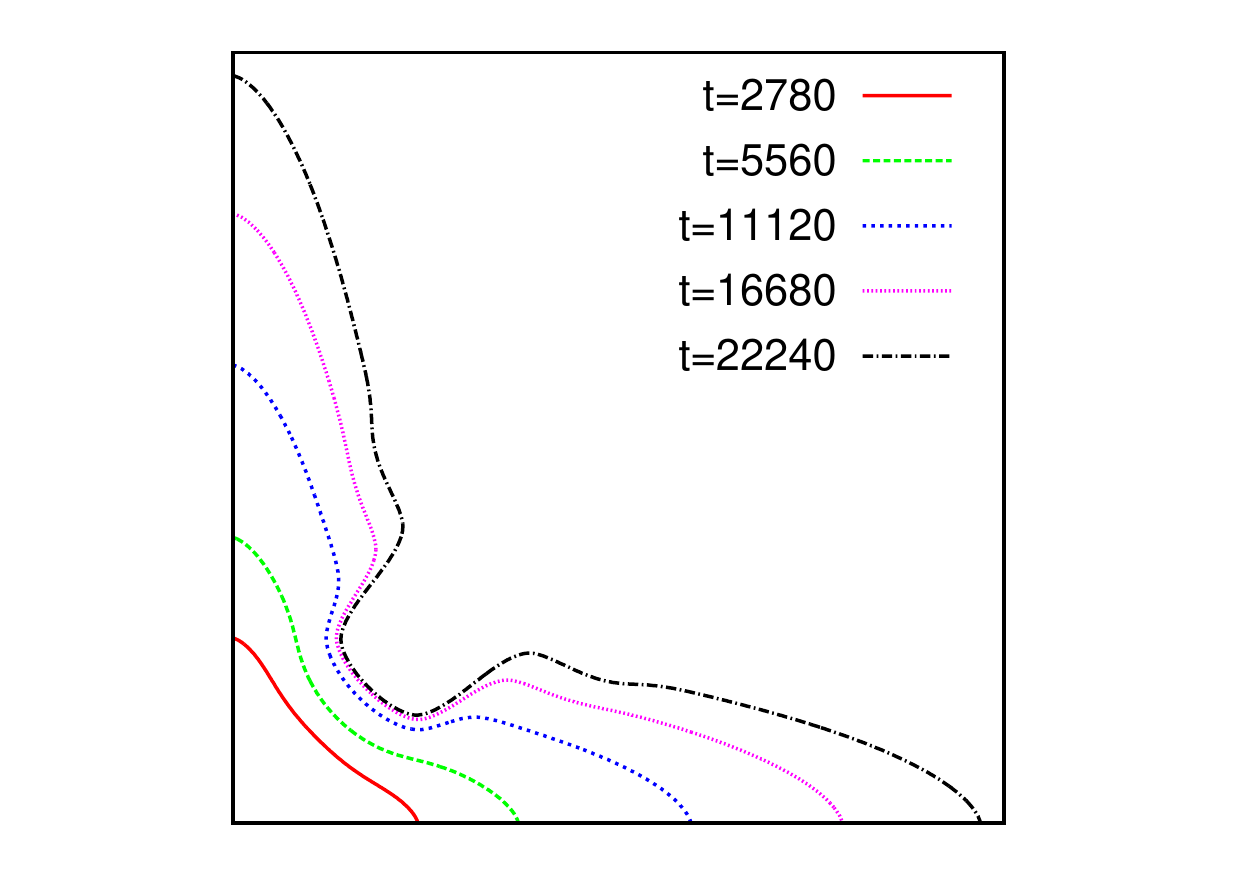}
 \caption{Contours of $\phi = 0.5$ in a one-fourth section at 
 normalized times $t = 2780$, $5560$, $11120$, $16680$, $22240$ 
 showing dendritic growth along  $\langle 10 \rangle$ directions. 
 Here, Zener anisotropy parameter is $0.5$, strength of anisotropy in 
 interfacial energy is $0.03$, supersaturation is $53\%$, and misfit 
 strain is $1\%$. }
   \label{dendrits_evol_sf03}
\end{figure}

\begin{figure}[!htbp]
 \centering
 \includegraphics[width=0.3\linewidth]{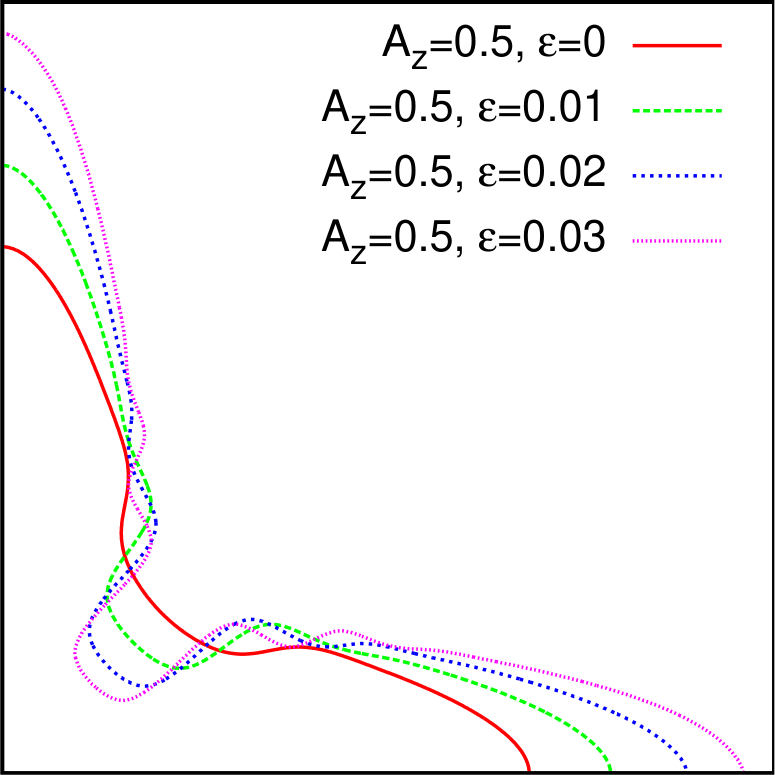}
 \caption{Contours of $\phi=0.5$ in a one-fourth section at a same 
 time showing dendritic structure for different strength 
 of anisotropies in interfacial energy. Here, the Zener anisotropy 
 parameter is $0.5$, the misfit strain is $1\%$, 
 and the supersaturation is $53\%$. The increase in $\varepsilon$
 accelerates the dendritic growth along $\langle 10 \rangle$.}
  \label{dendrit_compare_az0.05_sf}
\end{figure}

\begin{figure}[!htbp]
 \centering
 \subfigure[]{\includegraphics[width=0.48\linewidth]{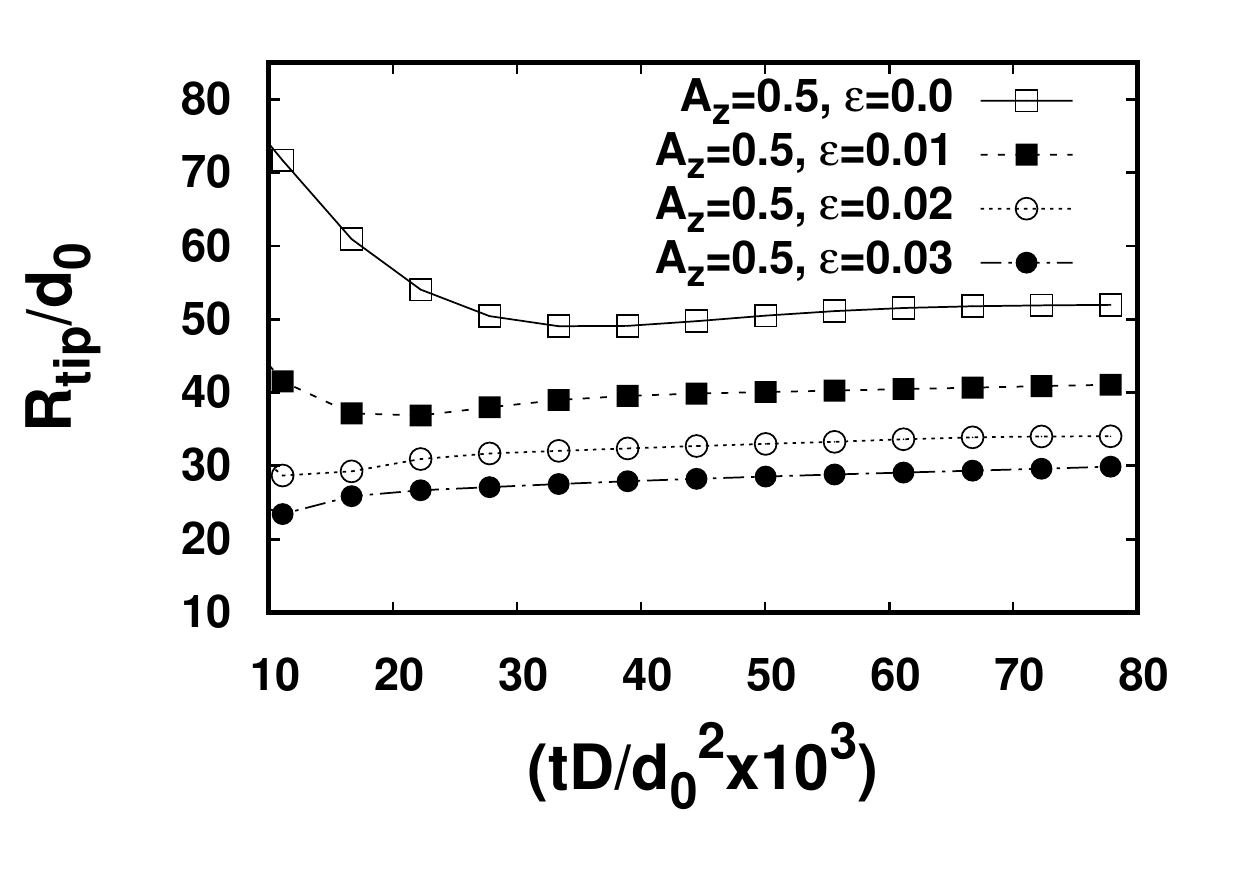}
 \label{radius_vary_inter_aniso_az_less_than_one}
 }%
  \centering
 \subfigure[]{\includegraphics[width=0.48\linewidth]{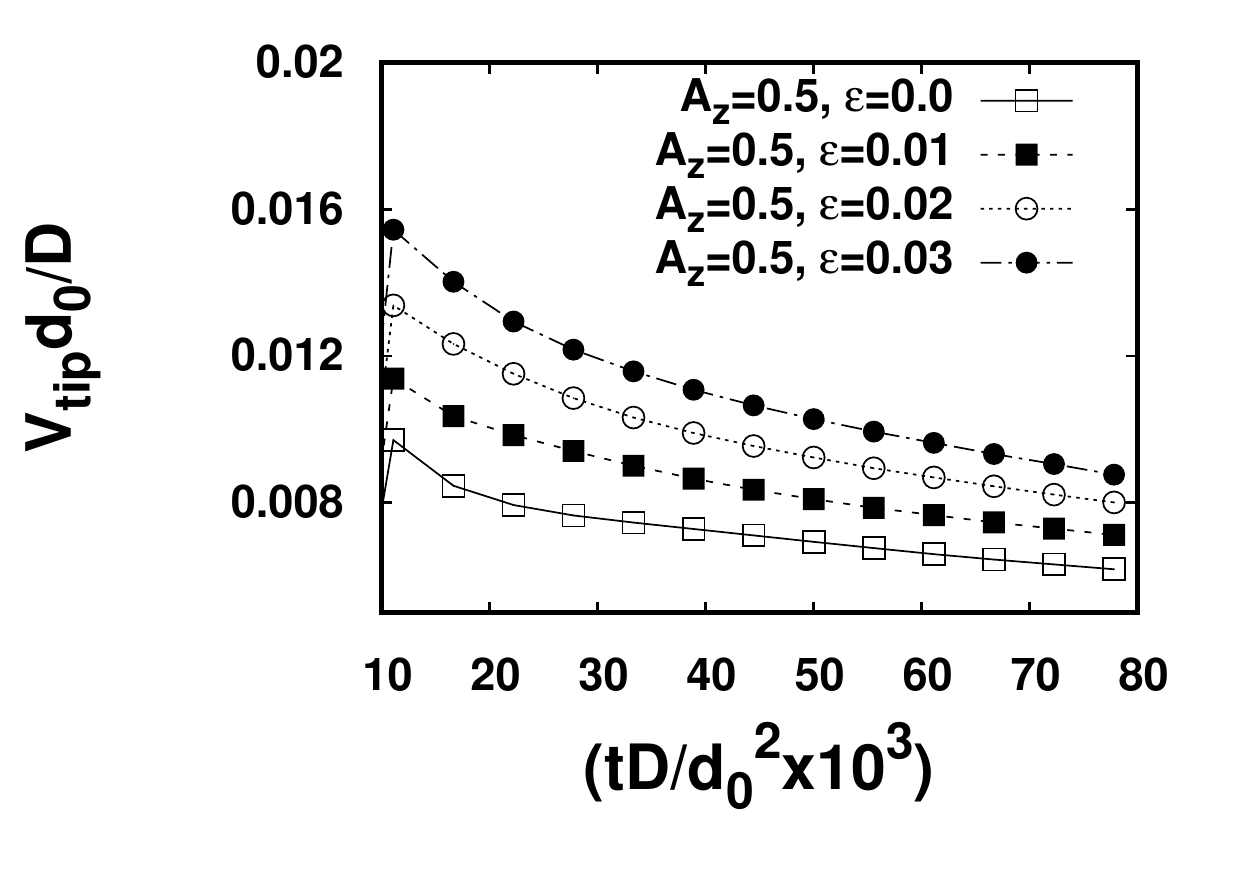}
 \label{velocity_vary_inter_aniso_az_less_than_one}
 }
 \subfigure[]{\includegraphics[width=0.48\linewidth]{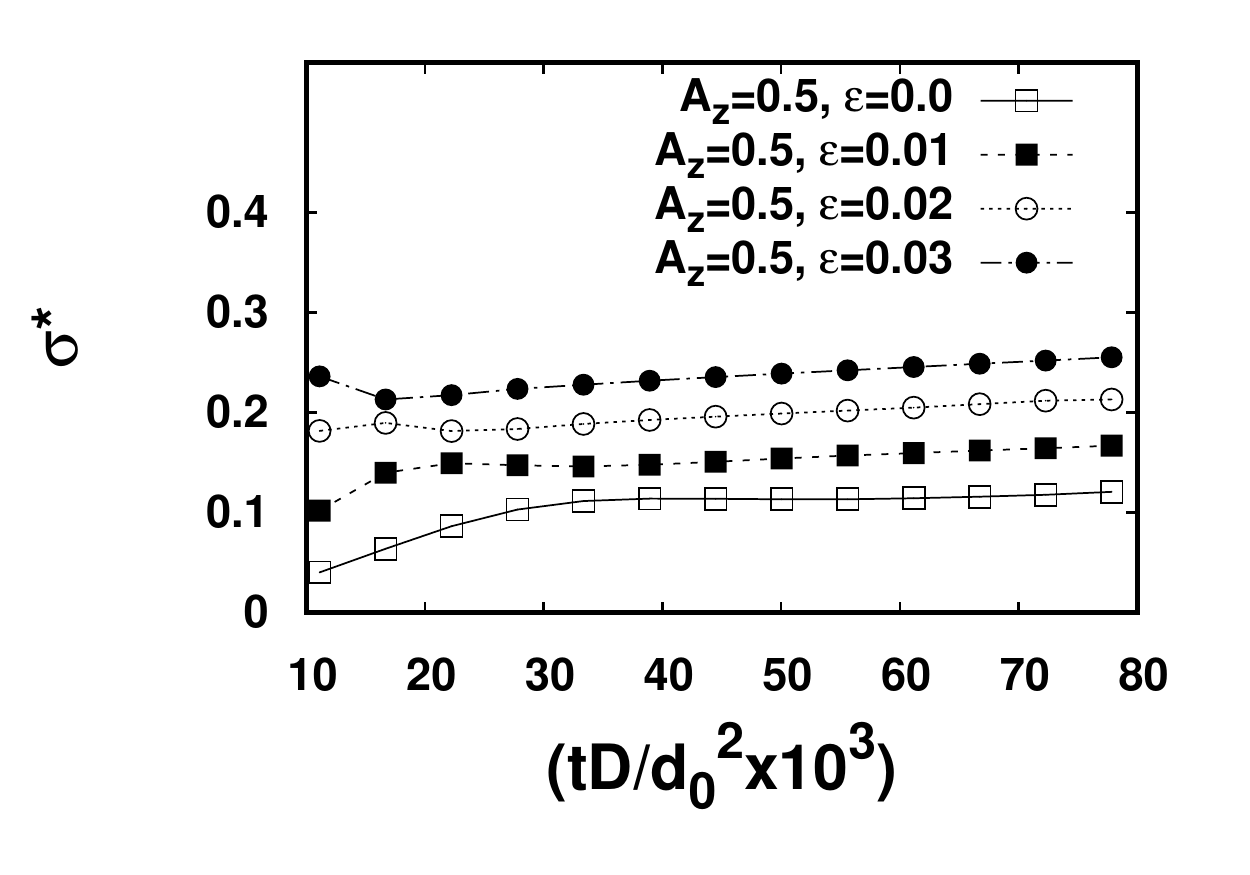}
 \label{sigma_az0.5_diff_sf_aniso}
 }
\caption{ Effect of anisotropy in interfacial energy on the 
temporal variation of (a) $R_{tip}$ (b) $V_{tip}$, and (c) $\sigma^*$ 
at $A_z=0.5$. Here, the misfit strain is 1\% and supersaturation is 53\%. 
The increase in $\varepsilon$ leads to sharper dendritic tip and 
faster tip velocity.}
\label{rtip_vtip_vary_inter_aniso_az_less_than_one}
\end{figure}

\begin{figure}[!htbp]
 \centering
 \includegraphics[width=0.5\linewidth]{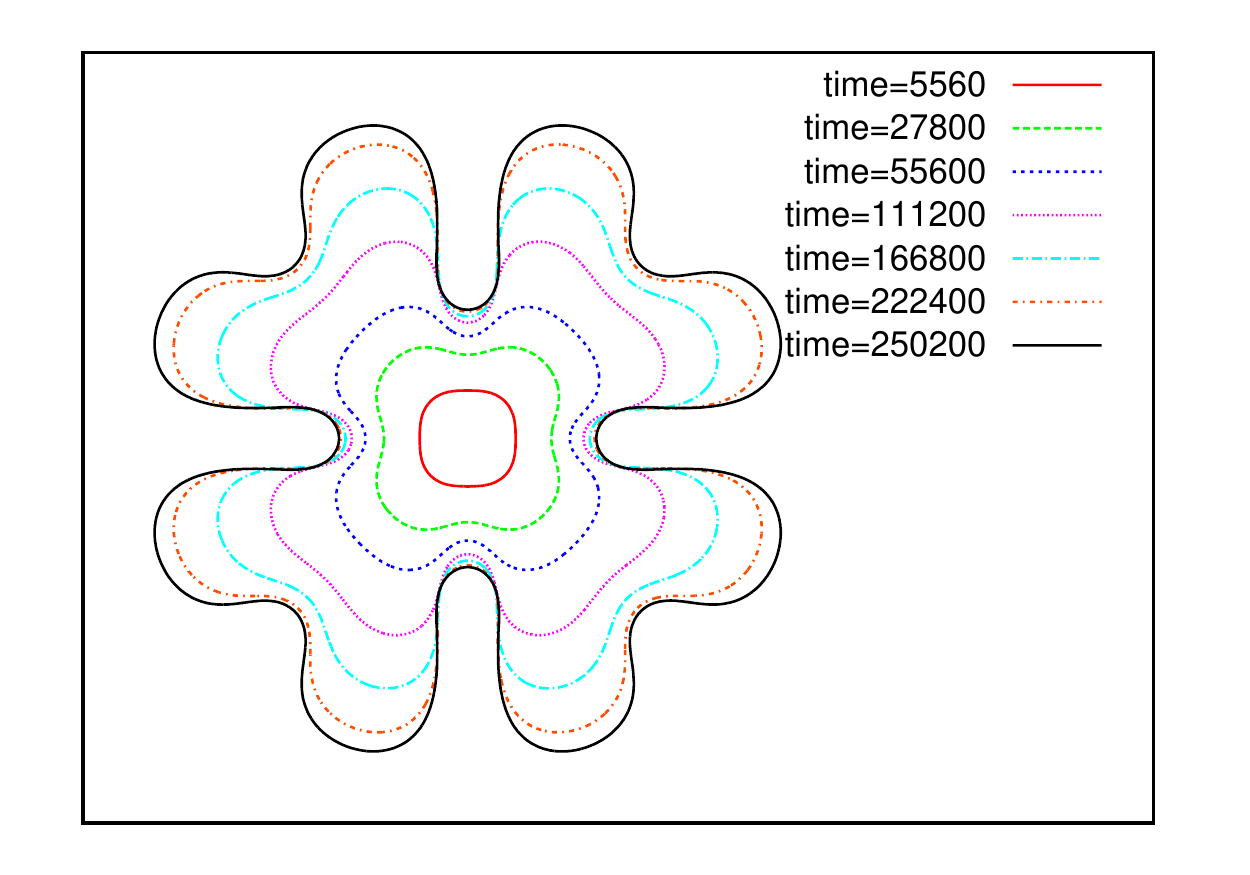}
 \caption{Contours of $\phi=0.5$ at different normalized times 
 showing the development of  a seaweed structure in a system
 with isotropic elastic energy and  interfacial energy. Here, the 
 misfit strain is 1\% and supersaturation is 53\%.}
   \label{sea_weed}
\end{figure}

In the previous section, the effect of varying the anisotropy in 
interfacial energy at 
constant elastic anisotropy i.e. $A_z=3.0$ is elaborated. Here, we 
explore the results upon varying the interfacial energy anisotropy 
while keeping the elastic energy anisotropy having the magnitude 
below one i.e. $A_z=0.5$. An exemplary simulation with the 
combination of both anisotropies is depicted in 
Fig.~\ref{dendrits_evol_sf03}, where the arms of the dendrite are 
oriented along $\langle 10 \rangle$ directions. 
The change in the dendrite shapes upon variation in the strength of 
anisotropy in the interfacial energy is highlighted in 
Fig.~\ref{dendrit_compare_az0.05_sf}. The superimposition of the two 
anisotropies leads to an effective increase in the anisotropy in the 
system.
Thus, with an increase in the magnitude of interfacial energy 
anisotropy, the precipitate tip becomes more sharper and elongated 
along $\langle 10 \rangle$ directions, as highlighted in 
Fig.~\ref{radius_vary_inter_aniso_az_less_than_one}, 
while the dendrite tip velocity increases as the strength in 
anisotropy in the interfacial energy becomes
larger (see Fig.~\ref{velocity_vary_inter_aniso_az_less_than_one}).
The variation of $\sigma^*$ with a combination of different 
anisotropies is portrayed in Fig.~\ref{sigma_az0.5_diff_sf_aniso}, 
where for a given time of evolution, with increase in the strength 
of anisotropy in the interfacial energy the magnitude of $\sigma^*$ 
increases. 
Finally, in order to show that anisotropy in either the interfacial 
energy or the elastic energy is required for the formation of 
dendrite-like structures, we consider the case of isotropic elastic 
and interfacial energies with the supersaturation at 
$\omega=53\%$. As the precipitate grows in size, the instabilities 
at the interface trigger to give rise to a seaweed type structure 
(see Fig.~\ref{sea_weed}), without the selection of a unique tip 
direction or shape. This situation can also occur for cases where 
the influences of the elastic energy anisotropy and the interfacial 
energy anisotropy cancel each other for a certain combination of their
respective strengths.

\section{Conclusions and Outlook}
We have systematically characterized the 
evolution of dendrite-like shapes as a function of elastic 
parameters such as the misfit strain, Zener anisotropy parameter 
, supersaturation in two and three dimensions. 
Although we notice the occurrence of solid-state morphologies 
that resemble dendrites typically occurring during solidification, 
in the presence of coherency stresses, 
the shapes of the tip as well the tip velocity do not achieve a steady 
state. Moreover, the selection constant $\sigma^{*}=2d_0D/R^{2}V$ 
increases linearly with simulation time for all the simulation 
conditions, which is in contrast to the dendrites derived just in the
presence of interfacial energy anisotropy. This lack of steady state is
due to a continuous change in the value of the jump in the elastic 
energy at the tip of the dendrite-like morphology, that increases as
the shape of the tip evolves with time. Consequently, 
the interfacial compositions as well as the Peclet number do not 
saturate. Thus, in the classical sense, in the presence of coherency 
stresses, while the presence of anisotropy leads to the propagation 
of instabilities in well defined directions, there
is no selection of a unique tip shape as in the case of 
solidification. Therefore, 
structures derived in solid-state in the presence of elastic 
anisotropy may only be referred to as dendrite-like. 

Future directions of study involve understanding the 
influence of inhomogeneity in elastic moduli, 
combination of anisotropies in both the misfit strain
and the elastic energy on the tip dynamics and shape. 
Additionally, in multi-component alloys, 
the relative diffusivities of the different elements 
can lead to widely different 
dendritic shapes as the effective capillary length 
may change appreciably. Therefore, the relative ratio 
of the diffusivities becomes an important parameter 
whose influence on the instability 
needs to be established in alloys with three or more components. 
Finally, while we have discussed only precipitate growth, the model 
is generic and may be utilized for studies of late-stage 
coarsening, where an extension to multi-component alloys will 
again bring in exciting new possibilities. 

\section*{Data availability}
The data that supports the results of this study are available 
from the corresponding author upon reasonable request.

\section*{Acknowledgement}
We (BB, TJ, SB, AC) thank the financial support from Department of Science 
and Technology (DST), Government of India (GOI), under the project 
TMD/CERI/Clean Coal/2017/034. 
TJ and SB also acknowledge financial grants and computational support under the project 
S\&T/15-16/DMR-309.01 from DMRL, DRDO, GOI.

\appendix
\section{Elastic free energy density}
\label{appendix_f_elast}
Eqn.~\ref{fel_phi_dend} gives the elastic free energy density, which includes several prefactors, i.e., $Z_3, Z_2, Z_1, Z_0$. 
These prefactors are dependent on particular values of elastic constant in respective phases, i.e., the precipitate and matrix. Their expressions are as follows:
Here $C^{\alpha,\beta}_{11} = C^{\alpha,\beta}_{1111}, C^{\alpha,\beta}_{22} = C^{\alpha,\beta}_{2222}, 
C^{\alpha,\beta}_{44} = C^{\alpha,\beta}_{1212}, C^{\alpha,\beta}_{12} = C^{\alpha,\beta}_{1122}$.
\begin{align*}
 Z_3 &= \left(C^{\alpha}_{11}-C^{\beta}_{11}+C^{\alpha}_{12}-C^{\beta}_{12}\right)\epsilon^{*2},\\
 Z_2 &= (C^{\beta}_{11}-C^{\alpha}_{11})(\epsilon_{xx}+\epsilon_{yy})\epsilon^* 
      +(C^{\beta}_{12}-C^{\alpha}_{12})(\epsilon_{xx}+\epsilon_{yy})\epsilon^* \\
      &+(C^{\beta}_{11}+C^{\beta}_{12})\epsilon^{*2},			     \\
 Z_1 &= \dfrac{1}{2}(C^{\alpha}_{11}-C^{\beta}_{11})(\epsilon_{xx}+\epsilon_{yy})\epsilon^* 
      - C^{\beta}_{11}(\epsilon_{xx}+\epsilon_{yy})\epsilon^*		\\
      &+ (C^{\alpha}_{12}-C^{\beta}_{12})\epsilon_{xx}\epsilon_{yy}
      - \dfrac{1}{2}C^{\beta}_{12}(\epsilon_{xx}+\epsilon_{yy})\epsilon_{yy}\\
      &+ 2(C^{\alpha}_{44}-C^{\beta}_{44})\epsilon^2_{xy},\\
 Z_0 &= \dfrac{1}{2}(C^{\beta}_{11}(\epsilon^2_{xx}+\epsilon^2_{yy})
      + C^{\beta}_{44}\epsilon_{xy} + C^{\beta}_{12}\epsilon_{xx}\epsilon_{yy}).\\
\end{align*}

\bibliography{reference_ssd.bib}

\end{document}